\documentstyle[11pt,aaspp4,psfig,epsf,flushrt]{article}  
\received{??? 1997}
\revised{??? 1997}
\accepted{??? 1997}
\journalid{VOL}{JOURNAL DATE 1997}
\articleid{START PAGE}{END PAGE}
\def\teq#1{$\, #1\,$}                         
                                 
\font\tinyfiverm=cmr5 scaled \magstep0
\font\fiverm=cmr5 scaled \magstephalf
\font\sevenrm=cmr7 scaled \magstephalf
\font\eightrm=cmr8 scaled \magstephalf

{\catcode`\@=11
\gdef\SchlangeUnter#1#2{\lower2pt\vbox{\baselineskip 0pt \lineskip0pt
  \ialign{$\m@th#1\hfil##\hfil$\crcr#2\crcr\sim\crcr}}}}

\def\erg{\varepsilon}
\def\emax{\varepsilon_{\hbox{\fiverm MAX}}}
                                
\def\sigt{\sigma_{\hbox{\fiverm T}}}
\def\taupp{\tau_{\gamma\gamma}}
\def\sigpp{\sigma_{\gamma\gamma}}

\def\flux{{\cal F}}
\def\pprod{\gamma\gamma\to e^+e^-}
\def\gammin{\Gamma_{\hbox{\fiverm MIN}}}   
\def\gamb{\Gamma}   
\def\betab{\beta}   
\def\Thetab{\Theta_{\hbox{\fiverm B}}}   
\def\dover#1#2{\hbox{${{\displaystyle#1 \vphantom{(} }\over{
   \displaystyle #2 \vphantom{(} }}$}}
\def\er{\rm}      
\def\apj{\it Ap. J. \rm}   \def\apjl{\it Astrophys. J. (Lett.) \rm}
\def\apjs{\it Ap. J. Suppl. \rm}
\def\apss{\it Astr. Sp. Sci. \rm}   
\def\asr{\it Adv. Space Res. \rm}   
\def\aap{\it Astron. Astr. \rm}  
\def\aapl{\it Astron. Astr. (Lett.) \rm}
\def\aaps{\it Astron. Astr. Supp.\rm}

\def\mnras{\it Mon. Not. R. astr. Soc. \rm}
\def\nat{\it Nature \rm}

\def\pr{\it Phys. Rev. \rm}    
\def\pra{\it Phys. Rev. A\rm}  
  
\def\ssr{\it Space Sci. Rev.\rm}                         
\begin{document}
%
%
\newcommand{\vol}[2]{$\,$\rm #1\rm , #2.}                 
\newcommand{\figureout}[2]{\centerline{} \vskip 0.0truein
   \centerline{\psfig{figure=#1,width=5.9in}}
    \figcaption{#2}\clearpage } 
\newcommand{\figureoutsmall}[2]{\centerline{} \vskip 0.0truein
   \centerline{\psfig{figure=#1,width=5.5in}} 
    \figcaption{#2}\clearpage } 
\newcommand{\figureouttwo}[3]{\centerline{}  \vskip 1.0truein
   \centerline{\psfig{figure=#1,width=3.0in}\hskip 0.3in
    \psfig{figure=#2,width=3.0in}} \figcaption{#3}\clearpage } 
\newcommand{\figureoutland}[2]{\centerline{}
   \centerline{\psfig{figure=#1,angle=180,width=3.5in}}
    \figcaption{#2}\clearpage } 
\newcommand{\tableout}[4]{
   \vskip 0.3truecm \centerline{\eightrm TABLE #1\rm}
   \vskip 0.2truecm\centerline{\sevenrm #2\rm}
   \vskip -0.3truecm  \begin{displaymath} #3 \end{displaymath} 
   {\parindent=20pt\narrower\noindent \er #4\rm\vskip 0.1truecm } } 
%
%
\title{THE ESCAPE OF HIGH-ENERGY PHOTONS FROM \\
       GAMMA-RAY BURSTS}
   \author{Matthew G. Baring\altaffilmark{1} and Alice K. Harding}
   \affil{Laboratory for High Energy Astrophysics, Code 661, \\
      NASA Goddard Space Flight Center, Greenbelt, MD 20771, U.S.A.\\
      \it baring@lheavx.gsfc.nasa.gov, harding@twinkie.gsfc.nasa.gov\rm}
   \altaffiltext{1}{Compton Fellow, Universities Space Research Association}
   \authoraddr{Laboratory for High Energy Astrophysics, Code 661,
      NASA Goddard Space Flight Center, Greenbelt, MD 20771, U.S.A.}
%
%
%
\begin{abstract}
Eleven bright gamma-ray bursts (GRBs) detected by BATSE have also been
seen at much higher energies by EGRET,  six at energies above 10 MeV.
Most distinctive among these is GRB940217, which includes long
duration, hard gamma-ray emission and the most energetic GRB photon
detection to date, around 18 GeV.  Such observations imply that these
bursts are optically thin to photon-photon pair production at all
observed energies, for target photons both internal and external to the
source.  For bursts more than about 30pc away, internal transparency
can be achieved only if the source is moving with a relativistic bulk
Lorentz factor $\Gamma\gg 1$, or if the radiation is highly beamed.
Early calculations of $\gamma\gamma\to e^+e^-$ considerations for GRBs
were limited to cases of a beam with opening half-angle $\Thetab\sim
1/\Gamma$, or expansions of infinitely thin spherical shells.  This
paper presents our extension of pair production optical depth
calculations in relativistically expanding sources to more general
geometries, including shells of finite thickness and arbitrary opening
angle.  The problem is reduced analytically to a single integral in the
special, but quite broadly applicable case, of observing photons only
along the axis of the expansion.  We find that the minimum bulk Lorentz
factor for the EGRET sources to be optically thin, i.e.  display no
spectral attenuation, is only moderately dependent on the shell
thickness and virtually independent of its opening solid angle if
$\Thetab\gtrsim 1/\Gamma$.  This insensitivity to $\Thetab$ relieves
the commonly-perceived number problem for non-repeating sources at
cosmological distances, i.e. it is not necessary to invoke small
$\Thetab$ to effect photon escape.  The values of \teq{\Gamma}
obtained, typically of the order of 10 for halo bursts and \teq{\gtrsim
100} for sources of cosmological origin, depend somewhat on the choice
of GRB timescale used to determine the expansion size.  Our new limits
on required velocity for given source geometries will aid in placing
realistic constraints on GRB source models.
\end{abstract}
\clearpage

\section{INTRODUCTION}
\label{sec:intro}

Gamma-ray bursts (GRBs) are the brightest sources in the gamma-ray sky
and they may also be among the most distant sources in the Universe.
The discovery by the BATSE detector on the Compton Gamma-Ray
Observatory (CGRO) that the spatial distribution of GRBs is isotropic
and inhomogeneous (Meegan et al. 1992, 1996) suggests that the sources
are either in an extended galactic halo or at cosmological distances.
The level of isotropy of the GRB spatial distribution limits halo
models to core radii of around 50--80 kpc (Hakkila et al. 1995);
tighter constraints are expected for more recent data accumulations,
and Briggs et al. (1996) suggest that a galactic halo shell
distribution must be at least 120 kpc distant.  The observed average
fluxes of GRBs at Earth therefore imply high luminosities for
isotropically emitting sources:  $L \simeq 10^{42 - 43}\,\rm
erg\,s^{-1}$ at a distance of $d = 100$ kpc and $L \simeq 10^{50 -
51}\,\rm erg\,s^{-1}$ at a distance of $d = 1$ Gpc.  In addition, rapid
time variability ($\Delta t \sim$ several ms) is observed in GRB light
curves, whose structural diversity is illustrated, for example in the
BATSE 1B catalogue in Fishman et al. (1994).  This variability implies
a compact source size, which in combination with the high luminosities
yields photon densities that are high enough to make galactic halo or
cosmological GRBs optically thick to photon-photon pair production by
many orders of magnitude.  One would then expect attenuation of the
observed spectrum (perhaps as a quasi-exponential cutoff, trough or
shelf: examples are depicted in Baring and Harding 1997) around the pair
production threshold of 1 MeV if the GRB sources are more distant than
a few kpc and have quasi-isotropic radiation fields (Schmidt 1978,
Epstein 1985).

Yet GRB spectra are observed to extend well beyond 1 MeV and into the
GeV range.  The GRS detector on the SMM satellite first measured
emission in GRB spectra significantly above 1 MeV, often extending up
to 10 MeV (e.g.  Nolan et al.  1983), and in one case up to 80 MeV
(Share et al. 1986).  BATSE routinely observes GRB spectra extending
up to and above 1 MeV.  While most bursts exhibit spectral steepening
at a variety of energies between 50 keV and few hundred keV (Band et
al. 1993; see also Schaefer et al. 1994 for the BATSE 1B spectroscopy
catalogue), a number of bursts display spectral breaks between 500 keV
and about 2 MeV (Schaefer et al. 1992), but no cutoffs.  The EGRET
instrument, also on CGRO, has detected emission above 50 MeV from four
of the brighter GRBs triggered by BATSE, a fifth up to 30 MeV and
another three up to a few MeV; all are consistent with power-law
spectra extending to as high as 1.2 GeV, in the case of GRB930131
(Sommer et al. 1994), and 3.4 GeV for GRB940217 (Hurley et al. 1994).
The GRB940217 source is best known for exhibiting delayed or prolonged
high energy emission, detected 80--100 minutes (i.e. more than one full
earth orbit of CGRO) after the initial trigger, including a photon of
energy 18 GeV (Hurley et al. 1994) that is not markedly inconsistent
with extrapolation of the power-law continuum.  In fact, some evidence
for delayed high energy emission pre-dated GRB940217, with the
observation (Dingus et al. 1994) of a single 10 GeV photon that could
have been associated with GRB910503.  It is clear that, in contrast to
soft gamma-ray repeaters (SGRs), which are now believed to probably be
a separate class of galactic sources (although the classical GRB
behaviour reported by Fenimore, Klebesadel and Laros 1996 of the 5th
March 1979 outburst at early times can support proponents of GRB-SGR
associations), there have been no attenuation-type turnovers or cutoffs
observed in a GRB spectrum.  High energy gamma-ray emission therefore
may be common in bursts, and the EGRET detection rate is consistent
(Dingus 1995, though this inference is subject to poor statistics) with
all bursts emitting above about 30 MeV.  Observed GRB spectra are
therefore in direct conflict with predicted pair production cutoffs for
isotropic emission.

An obvious solution (e.g. Krolik and Pier 1991, Fenimore, Epstein and Ho 1992)
is to allow some anisotropy of the emission, so that
the interaction angles $\theta_{ti}$ of the photons are restricted.  The
threshold for pair production, $\erg_{t} = 2/[\erg_i (1 - \cos
\theta_{ti})]$ where $\erg_t$ and $\erg_i$ are the energies of a
test photon and an interacting photon in units of $m_ec^2$, 
could therefore be increased above the maximum
observed energy.  Beaming of the radiation can be achieved through
relativistic motion: the radiation from a source that is isotropically
emitting in the comoving frame will be beamed 
mostly within an angle of the order of $1/\Gamma$
in the observer's frame, where $\Gamma$ is the bulk Lorentz factor.  For the
case of a small emitting blob moving relativistically, the pair production
optical depth \teq{\taupp} is reduced by a factor $\Gamma^{-(1+2\alpha)}$ below
the optical depth for isotropic radiation, where $\alpha$ is the
photon spectral index (Krolik \& Pier 1991, Baring 1993).   
The minimum bulk Lorentz factors required to make $\tau_{\gamma\gamma} < 1$
in the bright ``superbowl" burst (GRB930131) detected by EGRET 
(Sommer et al. 1994) up to an energy of \teq{\sim 1}GeV are $\Gamma \gtrsim 
10^3$ at a distance of 1 Gpc and $\Gamma \gtrsim 10$ at 30 kpc (Harding 1994, 
Harding \& Baring 1994).  In this case of relativistic beaming within angle 
$1/\Gamma$, the required luminosity \teq{L} at the source is smaller because 
the observed flux, $\phi \sim \Gamma^2 L/ 4\pi d^2$, 
is enhanced by a solid angle factor $\Gamma^2$ (Krolik \& Pier 1991).  
However, the number of sources must be a factor $\Gamma^2$ higher in order 
to account for the observed number of GRBs.  In the case of cosmological
GRBs, this factor could be as high as $10^6$ for the above limits on
$\Gamma$.  This is unacceptably large for many of the proposed models, 
including neutron star--neutron star or neutron star--black hole mergers
(Paczy\`nski 1986; Eichler et al. 1989; Narayan, Piran and Shemi 1991;
M\'esz\'aros \& Rees 1992), failed Type 1b supernovae (Woosley 1993)
and rapid spin-down of high-field millisecond pulsars (Usov 1992), and
hence defines the so-called ``number problem'' for beamed cosmological
bursts.

Source geometries with beaming angles larger than $1/\Gamma$ could ease
this problem if the high energy photons were able to escape.  In fact,
the radiation from GRB sources in the Galactic halo or at cosmological
distances is expected to involve a wind or fireball expanding
relativistically (Paczynski 1986, Goodman 1986, Piran and Shemi 1993).
Fenimore, Epstein \& Ho (1993) have shown that a
relativistically-expanding, thin spherical shell will allow escape of
high energy gamma-rays, because a test photon on the surface of the
shell will not be able to interact with all other emitted photons due
to causality limits.  This arises as a consequence of the transient
nature of the emission, since then only photons emitted within a
``look-back" surface around the test photon will interact to contribute
to the pair production optical depth.  The Fenimore, Epstein \& Ho
(1993) calculation was limited to the case of an infinitely thin shell.

In this study, we have extended the calculation of the pair production
optical depth in GRB sources to the full range of source geometries:
opening angles from $1/\Gamma$ to a spherical expansion, and shells of
arbitrary thickness.  The optical depth for test photons emitted within
the expanding shell will be limited to interaction with other photons
within a ``look-back"  volume.  We present analytic development and
simplification of the pair production optical depth, make detailed
numerical calculations, and derive analytic expressions in various
limits.  The intent is to provide a model-independent evaluation of the
pair production opacity of a relativistically expanding, transient
gamma-ray source whose emitted spectrum extends above observed
energies.  Using our results, we derive estimates for the minimum bulk
Lorentz factors required for source transparency in those GRBs detected
by EGRET at high energies.  These limits are largely insensitive to the
source opening angle provided that it exceeds \teq{1/\Gamma}.  This
reflects the strong impact of causality in determining the optical
depth, and clearly renders the number problem for cosmological source
models a non-issue: the total negation of this number problem for a
wide range of expansion geometries is a principal conclusion of this
paper.  A detailed description of the source geometry and the
derivation of an analytic form for the pair production optical depth
(and associated limiting cases) for infinite power-law source spectra
are presented in Section~2; there the general quintuple integral
expression for \teq{\taupp} [see Eq.~(\ref{eq:dtaupp1})] is expediently
reduced to a comparatively simple single integration in
Eq.~(\ref{eq:tauppfin}), a principal result of this research,
rendering our developments quite amenable to various observational and
theoretical applications.  Section~3 is devoted to the application of
these results to EGRET bursts and the estimation of minimum bulk
Lorentz factors in these sources, including a discussion of the
behaviour of our results in relevant parameter spaces and various
issues pertaining to our calculations.  Readers more interested in the
applications and implications of our calculations than in the detailed
derivations presented in Section~2 should note that
Eq.~(\ref{eq:tauppfin}) is the final form for the pair production
optical depth, which should be used in conjunction with the
normalization specified via the flux in Eq.~(\ref{eq:fluxasymp}).


\section{THE PAIR PRODUCTION OPTICAL DEPTH}
\label{sec:taupp}

The generic picture of a gamma-ray burst that is considered here
assumes the photon source (i.e. region of emission) to be expanding
with constant and homogeneous bulk Lorentz factor \teq{\Gamma}, with
opening angle \teq{2\Thetab} about some axis and thickness \teq{\Delta
R} (constant throughout) in the observer's frame, and with the initial
condition (at time \teq{t=0}) that the source's inner radius is
\teq{R_0}.  The expansion therefore traces a conical volume and can
assume a variety of geometries such as solid cones, solid spheres or
spherical shells, at any given time, depending on the values of the
input parameters \teq{\Thetab} and \teq{\Delta R/R_0}.  The constancy
of \teq{\Gamma} in time is a convenient assumption that is not strictly
valid during early epochs of the expansion if \teq{\Delta R/R_0} is not
much less than unity: when \teq{\Delta R/R_0\gtrsim 1}, the expansion
initially resembles a quasi-isotropic and almost stationary radiation
gas in the observer's frame.  The adiabatic redistribution of momenta
that naturally occurs in expanding (and ``inert'') photon gases is
therefore neglected.  We also opt to ignore the consideration of
possible dynamic acceleration or deceleration of the underlying plasma,
since such dynamics are quite model-dependent.

Suppose that a test (i.e. potentially observable) photon, with energy
\teq{\erg_t m_ec^2}, is emitted at time \teq{t=0} from the inner radius
\teq{R_0} of the shell and moves through the source to eventually
escape and reach the observer.  We opt for test photons originating at
the back of the expansion throughout this paper; starting these photons
closer to the outer surface will reduce the optical depth they
encounter by a factor of order unity, so that the results we obtain
will be qualitatively representative of general initial positions for
test photons within the source. If the angle cosine between the test
photon's momentum and position vectors is \teq{\mu_t}, then the radial
distance \teq{r_t} of the test photon from the center of the expansion,
at any time \teq{t}, is
\begin{equation}
   r_t\; =\; R_0+\mu_t ct\quad .
 \label{eq:rt}
\end{equation}
The overall geometry of this source expansion is depicted in
Fig.~\ref{fig:geometry}a and is discussed in more detail below.  Radial
propagation of the test photon along the axis of the expansion
corresponds to \teq{\mu_t=1}. An important assumption about the
expansion that is made in this paper is that no photons are present
prior to time \teq{t=0}.  This ``switch-on'' stipulation restricts the
photon population that can causally interact with the test photon at
early times, and indeed mimics burst temporal behaviour; it is
anticipated that the details of the switch-on will have only a
quantitative rather than a qualitative influence on the results
presented.  The objective of this section is to derive an analytic
expression for the pair production optical depth for this expanding
source geometry.  Consideration of the influence of the plasma that is
present in the emission region on the photons it generates will be
omitted from this analysis.

The optical depth for two-photon pair production \teq{\gamma\gamma\to
e^+e^-} can be obtained from well-known expressions for the reaction
rate \teq{R_{\gamma\gamma}} for interactions of photons in a single
population (e.g. see Eq.~(27) of Weaver 1976, or Eq.~(7) of Stepney and
Guilbert 1983).  In the case where one photon is a test photon of
dimensionless energy \teq{\erg_t}, the optical depth, differential in
the distance \teq{r_t} that the test photon travels, is (e.g. see
Eq.~(7) of Gould and Schreder 1967)
\begin{equation}
   \dover{d\taupp (\erg_t)}{dr_t}\; =\;\dover{1}{2}\int \sigpp (\chi )\,
   (1-\mu_{ti})\, n(\erg_i,\,\mu_{ti};\, r_t)\, d\erg_i\, d\mu_{ti}\quad .
 \label{eq:dtaudrt}
\end{equation}
Here subscripts \teq{i} denote quantities of the photon that interacts
with the test photon, \teq{\chi =\sqrt{\erg_t\erg_i(1-\mu_{ti})/2}} is
the center-of-momentum (CM) frame energy scaled by \teq{m_ec^2}, and
\begin{equation}
   \sigpp (\chi )\; =\;\dover{\pi r_e^2}{\chi^6}\,\biggl\{
   \bigl( 2\chi^4+2\chi^2-1\bigr)\,\log_e\Bigl\lbrack \chi+\sqrt{\chi^2-1}
   \;\Bigr\rbrack - \chi (1+\chi^2)\sqrt{\chi^2-1}\;\biggr\}\quad .
 \label{eq:sigpp}
\end{equation}
is the Lorentz-invariant pair production cross-section (e.g. see
Eq.~(13--40) of Jauch and Rohrlich, 1980), where \teq{r_e=e^2/m_ec^2}
is the classical electron radius.  Hereafter, all photon energies will
be assumed to be dimensionless, being scaled by \teq{m_ec^2}.  Also,
\teq{\mu_{ti}=\cos\theta_{ti}} is the angle between the momentum
vectors of the test and interacting photons, and
\teq{n(\erg_i,\,\mu_{ti};\, r_t)} is the source photon density
distribution at the position of the test photon.  The factor of 1/2 in
Eq.~(\ref{eq:dtaudrt}) is the standard correction for double-counting
in interactions of identical particles; it is omitted in the
calculations of Epstein (1985; see his Eq.~2.8) and Zdziarski (1984),
who treat the test photons as a separate population from the
interacting photons.

It is instructive to identify the typical energy \teq{\erg_i} of
photons that interact with test photons at \teq{\erg_t}, specifically
for an expansion of bulk Lorentz factor \teq{\Gamma} that spawns
power-law photon spectra, the conditions pertaining to the analysis of
this paper.  For such test photon energies, the minimum possible energy
of the interacting photons, defined by the pair production threshold,
is \teq{\sim\Gamma /\erg_t} in the rest frame of the expansion, and of
the order of \teq{\Gamma^2/\erg_t} in the stationary observer's
reference frame.  The pair production cross-section in
Eq.~(\ref{eq:sigpp}) peaks not far above threshold, and since the
optical depth is a convolution of this cross-section and the spectrum,
which is a strongly decreasing function of energy, it is clear that the
typical energy of an interacting photon is usually never far above
threshold, i.e. around \teq{\Gamma^2/\erg_t} in the observer's frame.
This result holds regardless of the source photon density provided
phase space near threshold is accessible, which always is the case for
infinite power-law spectra.

\subsection{Source Geometry}
\label{sec:geometry}

Before developing the calculation of the pair production optical depth
it is both necessary and elucidating to elaborate the details of the
expansion geometry and define useful spatial variables.  The general
picture of the expansion at any instant is given in
Fig.~\ref{fig:geometry}a; the definitions of the test and interacting
photons' spatial and angular variables are depicted in
Fig.~\ref{fig:geometry}b and are now enunciated.  The radius of the
test photon at any time is given by Eq.~(\ref{eq:rt}) and the angle
between the radius vector (OT in Fig.~\ref{fig:geometry}b) to the test
photon and the cone axis (OZ in Fig.~\ref{fig:geometry}b), which
bisects \teq{2\Thetab}, is defined as \teq{\Theta_t}
(\teq{\leq\Thetab}).  If the expansion has an inner radius \teq{R} at
time \teq{t} and has a thickness \teq{\Delta R} that is constant in
time, then the test photon remains within the expanding volume only
when
\begin{equation}
   R\; =\; R_0+\betab ct\;\leq\; r_t\;\leq\; R_0+\Delta R+\betab ct\quad ,
 \label{eq:rtlim}
\end{equation}
where \teq{\beta =\sqrt{1-1/\Gamma^2}} is the bulk velocity of the
expansion (in units of \teq{c}).  In general, the angle between the
test photon's position and momentum vectors is \teq{\theta_t
=\arccos\mu_t}; however, unless otherwise stated, in subsection~2.1 and
subsequent portions of the paper, \teq{\theta_t} will be assumed to be
zero so that \teq{r_t=R_0+ct}; this will be a specialization to the
most salient case of radial propagation of test photons.
Eq.~(\ref{eq:rtlim}) leads to the determination of the time \teq{t_e}
the test photon takes to escape the expanding plasma:
\begin{equation}
   t_e\; =\;\dover{\Delta R}{c(1-\betab )}\quad .
 \label{eq:te}
\end{equation}
Of course, calculation of the pair production optical depth will
involve an integration over times \teq{0\leq t <\infty}, including when
the test photon has escaped the expanding plasma.

\placefigure{fig:geometry}

The test photon interacts with photons at positions within some
causally-connected look-back volume.  Detailed considerations of such
look-back regions for relativistic expansions are presented in Rees
(1966), Fenimore, Epstein and Ho (1993).  Suppose that a typical
interacting photon is located at a radius \teq{r_i} with the angle
between its position vector (OI) and the expansion axis (OZ) being
\teq{\Theta_i} (see Fig.~\ref{fig:geometry}b).  Such an interacting
photon was emitted at time \teq{t_i} (\teq{0<t_i<t}) and at a distance
\teq{r_{ti}=c(t-t_i)} from the test photon.  Let \teq{\Theta_{ti}} be
the angle between the radius vectors (OT and OI in
Fig.~\ref{fig:geometry}b) of the test and interacting photons.  Further
let the angle between the momentum vectors of the test and interacting
photons be \teq{\theta_{ti}}.  If the line (TI in
Fig.~\ref{fig:geometry}b) between the positions of the test and
interacting photons makes an angle \teq{\theta_i} with the radius
vector of the interacting photon, then simple geometrical analysis
gives
\begin{eqnarray}
   r_{ti}^2\; &=&\; r_t^2+r_i^2-2r_tr_i\cos\Theta_{ti}\quad ,\nonumber\\
   r_i^2\; &=&\; r_t^2 + r_{ti}^2 - 2r_t r_{ti}
   \cos (\theta_{i}-\Theta_{ti})\quad , \vphantom{\Bigl(} \\
   r_t^2\; &=&\; r_i^2 + r_{ti}^2 + 2r_i r_{ti}\cos \theta_i
   \quad . \nonumber
 \label{eq:cosrule}
\end{eqnarray}
These relationships will be used to develop the integrations in the
expression for the optical depth that is derived in the following
subsection.  In general, if \teq{\phi} is the angle between the planes
defined by the test photon momentum and position vectors and the
momentum vectors of the test and interacting photons, then spherical
trigonometry yields
\begin{equation}
   \cos (\theta_i-\Theta_{ti})\; =\;\cos\theta_t\cos\theta_{ti}-
   \sin\theta_t\sin\theta_{ti}\cos\phi\quad .
 \label{eq:cosines}
\end{equation}
However, when specializing to the case of radial propagation of test
photons, the photon momenta lie in the OTI plane so that \teq{\phi
=\pi} and \teq{\theta_i=\theta_{ti}+\Theta_{ti}}; this simplification
will be used in subsequent sections.

The geometry of the source defined above, restricts the values of the
variables \teq{r_{ti}} and \teq{\theta_{ti}} that prescribe the
position of the interacting photons.  These restrictions arise because
the interacting photons can only be emitted from the portion of the
region that the expanding plasma occupied at the time of emission that
is causally connected to the test photon.  In the radial direction, the
volume that the emission region occupies at time \teq{t=t_i} is
specified simply by
\begin{equation}
   R_i\;\leq\; r_i\;\leq\; R_i+\Delta R\quad ,\quad R_i\, =\, 
   R_0+\betab ct_i \quad ,
 \label{eq:rilim}
\end{equation}
where the relation \teq{r_{ti}=c(t-t_i)} implies that
\teq{R_i=(1-\betab )R_0+\betab (r_t-r_{ti})}.
With the aid of Eq.~(\ref{eq:cosrule}), this radial constraint becomes
\begin{equation}
   \dover{r_t^2+r_{ti}^2-(R_i+\Delta R)^2}{2r_tr_{ti}}\;\leq\;
   \cos (\theta_i-\Theta_{ti})
   \;\leq\;\dover{r_t^2+r_{ti}^2-R_i^2}{2r_tr_{ti}}\quad ,
 \label{eq:cosbound}
\end{equation}
which, for the case of radially-propagating test photons (i.e. when
\teq{ \theta_i-\Theta_{ti}=\theta_{ti}}), is a compact representation
of the limits to the \teq{\mu_{ti}} integration in
Eq.~(\ref{eq:dtaudrt}).  The values of \teq{r_{ti}} that are achievable
are further constrained by the causality condition
\begin{equation}
   0\;\leq\; r_{ti}\;\leq\; ct=\dover{r_t-R_0}{\mu_t}\quad .
 \label{eq:rtilim}
\end{equation}
In fact, this restriction automatically guarantees that
\teq{(r_t^2+r_{ti}^2-R_i^2)/(2r_tr_{ti})\geq 1} and therefore that the
right hand inequality of Eq.~(\ref{eq:cosbound}) is always satisfied.
Physically this occurs because interacting photons emitted at the rear
of the expanding volume can never catch the test photon.

The constraints that the emission volume places on angles are simply
enunciated.  The requirement that the interacting photon be within the
expanding shell imposes no restriction on the radial variables, but
does constrain \teq{\Theta_i} according to
\begin{equation}
   0\;\leq\;\Theta_i\;\leq\;\Thetab\quad .
 \label{eq:thetailim}
\end{equation}
A similar condition limits the values of \teq{\Theta_t} at \teq{t=0}.
Eq.~(\ref{eq:thetailim}) restricts the allowable azimuthal angles of
the interacting photon for off-axis propagation of test photons;
discussion of this restriction is deferred to section 2.2.2 below,
specifically Eq.~(\ref{eq:etadef}).  In addition, the range of
\teq{\Theta_{ti}} is clearly bounded by the expansion geometry.
Inspection of Fig.~\ref{fig:geometry}b reveals that the maximum
possible value of \teq{\Theta_{ti}} is \teq{\Thetab +\Theta_t}; with
the aid of Eq.~(\ref{eq:cosrule}) this becomes the global
\teq{\Theta_{ti}} constraint
\begin{equation}
   \cos\Theta_{ti}\; 
   =\;\dover{1-\rho\mu_{ti}}{\sqrt{1-2\rho\mu_{ti}+\rho^2}}
   \;\geq\;\zeta \;\equiv\;\cos\Bigl(\min\{\pi ,\;\Thetab 
   +\Theta_t\}\Bigr)\quad ,
 \label{eq:costhetati}
\end{equation}
where \teq{\rho =r_{ti}/r_t} is a scaling of \teq{r_{ti}} that proves
convenient in the algebraic manipulations of this paper (see
immediately below). This can be inverted to find the ranges of
acceptable values of \teq{\mu_{ti}=\cos\theta_{ti}}, as is outlined
below in Eq.~(\ref{eq:mulim}).  This concludes the presentation of the
general forms for the constraints the source geometry places on the
spatial variables defined; specific developments in subsequent sections
are made according to algebraic need.

It shall prove convenient to define three dimensionless variables that
will facilitate the algebraic developments of the optical depth that
are performed in this paper:
\begin{equation}
   \rho\; =\;\dover{r_{ti}}{r_t}\quad ,\quad \psi\; =\;\dover{r_t}{R_0}
   \quad ,\quad s\; =\;\dover{1}{1-\rho}\;\sqrt{1-2\rho\mu_{ti}+\rho^2}
   \quad .
 \label{eq:variables}
\end{equation}
The first two of these are scaling transformations that define the
coupling of length scales in the expansion; they are used to reduce the
number of integrations in the optical depth over spatial variables by
one [see Eq.~(\ref{eq:taupp2})].  Note that the causality condition in
Eq.~(\ref{eq:rtilim}) yields \teq{0\leq\rho\leq 1} for \teq{\mu_t=1}.
The definition of \teq{s} is effectively an alternative to the angular
variable \teq{\mu_{ti}} that proves convenient in reducing the
integration of interacting photon angles analytically (see
Section~2.3).  These three variables will be referred to extensively in
subsequent equations.

Hereafter, this paper addresses the special case of radial propagation
of test photons along the axis of the expansion, so that
\teq{\mu_t=\cos\theta_t=1} and \teq{\Theta_t=0}.  While this choice is
motivated by the simplifications it introduces to the analysis, it is
concordant with the goal of obtaining representative estimates of
source bulk Lorentz factors that are consistent with GRB observations.
Relativistically expanding sources contribute most of their observable
emission along the direction of motion, corresponding to radial and
on-axis propagation of the test photons.  Off-axis (i.e. non-radial)
emission will mostly be outside the peak of beamed radiation, and
therefore form only a minor part of the observable flux of gamma-rays.
Hence we expect that photons produced somewhat off-axis will contribute
minimally to the observed flux, and therefore be largely irrelevant to
the determination of minimum bulk Lorentz factors.  A brief discussion
of this specialization, in the light of the results obtained, is
presented in Section~4. Note that while ``limb'' photons move on
average at large angles to the beamed photon population in their
locale, the phase space that connects them causally to the remaining
photon population is small; it is not clear whether or not limb photons
will have enhanced pair production optical depths relative to
line-of-sight radiation.

\subsection{Analytic Reduction of The Optical Depth}
\label{sec:reduction}

The differential optical depth in Eq.~(\ref{eq:dtaudrt}) can be
developed once the photon distribution function
\teq{n(\erg_i,\,\mu_{ti};\, r_t)} is known, and eventually an
integration over \teq{r_t} will be performed to obtain the total
optical depth \teq{\taupp (\erg_t)}.  The form that
\teq{n(\erg_i,\,\mu_{ti};\, r_t)} takes depends on basic assumptions
about the expanding photon gas.  In this paper, the \it rate of photon
emission \rm is taken to be constant in time and space after time
\teq{t=0} (following, for example, Fenimore et al. 1993, but in
contrast to the uniform density choice made by Harding and Baring
1994), but is zero for \teq{t<0}, and is isotropic only in the comoving
frame of the expansion.  This simplifying assumption is made for its
convenience; it is unrealistic since it may be acausal for some initial
conditions (e.g. the instantaneous ``switch-on'' over a finite
volume).  Observed temporal behaviour in individual sources is somewhat
chaotic (e.g. see the BATSE 1B catalogue, Fishman et al. 1994), so the
assumption of constant emissivity is not truly accurate, depending on
the timescale of specific consideration.  However, in conjunction with
the source geometry prescribed here, a constant emissivity may be able
to approximately produce  global properties of bursts, such as longer
average decay timescales than rise times (Nemiroff et al., 1994;
Norris, et al. 1994; Mitrofanov 1995).  Our assumption of a
temporally-invariant emissivity after switch-on is an appropriate
approximation for the objectives of this paper, since only estimates of
the bulk relativistic motion in gamma-ray bursts are at present
obtainable as the origin of bursts is still uncertain.  Note that the
rate of emission might be expected to decline in time due to expansion
effects such as adiabatic cooling and a decrease in plasma density.

An immediate consequence of this approximation is that the photon
distribution in the comoving frame of the expansion is anisotropic even
for isotropic photon emission, due to radiative transfer effects in
finite source volumes.  This assumption is clearly different from the
premise of Gould and Schreder (1967), from which the bulk motion
analyses of Krolik and Pier (1991) and Baring (1993) are derived, who
all effectively assumed that the photon \it distribution \rm is
everywhere and at all times isotropic (Krolik and Pier 1991 actually
invoked the equivalent assumption of isotropy of the photon intensity
in the comoving frame).  Isotropy of the photon distribution is an even
more elementary approximation: it is perhaps less realistic than the
assumption of isotropic emission rate that is made here, given that
isotropic radiation fields are usually best generated in optically
(Thomson) thick media, contrary to the basis of these pair-production
calculations.  Isotropic emissivities require only that the supporting
particle population is isotropic, and that there is no other preferred
direction in the emission region, such as that imposed by the presence
of a magnetic field.

\subsubsection{Photon density and the observed flux}

For a constant rate of photon emission, the photon distribution at any
point is simply obtained by adapting standard radiative transfer
results (e.g. see Rybicki and Lightman, 1979) to the consideration of
photon densities.  The conservation of photon numbers and volume
elements along light rays in optically thin environments automatically
yields conservation of the number density, from which the transfer
equation yields
\begin{equation}
   n(\erg_i,\,\mu_{ti};\, r_t)\; =\;\dover{1}{c}\int {\dot n}(\erg_i,\,
   \mu_i;\, r_i)\,dr_{ti}\, \dover{d\phi_{ti}}{2\pi}\quad .
 \label{eq:ndef}
\end{equation}
Here \teq{r_{ti}} traces out the photon path through the look-back
volume and \teq{{\dot n}(\erg_i,\,\mu_i;\,r_i)/2\pi} is the rate of
photon emission in the observer's frame, per unit solid angle, at the
position of the interacting photon (the point labelled I in
Fig.~\ref{fig:geometry}b).  The arguments of the emission rate in the
integrand are implicitly functions of the spatial variables relating to
the test photon position, i.e.  \teq{\mu_i=\cos\theta_i=\mu_i(r_t,\,
t_{ti},\,\mu_{ti})} and \teq{r_i=r_i(r_t,\, t_{ti},\,\mu_{ti})} are
defined by Eq.~(\ref{eq:cosrule}), and determined by the geometry in
Fig.~\ref{fig:geometry}b.  Here \teq{\mu_{ti}=\cos\theta_{ti}}.
Throughout the following analysis, it is assumed that
\teq{\theta_t=0}.  The angular dependence of the distribution and
emission rate has been retained because of the highly anisotropic
conditions encountered in this calculation.  The azimuthal angle
\teq{\phi_{ti}} is defined to be the angle between the OZT and OZI
planes in Fig.~\ref{fig:geometry}b.  Since the emission rate is
independent of position within the expanding volume, no azimuthal
dependence appears in the arguments of \teq{{\dot
n}(\erg_i,\,\mu_i;\,r_i)}.  Note that Harding and Baring (1994) combine
Eqs.~(\ref{eq:dtaudrt}) and~(\ref{eq:ndef}) into a single expression
for the optical depth in their Eq.~2; their result is mildly erroneous,
being a factor of \teq{2/\pi} too small.

If the source generates isotropic radiation in the comoving frame of
the expansion with a power-law emission spectrum \teq{{\dot
n}_c(\erg_c,\,\mu_c) \propto \erg_c^{-\alpha}}, where \teq{\erg_c} and
\teq{\mu_c} are the photon energy and emission angle in the comoving
frame, then it follows that the photon emission rate in the observer's
frame takes the form
\begin{equation}
   {\dot n}(\erg ,\,\mu ,\; r)\; =\; {\dot {\cal N}}\,\erg^{-\alpha}\, 
   (1-\betab\mu )^{-(1+\alpha )}\quad ,\quad \erg_-\,\leq\,\erg\,\leq\,
   \erg_+\quad ,
 \label{eq:ndotdef}
\end{equation}
where \teq{\erg_{\pm}} define the bounds to the observed source
spectrum.  This form is derived from the Lorentz transformation
relationships \teq{\erg_c =\gamb\erg (1-\betab \mu )} and \teq{\mu_c
=(\mu -\betab )/ (1-\betab\mu )} and their associated Jacobian
\teq{d\erg_c\, d\mu_c/ d\erg\, d\mu =\lbrack\gamb (1-\betab \mu
)\rbrack^{-1}}, given that the total photon number \teq{{\dot n}(\erg
,\,\mu ,\; r)d\erg d\mu\, dVdt} is a Lorentz invariant.  Note that in
Eq.~(\ref{eq:ndotdef}), a factor of \teq{1/\gamb} has been absorbed in
the definition of \teq{\dot {\cal N}}.

The value of the coefficient \teq{{\dot {\cal N}}} in
Eq.~(\ref{eq:ndotdef}) can be determined by computing the photon flux
\teq{\flux} at large distances from the source and equating the result
to the observed flux in individual GRBs.  Specifically, the flux at
test photon energy \teq{\erg_t =1} (i.e. 511 keV) is
\begin{equation}
   \flux\; =\; c\int d\mu_{ti}\,\mu_{ti}\, n(\erg_t\! =\! 1,\,\mu_{ti} ,
   \; r_t) \; =\; {\dot {\cal N}} \int \dover{\mu_{ti}}{
   (1-\betab\mu_i )^{\alpha +1}}\, d\mu_{ti}\, dr_{ti}\quad ,
 \label{eq:fluxdef}
\end{equation}
after integrating Eq.~(\ref{eq:ndef}) over azimuthal angles.  The units
of \teq{{\cal F}} are photons per square centimetre per second.  For
the moment assume that there are no angular restrictions to the phase
space, i.e. \teq{\Thetab =\pi/2}.  Further, note that in this integral
the test photon acts purely as a position marker and can be taken to be
on-axis without loss of generality, i.e. \teq{\Theta_t=0}.  The angle
cosine \teq{\mu_i} in the distribution is given by
Eq.~(\ref{eq:muidef}) below, and the limits on the integrals are
defined by the radial and causality constraints in
Eqs.~(\ref{eq:cosbound}) and~(\ref{eq:rtilim}).  At large distances
from the source, these restrictions imply that \teq{r_{ti}/r_t\approx
1} and \teq{\mu_{ti}\approx 1}, as is obvious from the description of
the geometry in Fig.~\ref{fig:geometry}a.  In fact, defining \teq{\psi
=r_t/R_0}, then Eqs.~(\ref{eq:cosbound}) and~(\ref{eq:rtilim}) can be
expressed as \teq{\psi_-\leq\psi\leq\psi_+}, where \teq{\psi_{\pm}} are
given in Eq.~(\ref{eq:psi+-}).  The evaluation of the integrals in
Eq.~(\ref{eq:fluxdef}) can be facilitated by changing variables thus:
the \teq{\mu_{ti}} integration is performed using the variable
\teq{s=\sqrt{1-2\rho\mu_{ti}+\rho^2}\, /(1-\rho )} for \teq{\rho
=r_{ti}/r_t}(\teq{\approx 1}), and the \teq{\rho} integration is
calculated using the variable \teq{t=\psi (1-\rho )}.  For \teq{\psi\gg
1}, then \teq{\mu_i\approx 1/s}.  Reversing the order of integration
yields \teq{1\leq t\leq (1-\beta + \Delta R/R_0)/(s-\beta )} and a
trivial result for the \teq{t} integral, so that
\begin{equation}
   \flux\; =\;\dover{{\dot {\cal N}}}{3}\,\dover{R_0^3}{d^2} 
   \int^{s_+}_1 \dover{s^{\alpha + 2}\,ds}{(s-\beta )^{\alpha + 4}}\,
   \Biggl\{\;\biggl(1-\beta +\dover{\Delta R}{R_0}\biggr)^3
   -(s-\beta    )^3\;\Biggr\}
   \quad ,\quad s_+\, =\, 1+\dover{\Delta R}{R_0}\quad .
 \label{eq:flux}
\end{equation}
Here \teq{r_t} is set equal to the distance \teq{d} between the source
and the observer, and the flux naturally obeys an inverse-square law:
\teq{\flux \propto d^{-2}}.  Note that the algebraic manipulations here
closely resemble those applied to the expression for the optical depth;
for this reason, detail is minimized here and is deferred to section
2.2.2 below.

In general, the result in Eq.~(\ref{eq:flux}) can be expressed in terms
of a hypergeometric function of two variables, however it is simple to
evaluate it directly by numerical integration.  In the special cases
where the expansion is, in the comoving frame, a thin spherical shell
with \teq{\Delta R/R_0\ll 1-\beta}, or it is a filled shell with
\teq{\Delta R/R_0\gg 1-\beta}, it becomes analytically tractable,
giving
\begin{equation}
   \flux\; \approx\;\dover{{\dot {\cal N}}}{2}\,\dover{(\Delta R)^2}{d^2}\,
   \cases{\dover{R_0}{(1-\beta )^{\alpha + 2}}\;\; ,&$\quad \dover{\Delta
   R}{R_0} \ll 1-\beta\quad .\vphantom{\Biggl\{}$\cr
   \dover{2\Delta R}{3(\alpha + 3)\beta}\; \biggl\{ \dover{1}{(1-\beta 
   )^{\alpha +3}} -1\biggr\}\;\; ,&$\quad \dover{\Delta R}{R_0}
   \gg 1-\beta\quad .\vphantom{\Biggl\{}$\cr}
 \label{eq:fluxasymp}
\end{equation}
The quadratic dependence of \teq{{\cal F}} in \teq{\Delta R} when
\teq{\Delta R/R_0\ll 1} reflects the two dimensions of the integration
in Eq.~(\ref{eq:fluxdef}).  At the same time, \teq{{\cal F}} is
independent of \teq{R_0} when \teq{R_0\ll\Delta R} since the inner
radius contributes negligibly to the source volume.  The value of
\teq{\dot {\cal N}} is therefore determined by equating the flux in
Eq.~(\ref{eq:flux}) to that observed at 511 keV for sources with
unbroken power-law spectra, or by a power-law extrapolation of the high
energy spectrum down to 511 keV for those sources with spectral breaks
above this energy (e.g. GRB940217; see the discussion in Section~3).

The modification to the expression in Eq.~(\ref{eq:flux}) for the flux
that is induced by reduction of \teq{\Thetab} below \teq{\pi /2} is
straightforward.  The considerations of angular constraints in
subsection 2.1 lead to the simple expression for the restriction of the
(\teq{\mu_{ti}}, \teq{\rho}) phase space in Eq.~(\ref{eq:costhetati}).
Since the flux is observed at infinity, \teq{\mu_{ti} \approx 1} and we
take \teq{\Theta_t=0} for the flux calculation.  Then
Eq.~(\ref{eq:costhetati}) immediately implies that \teq{s\leq 1/\zeta},
for \teq{\zeta} defined just below in Eq.~(\ref{eq:fluxangle}), and it
quickly follows that sub-spherical expansions generate a flux given by
Eq.~(\ref{eq:flux}) but with
\begin{equation}
   s_+\; =\; \min\biggl\{ 1+\dover{\Delta R}{R_0}\,
   ,\;\dover{1}{\zeta}\biggr\}
   \quad ,\quad \zeta\; =\;\cos\Bigl(\min\{\pi,\;\Thetab\}\Bigr)
 \label{eq:fluxangle}
\end{equation}
substituted as the upper limit to the integral.  Clearly then, the
angular restrictions play no role in determining the flux until the
solid angle [\teq{2\pi (1-\cos\Thetab )}] of the expansion becomes
comparable to the fractional shell thickness \teq{\Delta R/R_0}.

\subsubsection{Optical depth for radial propagation of test photons}

The differential optical depth in Eq.~(\ref{eq:dtaudrt}) can now be
expressed in more explicit form using Eq.~(\ref{eq:ndef}) and the form
of the photon emission rate in Eq.~(\ref{eq:ndotdef}), evaluated at the
position of the interacting photon:
\begin{equation}
   \dover{d\taupp (\erg_t)}{dr_t}\; =\;\dover{{\dot {\cal N}}}{4\pi c}
   \int \sigpp (\chi )\,\erg_i^{-\alpha}\,\dover{1-\mu_{ti}}{
  (1-\betab\mu_i)^{\alpha +1}}\, d\erg_i\, d\mu_{ti}\, dr_{ti}\,
   d\phi_{ti}\quad .
 \label{eq:dtaupp1}
\end{equation}
Here \teq{{\dot {\cal N}}} has been removed from the integration
because it is assumed to depend only on \teq{\Thetab} and be
independent of the position within the source.  Hereafter it will be
assumed that the emission spectrum in Eq.~(\ref{eq:ndotdef}) has a
large or infinite range (\teq{\erg_+\gg\erg_-}), for which it is
possible evaluate the \teq{\erg_i} integration separately, and
analytically.  Specifically, it is permissible to change variables in
Eq.~(\ref{eq:dtaupp1}) to the CM frame energy variable \teq{\chi
=\sqrt{\erg_t\erg_i(1-\mu_{ti})/2}} via \teq{4\chi d\chi=\erg_t
(1-\mu_{ti}) d\erg_i}, following the procedure of Gould and Schreder
(1967; see also Baring, 1993), and perform the integration of the
cross-section separately.  Consequently, the differential optical depth
assumes the form
\begin{equation}
   \dover{d\taupp (\erg_t)}{dr_t}\; =\; {\dot {\cal N}}\,
   \dover{\sigt}{\pi c}\,
   \erg_t^{\alpha -1}\, \dover{{\cal H}(\alpha )}{2^{\alpha+2}}
   \int \dover{ (1-\mu_{ti})^{\alpha}}{(1-\betab\mu_i)^{\alpha +1}}
   \, d\mu_{ti}\, dr_{ti}\, d\phi_{ti}\quad ,
 \label{eq:dtaupp2}
\end{equation}
where the integration of the cross-section over \teq{\chi} is
\begin{equation}
   {\cal H}(\alpha )\; =\;\dover{4}{\sigt}\int^{\infty}_1\chi^{1-2\alpha}
   \sigma_{\gamma\gamma}(\chi )\, d\chi\;\approx\; \dover{7}{6\alpha^{5/3}}
   \quad .
 \label{eq:Hdef}
\end{equation}
The approximation in Eq.~(\ref{eq:Hdef}) was obtained (see Baring,
1993) from Eq.~B6 of Svensson (1987), who also gave the exact analytic
expression for the integral; it is accurate to better than 1\% for
\teq{1 <\alpha <7}.  The angle cosine \teq{\mu_i=\cos\theta_i} of the
interacting photon that appears in Eq.~(\ref{eq:dtaupp2}) can be
determined explicitly from the geometry in Fig.~\ref{fig:geometry}b
using Eq.~(\ref{eq:cosrule}); eliminating the variables \teq{r_i} and
\teq{\Theta_{ti}} gives (for \teq{\theta_t=0})
\begin{equation}
   \mu_i\; =\;\dover{\mu_{ti}-\rho}{\sqrt{1-2\rho\mu_{ti}+\rho^2}}\quad .
 \label{eq:muidef}
\end{equation}
The scaling variable \teq{\rho =r_{ti}/r_t} (\teq{<1} for \teq{\mu_t=1})
will be of use in the development of the optical depth integration.

In applications where the emission spectrum is of finite energy range,
the modification for performing the \teq{\erg_i} integration has been
developed by Gould and Schreder (1967) and Krolik and Pier (1991).  Low
energy spectral turnovers or cutoffs are unlikely to be influential in
pair production continuum attenuation calculations applied to gamma-ray
bursts that have bulk motions with large Lorentz factors \teq{\Gamma}
and maximum energies under 100 MeV (Baring, 1994).  While turnovers are
observed as photon energies drop into the BATSE range, sharp cutoffs
can presumably be only in the soft X-ray range where photons interact
with gamma-rays of energy much more than 1 GeV to produce pairs.
Therefore any suppression of \teq{\pprod} continuum attenuation that is
introduced by a low-energy cutoff is unlikely to be observed by EGRET.
However, for observations in the super-GeV range, spectral structure in
the BATSE range becomes quite relevant to opacity determinations
(Baring and Harding 1997), and is discussed briefly at the end of
Section~3.  Introduction of high energy cutoffs to the spectrum of
interacting photons is also largely irrelevant to this investigation
because the most energetic EGRET source photons predominantly interact
with photons at energies considerably below the maximum detected.

The azimuthal integration in Eq.~(\ref{eq:dtaupp2}) can be performed
after establishing the restrictions the source geometry places on
\teq{\phi_{ti}}.  For the interacting photon to be in the expanding
``conical shell,'' the radial restriction in Eq.~(\ref{eq:cosbound}) is
independent of \teq{\phi_{ti}}.  In contrast, the angular constraint
\teq{0\leq\Theta_i\leq\Thetab} in Eq.~(\ref{eq:thetailim}), which is
independent of time, does restrict the allowable azimuthal angles.
Assuming that the plasma emits uniformly at any one time, this angular
constraint results in an analytic determination of the azimuthal
integration, because at a given radius each azimuthal angle within the
cone of expansion \it contributes equally.  \rm  Since the azimuthal
angle \teq{\phi_{ti}} is defined to be the angle between the planes OZT
and OTI (see Fig.~\ref{fig:geometry}b), then considerations of
spherical trigonometry yield
\begin{equation}
   \cos\Theta_i\; =\;\cos\Theta_t\cos\Theta_{ti} 
   +\sin\Theta_t\sin\Theta_{ti}
   \cos\phi_{ti}\;\geq\;\cos\Thetab\quad 
 \label{eq:costhetai}
\end{equation}
when \teq{\vert\Theta_t\pm\Theta_{ti}\vert\leq\Thetab}.  When this
condition is not satisfied, the interacting photon is always within the
cone defined by the expansion and all values \teq{0\leq\phi_{ti}\leq
2\pi} are permitted.  It follows that the \teq{\phi_{ti}} integration
has limits defined by \teq{\vert\cos\phi_{ti}\vert\leq\eta}, where
\begin{equation}
   \eta\; =\;\cases{ \dover{\cos\Thetab 
   -\cos\Theta_t\cos\Theta_{ti}}{\sin\Theta_t\sin\Theta_{ti}}
   \;\; ,& $\quad \Thetab -\Theta_t\leq\Theta_{ti}\leq\Thetab 
   +\Theta_t\;\;$,\cr
   \vphantom{\Bigl(} -1\;\; ,& $\quad 0\leq\Theta_{ti}\leq\Thetab 
   -\Theta_t\;\;$. \cr}
 \label{eq:etadef}
\end{equation}
The restriction \teq{\Theta_{ti} >\Thetab -\Theta_t} is necessary to
achieve \teq{\eta > -1}, and as \teq{\Theta_{ti} \to\Thetab +\Theta_t}
then \teq{\eta\to 1} and the permitted \teq{\phi_{ti}} phase space
shrinks to zero.  When \teq{\Theta_t\to 0}, (i.e. the test photon is on
the axis of the expansion), \teq{\eta\to -1} for all permissible
\teq{\Theta_{ti}}. This simple special case will be assumed throughout
subsequent sections of the paper.  The \teq{\phi_{ti}} integration in
Eq.~(\ref{eq:dtaupp2}) is then trivially evaluated to give
\teq{2\arccos\eta}.  Equation~(\ref{eq:dtaupp2}) can then be integrated
over the test photon position \teq{r_t} to give the total optical
depth:
\begin{equation}
   \taupp (\erg_t)\; =\; {\dot {\cal N}}\,\dover{\sigt}{\pi c}\,
   \erg_t^{\alpha -1}\,\dover{{\cal H}(\alpha )}{2^{\alpha+1}} \int 
   \arccos\eta\,\dover{ (1-\mu_{ti})^{\alpha}}{(1-\betab\mu_i)^{\alpha +1}}
   \, d\mu_{ti}\, dr_{ti}\, dr_t\, \quad ,
 \label{eq:taupp1}
\end{equation}
and the value for \teq{\eta} when \teq{\vert\Theta_t\pm\Theta_{ti}\vert
\leq\Thetab} becomes (for \teq{\theta_t=0})
\begin{equation}
   \eta\; =\;\dover{\cos\Thetab \sqrt{1-2\rho\mu_{ti}+\rho^2}\;
   -\cos\Theta_t(1-\rho\mu_{ti})}{\rho\sin\Theta_t\sqrt{1-\mu_{ti}^2}}
   \quad ,
 \label{eq:eta2}
\end{equation}
where the substitution for \teq{\Theta_{ti}} has been effected using
Eq.~(\ref{eq:cosrule}), the variable \teq{\rho} is defined in
Eq.~(\ref{eq:variables}), and the sine rule applied to triangle OTI in
Fig.~\ref{fig:geometry}b.

For the moment, consider \teq{4\pi} steradian expansions, where
\teq{\Thetab =\pi} and angular restrictions to the interaction phase
space do not enter the analysis.  Evaluation of the triple integral in
Eq.~(\ref{eq:taupp1}) can be facilitated by changing variables via the
scaling transformations defined in Eq.~(\ref{eq:variables}), so that
\teq{\mu_i} and \teq{\eta} are rendered independent of \teq{r_t} (see
Eqs.~(\ref{eq:muidef}) and~(\ref{eq:eta2})).  A complete set of
dimensionless variables has therefore been chosen and reversing the
order of integration so that the \teq{\psi} integration is performed
first yields an analytic reduction of the optical depth to a double
integral:
\begin{equation}
   \taupp (\erg_t)\; =\; {\dot {\cal N}}\,\dover{\sigt}{\pi c}\, R_0^2\,
   \erg_t^{\alpha -1}\,\dover{{\cal H}(\alpha )}{2^{\alpha+2}} 
   \int^1_0 d\rho \int^1_{\mu_{\hbox{\tinyfiverm MIN}}}
   d\mu_{ti} \;\bigl(\psi_+^2-\psi_-^2\bigr)\;
   \arccos\eta\,\dover{ (1-\mu_{ti})^{\alpha}}{(1-\betab\mu_i)^{
   \alpha +1}}\quad .
 \label{eq:taupp2}
\end{equation}
The value of \teq{\mu_{\hbox{\fiverm MIN}}} is given in
Eq.~(\ref{eq:mumin}).  Here \teq{\psi_{\pm}} are the limits of the
\teq{\psi} integration (\teq{\psi_-\leq\psi\leq\psi_+}), and are
determined from the radial constraint in Eq.~(\ref{eq:cosbound}) and
the causality condition in Eq.~(\ref{eq:rtilim}):
\begin{equation}
   \psi_-\; =\;\dover{1}{1-\rho}\quad ,\quad \psi_+\; =\;\dover{ 1-\betab + 
   \Delta R/R_0}{\sqrt{1-2\rho\mu_{ti}+\rho^2}-\betab (1-\rho )}\quad ,
 \label{eq:psi+-}
\end{equation}
where \teq{0\leq\rho\leq 1} defines a ``look-back'' volume.  If the
expansion is at least fully hemispherical (\teq{\Omega\geq 2\pi}), then
\teq{\eta =-1}, and the \teq{\mu_{ti}} integration is over a range
determined by the condition \teq{\psi_+\geq\psi_-}.  It follows from
Eq.~(\ref{eq:psi+-}) that the range of \teq{\mu_{ti}} permitted in
Eq.~(\ref{eq:taupp2}) is \teq{\mu_{\hbox{\fiverm MIN}}\leq\mu_{ti}
\leq 1}, where, for \it any \rm \teq{\beta <1},
\begin{equation}
   \mu_{\hbox{\fiverm MIN}}\; =\;
   \cases{\dover{1}{2\rho}\biggl\{ 1+\rho^2-(1-\rho )^2\Bigl( 1+
   \dover{\Delta R}{R_0}\Bigr)^2\biggr\}\;\; ,& $\quad 
   \dover{\Delta R}{2R_0+\Delta R}\, <\,\rho\,\leq\, 1 \;\; ,$\cr
   -1\vphantom{\Bigl(}\;\; ,& $\quad$ otherwise.\cr}
 \label{eq:mumin}
\end{equation}
Clearly \teq{\mu_{\hbox{\fiverm MIN}}\leq 1} is always true.  This
restriction indicates that the look-back volume is not spherical
because of the presence of edges to the expansion in the radial
direction.  The form of the pair production optical depth in
Eq.~(\ref{eq:taupp2}) is not generally reducible to a simpler form, and
is ready for numerical evaluation in cases where \teq{\Thetab\geq\pi
/2}.  However, analytic development is possible in the special case of
propagation of test photons along the axis of the expansion (i.e.
\teq{\Theta_t=0}), which will be treated in the next subsection.

Consider now the additional restrictions on the integration phase space
due to a reduction in the expansion opening angle \teq{\Thetab}.  The
way the expansion has been defined automatically precludes any
necessity to treat cases where \teq{\Thetab >\pi /2}, since they reduce
to the \teq{\Thetab =\pi /2} case.  This arises because switching on
the expansion at \teq{t=0} implies that interacting photons from the
back hemisphere \teq{\Theta_i>\pi /2} (for \teq{\Theta_t=0}) can never
reach the test photon originating in the forward hemisphere, and
therefore cannot contribute to the optical depth.  This switch-on
stipulation is a reasonable approximation to a real burst, and the
contribution to the optical depth from photons originating in the back
hemisphere is expected to be strongly suppressed due to the
relativistic nature of the expansion.  When \teq{\Thetab <\pi}, both
\teq{\mu_{ti}} and the azimuthal angle \teq{\phi_{ti}} are restricted.
The constraint on \teq{\Theta_{ti}} in  Eq.~(\ref{eq:costhetati}) can
be inverted to find the ranges of acceptable values of  \teq{\mu_{ti}}
in terms of \teq{\rho}, which defines how the reduction of the opening
angle of the expansion restricts the \teq{(\mu_{ti},\,\rho )} phase
space:
\begin{mathletters}
\begin{equation}
   -1\;\leq\;\mu_{ti}\;\leq\;\mu_-\quad ,\quad \mu_+\;\leq\;\mu_{ti}
   \;\leq\; 1\quad ,
\end{equation}
where
\begin{equation}
   \mu_{\pm}\; =\;
   \cases{\dover{1}{\rho}\Bigl\{ 1-\zeta^2\pm\vert \zeta \vert
   \sqrt{\rho^2+\zeta^2-1}\Bigr\}\;\; ,& $\quad 
   \rho\geq\sqrt{1-\zeta^2}$\cr
   \sqrt{1-\zeta^2}\vphantom{\biggl(}\;\; ,& 
   $\quad\rho <\sqrt{1-\zeta^2}$\cr}
\end{equation}
\label{eq:mulim}
\end{mathletters}
Therefore when \teq{\rho <\sqrt{1-\zeta^2}} all values of
\teq{\mu_{ti}} are permitted, since then the angular boundary to the
expansion lies outside the look-back volume defined by \teq{\rho}. The
azimuthal restrictions are reflected in the value of \teq{\eta} in
Eq.~(\ref{eq:eta2}), and by a similar analysis, the boundary where
\teq{\eta} increases above \teq{-1} is also defined by
Eq.~(\ref{eq:mulim}), but with \teq{\zeta\to \cos (\Thetab
-\Theta_t)}.  The ranges in Eq.~(\ref{eq:mulim}) are, of course,
subject to the \teq{\mu_{\hbox{\fiverm MIN}}\leq\mu_{ti}} limitation
that is imposed by  radial considerations.

In the case where the test photon propagates radially along the axis of
the expansion (i.e. \teq{\Theta_t=0}),  which will be the focal point
of all subsequent developments in this paper, the double integral
expression for the optical depth in Eq.~(\ref{eq:taupp2}) can be
manipulated into a form that is more convenient for numerical
computation, where one integral can be expressed in terms of familiar
hypergeometric functions.  The analytical approach is similar to the
derivation of the photon flux in Eq.~(\ref{eq:flux}).  The
\teq{\mu_{ti}} integration is expressed in terms of the variable
\teq{s}, defined in Eq.~(\ref{eq:variables}), and the \teq{\rho}
integration can be performed first using the variable \teq{t=1-\rho}.
This change of variables leads to the range
\begin{equation}
   1\;\leq\; s\;\leq\; 1+\dover{\Delta R}{R_0}\quad ,
 \label{eq:slim}
\end{equation}
for the variable \teq{s}, which can be easily deduced from the
requirement that \teq{\mu_{ti}<1} and the condition that
\teq{\psi_+>\psi_-} in Eq.~(\ref{eq:psi+-}).  For specific \teq{s}
within this range, inversion of the restriction in Eq.~(\ref{eq:mumin})
leads to the upper bound \teq{t_+=2/(1+s)} for \teq{t}, which can also
be obtained equivalently from the radial constraint in
Eq.~(\ref{eq:cosbound}).  The lower bound to \teq{t} can be derived
directly from the angular constraint in Eq.~(\ref{eq:costhetati}); the
result gives
\begin{equation}
   \max\biggl\{ 0\; ,\dover{2(\zeta s-1)}{s^2-1}\,\biggr\}\; =\;
   t_-\;\leq\; t\;\leq\; t_+\; =\;\dover{2}{1+s}\quad ,
 \label{eq:tlim}
\end{equation}
where \teq{\zeta} is defined in Eq.~(\ref{eq:fluxangle}).  Since
\teq{\zeta \leq 1}, it follows that \teq{t_+\geq t_-} so that only one
range arises for the \teq{t}-integration.  This simplicity does not
arise if the \teq{s}-integration is performed first since \teq{t_-} has
a maximum of \teq{1-\sqrt{1-\zeta^2}} that can then yield two
integration ranges for \teq{s} for some \teq{t}.  Note that \teq{t_-}
exceeds zero only when \teq{s >1/\zeta}.  Hence, the angular constraint
only impacts the calculation of the optical depth when \teq{\zeta
>R_0/(R_0+\Delta R)}, a situation identical to that arising in the
treatment of the source flux.  Remembering that \teq{\arccos\eta =\pi}
for this case of radial and axial propagation of test photons, the
optical depth in Eq.~(\ref{eq:taupp2}) develops to the form
\begin{eqnarray}
   \vphantom{\biggl(}
   \taupp (\erg_t)\; &=&\; {\dot {\cal N}}\,\dover{\sigt}{c}\, R_0^2\,
   \erg_t^{\alpha -1}\,\dover{{\cal H}(\alpha )}{2^{\alpha+1}}
   \int_1^{1+\delta} ds\; \biggl\lbrack\,
   \biggl(\dover{1-\beta +\delta}{s-\beta}\biggr)^2\,
    -1\,\biggr\rbrack\nonumber\\
   &\times &\;\; s^{2+\alpha}\, {(s^2-1)}^{\alpha}\;
   \int^{t_+}_{t_-} dt\,\dover{t^{2\alpha}}{{\cal D}^{1+\alpha}}
   \quad ,\quad {\cal D}\, =\, 2(s-\beta )+t\lbrack (1+s^2)\beta 
   -2 s\rbrack\quad ,\vphantom{\Biggl(} 
 \label{eq:taupp3}
\end{eqnarray}
for fractional shell thickness \teq{\delta =\Delta R/R_0}.

The angular constraint in Eq.~(\ref{eq:costhetati}) is unimportant only
when \teq{t_-=0}.  In general, this is not so, and the
\teq{t}-integration in Eq.~(\ref{eq:taupp3}) can be written as the
difference between integrations over the ranges \teq{[0,\, t_+]} and
\teq{[0,\, t_-]}.  Therefore two terms appear, each of which can be
manipulated in a similar fashion.  Consider first the integration over
\teq{[0,\, t_-]}: the \teq{t}-integration can be rewritten via the
substitution \teq{t=t_-(1-q)}, leading to the transformation
\begin{equation}
   s^{2+\alpha}\, {(s^2-1)}^{\alpha}\;
   \int^{t_-}_0 dt\,\dover{t^{2\alpha}}{{\cal D}^{1+\alpha}}\; =\;
   \dover{2^{\alpha}\sigma^{\alpha + 1}}{(1+\beta )^{\alpha 
   +1}}\,\dover{s}{s-1}\,
   \biggl(\dover{\zeta s-1}{s-1}\biggr)^{\alpha}
   \int_0^1 dq\,\dover{(1-q)^{2\alpha}}{(1-\sigma\lambda q)^{\alpha +1}}
   \quad .
 \label{eq:trans}
\end{equation}
Here \teq{\lambda =\lambda (s,\,\beta )} and \teq{\sigma =
\sigma (s,\,\beta ,\,\zeta )} are given by
\begin{equation}
   \lambda\; =\; \dover{(1+s^2)\beta -2s}{s(s-1)(1+\beta )}\quad ,\quad
   \sigma\; =\;\dover{\zeta s-1}{(s-1)-(1-\zeta )s\lambda}\quad .
 \label{eq:lambdasig}
\end{equation}
It can be shown that \teq{\lambda} is a monotonically decreasing
function of \teq{s} with the range \teq{-\infty <\lambda <\beta
/(1+\beta )}, and furthermore that \teq{0\leq\sigma\leq 1}.  When
\teq{\zeta\to 1} (i.e. \teq{\Thetab\to 0}), then \teq{t_-\to t_+} and
\teq{\sigma\to 1}, so that the result for t-integration over the range
\teq{[0,\, t_+]} is recovered.  Putting the two terms together and
defining the Heaviside step function \teq{{\cal H}_{\hbox{\fiverm
S}}(x)} to be unity when \teq{x>0} and zero otherwise, the optical
depth can therefore be written in the form
\begin{mathletters}
\begin{equation}
   \taupp (\erg_t)\; =\; {\dot {\cal N}}\,\dover{\sigt}{2c}\, R_0^2\,
   \erg_t^{\alpha -1}\,\dover{{\cal H}(\alpha )}{(1+\beta)^{\alpha+1}}
   \int_1^{1+\delta} ds\, \dover{s}{s-1}\,\biggl\lbrack\,
   \biggl(\dover{1-\beta +\delta}{s-\beta}\biggr)^2\, -1\,\biggr\rbrack
   \; {\cal J}_{\alpha}(s;\,\beta ,\,\zeta )
\end{equation}
for \teq{\delta =\Delta R/R_0}, with
\begin{equation}
   {\cal J}_{\alpha}(s;\,\beta ,\,\zeta )\;\equiv\; {\cal G}_{\alpha}
   (\lambda )-
   {\cal H}_{\hbox{\fiverm S}}\Bigl(s-\dover{1}{\zeta}\Bigr)\;
   \biggl(\dover{\zeta s-1}{s-1}\biggr)^{\alpha}\,\sigma^{\alpha + 1}\,
   {\cal G}_{\alpha}(\sigma\lambda )
\end{equation}
\label{eq:tauppfin}
\end{mathletters}
\noindent\hskip -12pt
The function \teq{{\cal G}_{\alpha}(z)} is just the integral that
appears in Eq.~(\ref{eq:trans}), and is expressible in terms of the
standard hypergeometric function \teq{F(\alpha ,\,\beta ;\, \gamma;\,
z)} using the identity 3.197.3 in Gradshteyn and Ryzhik (1980):
\begin{equation}
   {\cal G}_{\alpha}(z)\;\equiv\;
   \int_0^1 dq\,\dover{(1-q)^{2\alpha}}{(1-z q)^{\alpha +1}}\; =\;
   \dover{1}{1+2\alpha}\, F(\alpha +1,\, 1 ;\, 2\alpha +2;\, z)\quad .
 \label{eq:Gdef}
\end{equation}
The numerical evaluation of \teq{{\cal G}_{\alpha}} is straightforward,
and is described in the Appendix.  An alternative form for the optical
depth can be derived by leaving the \teq{t}-integration in
Eq.~(\ref{eq:taupp3}) as one integral over the range \teq{[t_-,\, t_+]}
and rescaling the integration variable.  This second form yields
Eq.~(\ref{eq:tauppfin}a) but with an alternative representation of the
function \teq{{\cal J}_{\alpha}(s;\,\beta ,\,\zeta )}:
\begin{equation}
   {\cal J}_{\alpha}(s;\,\beta ,\,\zeta )\; =\; \int_0^{q_+} dq\,
   \dover{(1-q)^{2\alpha}}{(1-\lambda q)^{\alpha +1}}\quad ,\quad 
   q_+\, =\,\min \biggl\{ 1,\;\dover{s(1-\zeta )}{s-1}\biggr\}\quad .
 \label{eq:Jaltdef}
\end{equation}
While slightly less convenient than Eq.~(\ref{eq:tauppfin}) for
numerical evaluation of the optical depth, this second form is useful
when obtaining results in the limiting case of small opening angles,
treated in subsection 2.3.3 below.  Computationally, if the series
representation of \teq{{\cal G}_{\alpha}} described in the Appendix is
used, Eq.~(\ref{eq:tauppfin}) is a single integral that is relatively
simple to evaluate.  Note that the integrand does not diverge at
\teq{s=1} due to the behaviour of \teq{{\cal G}_{\alpha}(\lambda )}
there (see the Appendix).

It is important to emphasize that the optical depth in
Eq.~(\ref{eq:tauppfin}) was obtained under the assumptions that the
test photon originates at the rear of the expansion and propagates
radially outward.  With this specification, Eq.~(\ref{eq:tauppfin}) is
intended to approximate a variety of possibilities for test photon
initial conditions.  Non-radial test photon motion will increase the
optical above that in Eq.~(\ref{eq:tauppfin}), primarily because of
increased angles with interacting photons, however an observer's unique
perspective will strongly bias against such situations for relativistic
expansions.  On the other hand, test photons can plausibly originate
closer to the surface of the expansion than \teq{R_0}, diminishing the
optical depth accordingly.  Suppose that the test photon starts at
radius \teq{R_0+\nu\Delta R} at time \teq{t=0}.  Then all interacting
photons inside this radius are \it always \rm causally-disconnected
from the test photon because it propagates radially.  Hence the region
interior to \teq{R_0+\nu\Delta R} is irrelevant to the optical depth
calculation, and a new initial condition can be defined, with
\teq{R_0+\nu\Delta R} and \teq{R_0+\Delta R} denoting the relevant
inner and outer radii of the expansion.  The test photon is now at the
rear of this section of the conical shell, and the optical depth
computation can be repeated entirely with the aid of the substitution
\begin{equation}
   R_0\;\to\; R_0+\nu\Delta R\; ,\quad \Delta R\; \to\; (1-\nu )\Delta R
 \label{eq:nutrans}
\end{equation}
without any additional manipulation.  This elementary transformation
propagates all the way through the development so that the optical
depth for test photons starting at arbitrary positions
\teq{R_0+\nu\Delta R} in the expansion is simply from
Eq.~(\ref{eq:tauppfin}) by the substitution \teq{\delta\to (1-\nu
)\delta /(1+\nu\delta )}, a very attractive scheme of generalization.
Note that the normalizing flux in Eq.~(\ref{eq:flux}) is unaffected by
these considerations.

This concludes the analytic development of the optical depth formula,
which is ready for numerical computation and certainly is much more
amenable than the quintuple integral in Eq.~(\ref{eq:dtaupp1}).  Before
presenting such computations (in subsection 2.4 below), it is
instructive to examine the optical depth for some limiting cases of the
source geometry.

\clearpage

\subsection{Approximations in Limiting Cases}
\label{sec:limits}

There are four special cases where it is both possible and useful
to obtain analytic limits to the pair production optical depth: these
correspond to the thin-shell limit, thick-shell or filled-sphere
expansions, narrow beams, and a stationary photon gas.

\subsubsection{The thin-shell limit}

The expression in Eq.~(\ref{eq:tauppfin}) is in suitable form for the
derivation of the optical depth in certain special cases.  The first of
these is the limit of a thin, spherical shell for the expansion, where
\teq{\delta =\Delta R/R_0\ll 1-\beta} (i.e. the shell is also thin in
the comoving frame of the expansion), and \teq{\delta\ll 1-\zeta} so
that the angular constraints are immaterial.  The \teq{{\cal
G}_{\alpha}(\sigma \lambda)} term in Eq.~(\ref{eq:tauppfin}) is
therefore absent.  As noted in the Appendix, in the \teq{s\to 1}
limit,  \teq{(1+2\alpha ){\cal G}_{\alpha}(\lambda)} approaches
\teq{F(\alpha +1,\, 1 ;\, 2\alpha +2;\, 1)/(1-\lambda) }, which with
the aid of identity 9.122 of Gradshteyn and Ryzhik (1980) leads to the
limit \teq{{\cal G}_{\alpha}(\lambda)\to 1/[\alpha (1-\lambda )]
\approx (s-1)(1+\beta )/[2\alpha (1-\beta )]}.  It is then elementary
to derive the result \teq{\taupp (\erg_t)\propto {\dot {\cal N}}\,
(\Delta R)^2}.  The two powers of \teq{\Delta R} are due to the thin
shell severely restricting the spatial extent of the \teq{r_t} and
\teq{r_{ti}} integrations [e.g. see Eq.~(\ref{eq:dtaupp1})].  Such
behaviour is largely meaningless until the dependence of the formula
for the flux is factored in.  Remembering that in this limit, for a
fixed observed flux, Eq.~(\ref{eq:fluxasymp}) yields \teq{{\dot {\cal
N}}\propto 1/(\Delta R)^2}, the optical depth is virtually independent
of \teq{\Delta R}, as is expected.  Explicitly, we obtain
\begin{equation}
   \taupp (\erg_t)\;\approx\; \dover{\sigt}{2c}\,\dover{d^2}{R_0}\; 
   {\cal F}\; \dover{{\cal H}(\alpha )}{\alpha}\,\biggl( 
   \dover{1-\beta }{1+\beta} \biggr)^{\alpha} \,\erg_t^{\alpha -1}
   \quad ,\quad\dover{\Delta R}{R_0}\,\ll\,
   \min\bigl\{ 1-\beta ,\, 1-\zeta\bigr\}\quad .
 \label{eq:thinshell}
\end{equation}
This thin-shell limit displays a strong inverse dependence on the bulk
Lorentz factor \teq{\Gamma =1/\sqrt{1-\beta^2}} of the expansion, as is
evident from Fig.~\ref{fig:taupp}: typically \teq{\alpha\sim 2-3} for
EGRET bursts (e.g. see Table~\ref{Table2}).  This dependence is
enhanced by one or two powers of \teq{\Gamma} that appear in the
determination of \teq{R_0}.  This case is closest to the work of
Fenimore, Epstein and Ho (1993), who treat test photons coming from an
entire spherical shell (i.e.  including the limb regions).  As argued
in Section~2.1, the major contribution to the optical depth comes from
test photons originating in near-axis environs so that the differences
between conclusions made using a formula like Eq.~(\ref{eq:thinshell})
and the work of Fenimore et al. (1993) are marginal.

\subsubsection{Filled-sphere expansions}

The other extreme class of expansions from the point of view of the
radial dimension contains filled spheres initially, i.e. thick shells
(in the comoving frame) with \teq{\Delta R/R_0\gg 1-\beta}.  Again we
shall ignore the impact of narrowing the solid angle down and demand
\teq{\Thetab =\pi /2} here, so that only one term in
Eq.~(\ref{eq:tauppfin}) contributes to the optical depth.  In this
limit, inspection of Eq.~(\ref{eq:tauppfin}) soon reveals that the
dominant contribution to the integral is for \teq{s-1\lesssim
1-\beta}.  Then it follows that \teq{\lambda\approx -(1-\beta )/(s-1)},
since relativistic expansions with \teq{\beta\approx 1} are considered
here.  Choosing \teq{\lambda /(\lambda -1)} as the integration variable
and using the transformation of the hypergeometric function in
Eq.~(\ref{eq:Gdef2}) yields a result that is proportional to the
integral in Eq.~(\ref{eq:Gint1}).  As in the thin-shell limit, here
\teq{\taupp (\erg_t)\propto (\Delta R)^2}, reflecting the
dimensionality of the integrations.  The filled sphere limit of
Eq.~(\ref{eq:fluxasymp}) indicates that an observed flux gives a
volume-determined photon injection rate \teq{{\dot {\cal N}}\propto
1/(\Delta R)^3}, so that the overall expression for the optical depth
is a declining function of the expansion thickness: for
\teq{\beta\approx 1}, the asymptotic result
\begin{equation}
   \taupp (\erg_t)\;\approx\; \dover{3\sigt}{4c}\, \dover{d^2}{\Delta R}\,
   {\cal F}\, (1-\beta )^{\alpha +1}\, \erg_t^{\alpha -1}\, (\alpha +3)\,
   \dover{{\cal H}(\alpha )}{2^{\alpha}}\,\biggl[\,\dover{1}{\alpha} + 
   \dover{2}{\alpha -1}\,\Bigl\{\psi (2\alpha ) - \psi (\alpha )-1\Bigr\}
   \,\biggr]
 \label{eq:fullsphere}
\end{equation}
is derived, where \teq{\psi (x)} is the derivative of the logarithm of
the Gamma function, defined in Eq.~(\ref{eq:Gint1}).  Again, a strong
dependence on the bulk Lorentz factor of the expansion is evident.  The
moderate decline of \teq{\taupp (\erg_t)} with \teq{\Delta R} reflects
the fact that large regions are less compact for a given source
luminosity.  Note that extremely-filled spheres with \teq{\Delta R
>R_0} are not really discussed in this paper; these seem unlikely to be
realized in bursts and require an alternative coupling of length scale
to time variability, via \teq{\Delta R=c\Delta t}, as is mentioned
below.

\subsubsection{Narrow beam expansions}

The case of small opening angles of the expansion is also of interest.
Since the emphasis here is on axial viewing perspective, this limiting
case corresponds to \teq{\zeta\equiv\cos\Thetab\approx 1}.  Specifically,
this narrow beam limit satisfies \teq{1-\zeta\ll 1-\beta}, so that the
reduction in opening angle dominates the causality limitations, and
also \teq{1-\zeta\ll\delta\equiv\Delta R/R_0}.  However, such an
identification with small solid-angles is not sufficient to define
narrow beam cases; as will be evident shortly, the size of the opening
angle itself is also quite pertinent.  The most suitable form of the
optical depth for development here is using Eq.~(\ref{eq:tauppfin}a)
combined with the representation in Eq.~(\ref{eq:Jaltdef}).   The
\teq{s}-integration then consists of a range \teq{1\leq s\leq 1/\zeta}
over which the volume is not opening angle-limited, and this
contributes of the order of \teq{1-\zeta} to the optical depth
expression in Eq.~(\ref{eq:tauppfin}a).  However, it turns out that the
range \teq{1-\zeta\ll s-1/\zeta\ll 1} dominates the contribution to the
integral.  This leads to a simplification for the integral in
Eq.~(\ref{eq:Jaltdef}) for \teq{{\cal J}_{\alpha}(s;\,\beta ,\,\zeta
)}.  Yet the integral is not trivial since the parameter \teq{\lambda
q_+} is not necessarily small; in fact \teq{-\lambda q_+\approx
s(1-\zeta ) (1-\beta )/(s-1)^2}.  Reversing the order of integration
and using 3.194 of Gradshteyn and Ryzhik (1980) leads to the result
that \teq{\taupp (\erg_t) \propto {\dot {\cal N}}\sqrt{1-\zeta}}.  This
result is subject to the requirements that \teq{\sqrt{1-\zeta}\ll
1-\beta} and \teq{\sqrt{1-\zeta}\ll\delta}.  Clearly the reduction in
angular phase space for pair creation is not solid-angle limited, but
rather constrained purely by the maximum size of the angle between the
momentum vectors of the test and interacting photons --- this scales as
\teq{\Thetab}.  Such a linear dependence on \teq{\Thetab} is manifested
in both this asymptotic limit for Eq.~(\ref{eq:tauppfin}a) and the
range of its validity.  When \teq{\delta\ll\sqrt{1-\zeta}} then the
optical depth in Eq.~(\ref{eq:tauppfin}a) no longer behaves like a
narrow beam limit, but rather like a thin-shell limit as discussed
above, where \teq{\taupp (\erg_t)\propto {\dot {\cal N}}\,\delta^2}.

\placetable{Table1}

\newpage

The other part of the narrow beam calculation relates to the flux
normalization factor \teq{{\cal F}} in Eq.~(\ref{eq:flux}).  Remembering
the limitation in Eq.~(\ref{eq:fluxangle}), the computation of
\teq{{\cal F}} is straightforward, yielding a solid-angle limited flux:
\teq{{\cal F}\propto {\dot {\cal N}}(1-\zeta )}.  This intuitively
obvious result is applicable only when \teq{1-\zeta\ll 1-\beta} and
\teq{1-\zeta\ll\delta}.  Such a solid-angle-limited range differs from
the requirements imposed by Eq.~(\ref{eq:tauppfin}), thereby
complicating the consideration of the narrow beam limit.  To aid
understanding of this limit, the various dependences of
Eq.~(\ref{eq:flux}) and Eq.~(\ref{eq:tauppfin}), and the resulting
behaviour of the overall optical depth, as functions of \teq{\Thetab}
and \teq{\delta}, are listed in Table~\ref{Table1}.  There, four
parameter regimes with \teq{\sqrt{1-\zeta}\ll 1-\beta} are
identified, depending on \teq{\delta}.  Three of these regimes are
strictly narrow beam limits, while the fourth, for \teq{\delta\ll
1-\zeta}, corresponds to the thin-shell limit in
Eq.~(\ref{eq:thinshell}), and is independent of \teq{\Thetab}.  For the
first two regimes in Table~\ref{Table1}, the developments of
Eqs.~(\ref{eq:flux}) and~(\ref{eq:tauppfin}) just mentioned lead to an
approximate overall optical depth that can be written as one
expression:
\begin{eqnarray}
   \vphantom{\Biggl(} \taupp (\erg_t) &\approx &
   \dover{3\sigt}{2c}\,\dover{d^2}{R_0}\; {\cal F}\; 
   \sqrt{\pi}\;\dover{\Gamma (\alpha +1/2)}{\Gamma (\alpha +1)}\;
   \dover{{\cal H}(\alpha )\; [2(1-\beta )+\delta ]}{3(1-\beta )^2+3(1-\beta )
   \delta+\delta^2}\nonumber\\
  & &\qquad\times\quad \dover{(1-\beta )^{\alpha +3/2}}{
   (1+\beta )^{\alpha    +1}}
   \;\dover{\erg_t^{\alpha -1}}{\sqrt{1-\zeta}}\quad ,\quad\;\;\sqrt{1-\zeta}
   \,\ll\,\min\biggl\{ 1-\beta ,\, \dover{\Delta R}{R_0}\biggr\}\quad ,
   \vphantom{\Biggl(}
 \label{eq:beam}
\end{eqnarray}
for \teq{\delta =\Delta R/R_0}.  This formula encompasses both
thin-shell, narrow beam (\teq{\sqrt{1-\zeta}\ll\delta\ll 1-\beta}) and
thick-shell, narrow beam (\teq{\sqrt{1-\zeta}\ll 1-\beta\ll\delta})
regimes.  Surprisingly, the calculated optical depth actually \it
increases \rm when the opening angle closes down, reflecting the \it
explicit \rm dependence [see Eqs.~(\ref{eq:flux})
and~(\ref{eq:fluxangle})] of the observed flux on the solid-angle of
the expansion, combined with the pair production rate explicitly
depending only on the angle between the test and interacting photon
momenta.  Essentially, the photon density in the source increases for
constant observed flux as the expansion opening angle is reduced.  The
resulting optical depth varies only with the thickness of the shell
(\teq{\propto 1/\Delta R}) in the regime of thick shell expansions,
i.e. for \teq{\Delta R/R_0\gg 1-\beta}.  The third regime in
Table~\ref{Table1} requires use of the thin-shell evaluation of
\teq{\taupp /\dot{\cal N}}, and yields the asymptotic approximation
\begin{equation}
\taupp (\erg_t) \approx \dover{\sigt}{4c}\,\dover{d^2}{R_0}\, {\cal F}\, 
   \erg_t^{\alpha -1}\,\dover{{\cal H}(\alpha )}{\alpha}\;
   \biggl(\dover{1-\beta}{1+\beta}\biggl)^{\alpha}\;\dover{\delta}{1-\zeta}
   \;\; ,\quad\;\; 1-\zeta\ll\delta\ll\sqrt{1-\zeta}\ll 1-\beta .
 \label{eq:beam2}
\end{equation}
This bears an even stronger dependence on the opening angle
\teq{\Thetab\approx\sqrt{2(1-\zeta )}}, again rising with increased
narrowness of the beam, a property that corresponds to an enhanced mean
density of radiation in the expansion.  In this case, the phase-space
for pair production is not restricted by the opening angle, but rather
only by the thinness of the shell, while the flux is still solid-angle
limited.

Finally, in concluding the consideration of narrow beam cases,
observe that Eq.~(\ref{eq:beam}) approximately reproduces the
thin shell and filled-sphere forms in Eqs.~(\ref{eq:thinshell})
and~(\ref{eq:fullsphere}) when the beam is opened up to
\teq{1-\zeta\sim 1-\beta}.  For this intermediate (or critical) regime
of opening angles, a domain common to all three of the limiting cases
discussed so far is achieved when \teq{\Delta R/R_0\sim 1-\beta}.  This
locality in phase space corresponds to the so-called ``blob'' scenario
of earlier work (e.g. Krolik and Pier 1991; Baring 1993; Baring and
Harding 1993) on pair production transparency constraints in gamma-ray
bursts, a situation that is discussed in Section~3.

\subsubsection{Stationary radiation gas}

The remaining limiting case of the optical depth is for a \teq{\beta
=0} or non-relativistic expansion.  This is mostly of academic interest
as a check on the numerical evaluations that follow, and does not have
great physical import for the problem considered in this paper.  In
fact the result in Eq.~(\ref{eq:tauppfin}) does not aptly model
stationary gases since we have neglected the limb contributions to the
optical depth; these become significant in non-relativistic
expansions.  Set \teq{\Thetab =\pi /2} for simplicity.  In the
\teq{\beta\to 0} limit, \teq{\lambda\to -2/(s-1)}, which simplifies
Eq.~(\ref{eq:tauppfin}) somewhat.  However, analytic development of the
subsequent result is not fruitful, so it is convenient to consider
separately the thin and thick shell cases.  The limiting result in
Eq.~(\ref{eq:thinshell}) was derived without restriction on
\teq{\beta}, so the limit \teq{\beta\to 0} can be taken to obtain the
optical depth for stationary, thin shell sources.  For \teq{\Delta
R/R_0\gg 1-\beta} a derivation alternative to that in subsection 2.3.2
is requisite.  Then the dominant contribution to the integration comes
from \teq{s-1\ll\delta}.  The transformation relation in
Eq.~(\ref{eq:Gdef2}) can be used, together with a change of variable to
yield a result proportional to both \teq{\Delta R^2} and the integral
in Eq.~(\ref{eq:Gint2}).  The flux is simply evaluated when \teq{\beta
=0}, so that the optical depth for a thick stationary gas is
\begin{equation}
   \taupp (\erg_t)\;\approx\; \dover{3\sigt}{4c}\, \dover{d^2}{\Delta R}\,
   {\cal F}\,\dover{{\cal H}(\alpha )}{1+2\alpha}\;
   \Biggl\{ \, 1+(\pi -3)\biggl[\dover{3}{2(\alpha +1)}\biggr]^{9/8}\,
   \Biggr\} \;\erg_t^{\alpha -1}\quad ,
   \quad \dover{\Delta R}{R_0}\,\gg\, 1-\beta\quad .
 \label{eq:stationary}
\end{equation}
Note that this estimate differs somewhat from results derived for
isotropic photons (e.g. Schmidt 1978; Epstein 1985; in particular those
that use the formalism of Gould and Schreder 1967), because in this
paper we have assumed isotropic \it injection \rm in the comoving frame
(in this particular limit the observer's frame), which is not
equivalent to radiation isotropy due to radiative transfer in the
sphere.

\subsection{Numerical computation of the optical depth}
\label{sec:numerics}

The various limiting cases just explored guide the technique for
numerical determination of the optical depth, and further act as checks
on computational accuracy.  The numerics are generally straightforward,
and it is expedient to use \teq{s-1} as an integration variable in
Eqs.~(\ref{eq:flux}) and~(\ref{eq:tauppfin}), and scale quantities in
terms of \teq{1-\beta} to maintain good accuracy for large Lorentz
factors.  Since the range of integration variable contributing
significantly to the two integrals is sometimes quite large,
logarithmic sampling of \teq{s-1} is favoured.  Use of the functional
form in Eq.~(\ref{eq:tauppfin}b) comfortably produces smooth results
for \teq{\Thetab} down to even smaller than 0.01 degrees, so use of the
alternative representation in Eq.~(\ref{eq:Jaltdef}), or a series in
\teq{1-\zeta}, is unnecessary for the purposes of this paper.

\placefigure{fig:taupp}

Numerical determinations of the optical depth formed by the combination
of Eqs.~(\ref{eq:tauppfin}) and (\ref{eq:flux}) are presented in
Fig.~\ref{fig:taupp} for different fractional shell thicknesses
\teq{\Delta R/R_0} illustrating its strong dependence on \teq{\Gamma},
and significant dependence on source opening angle only for smaller
\teq{\Gamma}.  The inclusion of \teq{\beta\ll 1} cases is intended only
to provide a general guide to the behaviour of the optical depth
(perhaps to order of magnitude accuracy), but this is strictly
incorrect since they neglect limb contributions.  The quantity actually
plotted in Fig.~\ref{fig:taupp} is the scaled optical depth
\begin{equation}
   \dover{\taupp (\erg_t)}{{\cal F}}\;\dover{\Delta t\, (\hbox{ms})}{
   d^2_{\hbox{\sevenrm Gpc}}}\;\Bigl[ \erg_t (1+z)\Bigr]^{1-\alpha}
   \; =\; 1.06\times 10^{13}\;\dover{c\Delta t}{R_0}\;
   {\cal R}(\alpha ;\,\beta ,\, \Delta R/R_0)\quad ,
 \label{eq:numerical}
\end{equation}
which is dimensionless since \teq{{\cal F}} is measured in photons per
square centimeter per second, where \teq{{\cal R}(\alpha ;\,\beta
,\,\Delta R/R_0)} is just \teq{{\cal H}(\alpha )/(1+\beta )^{1+\alpha}}
times the ratio of the two integrals in Eqs.~(\ref{eq:tauppfin})
and~(\ref{eq:flux}).  Here \teq{\Delta t} is the observed source
variability timescale, typically in the range of \teq{10^{-3}}--1
seconds, inferred, for example, from time histories such as those
exhibited in the BATSE 1B catalogue (Fishman et al. 1994), and \teq{z}
is the cosmological redshift of the source (it is set to zero in
Fig.~\ref{fig:taupp}).  The coefficient of this equation clearly
defines the optical depth scale for cosmological bursts, and would be
8--10 orders of magnitude smaller for galactic halo sources.  A
``canonical'' spectral index of \teq{\alpha =2.5} (see
Table~\ref{Table2} for specific values) is chosen in
Fig.~\ref{fig:taupp} for simplicity; increasing (reducing) \teq{\alpha}
just increases (lowers) the slope of the curves in the \teq{\Gamma\gg
1} regimes.  Remembering that the energy \teq{\erg_t} is expressed in
units of \teq{m_ec^2}, it is evident from Fig.~\ref{fig:taupp} that the
maximum observed energies of EGRET sources (see Table~\ref{Table2})
imply that Lorentz factors in the range of approximately 50--500 are
required to render these bursts optical thin to pair production.

The curves in Fig.~\ref{fig:taupp}a clearly delineate three regimes of
parameter space in order of increasing bulk Lorentz factor: (i)
non-relativistic flows where the optical depth is independent of the
expansion speed, (ii) thin-shell expansions (\teq{\Delta R/R_0\ll
1-\beta} and \teq{\Gamma\gg 1}), which is the portion of phase space
sampled by the work of Fenimore, Epstein and Ho (1993), yielding a
strong inverse dependence of \teq{\taupp} on \teq{\Gamma}, and (iii) at
the highest Lorentz factors (i.e. above the break at \teq{(\Delta
R/R_0)/(1-\beta )\sim 1}), thick-shell expansions, where the inverse
dependence on \teq{\Gamma} is slightly stronger (by two powers of
\teq{\Gamma}: compare Eqs.~(\ref{eq:thinshell}) where
\teq{\taupp\propto (1-\beta )^{\alpha}/R_0} and~(\ref{eq:fullsphere})
where \teq{\taupp\propto (1-\beta )^{\alpha +1}/\Delta R}).  
We note that Fenimore, Epstein \& Ho (1993) produced an
optical depth vs. \teq{\log\Gamma} plot for their infinitely thin shell
analysis that exhibited a dramatic reduction of \teq{\taupp} above some
critical Lorentz factor.  These turnovers were found to be artificial,
being caused by a coding error (Fenimore, private communication).
The ratio \teq{\Delta R/R_0} is independent of \teq{\Gamma} in 
Fig.~\ref{fig:taupp}a, though other choices are quite plausible.
For opening angles \teq{\Thetab\gtrsim 1/\Gamma}, the ratio
\teq{\chi\equiv (\Delta R/R_0)/(1-\beta )} is the only critical
parameter delineating the thin-shell and thick-shell cases, and is
intimately related to the the portion of the emission region that is
causally-connected to the test photon.  As is evident from
Eqs.~(\ref{eq:thinshell}) and~(\ref{eq:fullsphere}), the optical depth
is independent of the thickness of the shell when this parameter is
much less than unity, and inversely proportional to \teq{\Delta R} when
\teq{\chi\gg 1}; the transition between these regimes is quite
gradual.  The curves exhibit a lack of dependence on the opening angle
\teq{\Thetab} when \teq{\Thetab\gtrsim\min\{ 1/\Gamma, (\Delta
R/R_0)^{1/2}\,\} } (the curves in Fig.~\ref{fig:taupp}a are coincident
for \teq{\Thetab =90^\circ, 10^\circ}, and \teq{1^\circ}):  this is a
consequence of causality dominating geometry in the restriction of the
integration phase space.  When \teq{\Thetab\lesssim 1/\Gamma}, the
optical depth (\teq{\propto (1-\beta )^{\alpha +3/2}/[\Thetab\, R_0]})
is actually increased by the angular reduction of the
available phase space, an effect that is particularly evident for
non-relativistic expansions.  As mentioned in Section~2.3.3, the reason
for this increase in optical depth with declining \teq{\Thetab} is that
the pair production rate is proportional to the angle between the test
and interacting photons (which is limited linearly by \teq{\Thetab}),
while the flux scales as the solid angle (i.e. \teq{\Thetab^2} when
\teq{\Thetab\ll 1}) of the expansion.  Hence the density of photons in
the source that is inferred for a given flux actually increases as
\teq{1/\Thetab} when \teq{\Thetab} declines below \teq{1/\Gamma}.  In
Fig.~\ref{fig:taupp}b, where \teq{\Delta R/R_0} is approaching
unity, the thin-shell regime is no longer distinct from the
non-relativistic domain so that all relativistic expansions are
thick-shell.  

The specific choice of \teq{R_0=\Gamma c\Delta t} is made in
Fig.~\ref{fig:taupp} for simplicity; alternative dependences on
\teq{\Gamma} are possible, corresponding to different interpretations
of source timescales (as discussed in Section~3 below), and these
result in only a slightly different appearance from the curves in
Fig.~\ref{fig:taupp}.  The results presented in this subsection are
obtained under the assumption that the test photon starts its life at
the rear of the expansion; Eq.~(\ref{eq:nutrans}) provides a simple
scheme of substitution for Eq.~(\ref{eq:tauppfin}) that yields optical
depths for arbitrary initial positions of the test photon within the
expansion.  It is evident that imposing a \teq{\taupp =1} condition on
EGRET bursts potentially can map over into both thin and thick shell
regimes, depending on the assumed source distance, measured flux and
maximum photon energy observed.  We note that using a \teq{\taupp =1}
criterion for source transparency actually leads to conservative lower
bounds for \teq{\Gamma}, since significant attenuation is already
present when \teq{\taupp =1}.  This can be easily seen if the spectrum
is attenuated by an exponential factor \teq{\exp\{-\taupp\} }, a common
choice.  A central consequence of the assumption of an infinite
power-law burst spectrum is that the optical depth is an {\it
increasing} (and power-law) function of energy, so that spectral
attenuation arises only at high energies; departures from this
behaviour will be discussed briefly at the end of the next Section.  We
remark also that opacity skin effects (which depend on the spatial
distribution of photons and therefore are model-dependent) can
sometimes render the exponential \teq{\exp\{-\taupp\} } a poor
descriptor of attenuation, with \teq{1/(1+\taupp )} perhaps being an
improvement for uniformly-distributed photons, leading to broken
power-laws rather than exponential turnovers.  A variety of signatures
of spectral attenuation are possible, particularly if pair cascading is
involved, and some of these are illustrated in the work of Baring and
Harding (1997).

Undoubtedly the most crucial piece of information to be gleaned from
Fig.~\ref{fig:taupp} is that causality minimizes the role of the
opening angle \teq{\Thetab} of the expansion in faster expansions, and
specifically that the \teq{\taupp=1} condition will be virtually
independent of \teq{\Thetab} in the range
\teq{90^{\circ}\geq\Thetab\gtrsim 1/\Gamma} for burst sources.  This
insensitivity to the angular extent of the expansion is the keystone to
elimination of the number problem for cosmological bursts, as is
discussed below.

\newpage

\section{BULK LORENTZ FACTORS FOR EGRET SOURCES}
\label{sec:egret}

The calculations for \teq{\taupp} we have performed can readily by
applied to gamma-ray bursts detected by EGRET.  The pair production
optical depth in Eq.~(\ref{eq:tauppfin}) depends on the free
parameters: \teq{\Gamma}, \teq{\Delta R/R_0}, \teq{\Thetab}, \teq{d},
and on the observed parameters: source flux \teq{{\cal F}} [defined in
Eq.~(\ref{eq:fluxdef})], high energy spectral index \teq{\alpha} [see
Eq.~(\ref{eq:ndotdef})], test photon energy \teq{\erg_t}
(\teq{=\emax}), and \teq{\Delta t}.  The burst variability timescale
\teq{\Delta t} can be used to infer an upper limit on the source size,
which we take for the moment to be \teq{R_0 = \Gamma c\Delta t}, based
on the apparent size of the expanding shell perpendicular to the light
of sight, as seen by a stationary observer (e.g. see Rees, 1966).
Alternative size-estimates, such as those that couple the source
variability to dimensions along the line of sight to the source, are
possible.  In addition, measured minimum variability timescales for
bursts range from milliseconds in the BATSE range to supersecond values
in EGRET data.  Motivations for choosing either of these \teq{\Delta
t}, and also different source size determinations, are discussed
below.  Both variability timescales are addressed in the results
presented here, for the sake of completeness.

The key observable parameters, to be used in the pair production
opacity calculations of this paper, are displayed in Table~\ref{Table2}
for six of the burst detections by EGRET.  Note that there are eleven
EGRET bursts in total (Schneid et al. 1996), three of which have
insufficient (published) data for the purposes of our analyses; GRB
920622 and GRB 940301 have the required observational parameters in
Schneid et al. (1995), but suffer from poor statistics above about 2
MeV.  We therefore conservatively opt to study just the most
significant six sources of the EGRET population.  In Table~\ref{Table2}, the
fluxes are expressed in observer-friendly units, via
\teq{f(1\,\hbox{MeV})}, which is just the flux, evaluated at 1 MeV, per
MeV energy interval (the flux \teq{{\cal F}} is per \teq{m_ec^2} at 511
keV).  These fluxes are obtained via extrapolations down to 1 MeV of
the best-fit power-laws to the time-integrated super-MeV spectral data,
and are not necessarily the actual fluxes measured at 1 MeV.  Note that
both EGRET and COMPTEL parameters are listed for GRB 910601 since this
burst was relatively soft and actually had a slightly more significant
detection by COMPTEL than by the EGRET TASC in the 3--5 MeV range.  We
remark that the COMPTEL listings for GRB 910601 (based on Hanlon et al.
1994) differ slightly from those quoted by Baring and Harding (1993)
that were obtained from Winkler et al. (1993).  As noted in the Table
caption, in the computations of this paper we neglect the highest
energy (18 GeV) photon detected for GRB 940217.  This conservative step
is taken because the statistically-limited sample provided by a single
photon leads to a large uncertainty in the spectral form at these
energies, which is compounded by the lack of contemporaneous spectral
information at lower (i.e. sub-MeV) energies (e.g. Hurley et al.
1994); time-resolved spectra of good statistical quality in the
super-100 MeV range await future generations of instrumentation.  A
nice depiction of the relative fluxes and spectra of four of these
bursts is given in Hurley (1996).

\placetable{Table2}

\subsection{Geometries with $\Thetab\sim 1/\Gamma$}

Before presenting the results for the bulk Lorentz factor constraints
inferred from our optical depth calculations here, it is instructive to
first review the results of previous, more primitive bulk motion
determinations.  As outlined in the Introduction, the earlier work of
Krolik and Pier (1991) and Baring (1993), and subsequent papers,
considered ``blobs'' of radiation-emitting material moving at
relativistic speeds more-or-less along the line of sight to an
observer.  The angular extent of these blobs, i.e. the width of the
angular distribution of photons as measured in the observer's frame,
was assumed to be comparable to \teq{1/\Gamma}, where \teq{\Gamma} was
the Lorentz factor of the blob (\teq{\Gamma\gg 1}).  For such
relativistically-moving blobs, the minimum bulk Lorentz factors at
redshift \teq{z} are obtained (i.e. for \teq{\taupp =1}) from the result
derived by Baring and Harding (1993, corrected in Harding 1994):
\begin{equation}
   \Gamma^{1+2\alpha}\;\gtrsim\;\dover{(3.83)^{\alpha}\, 
   (1+z)^{\alpha -1}}{3\alpha^{5/3} (4/3+\alpha )^{27/11}}\;
   \dover{d_{\hbox{\sevenrm kpc}}^2}{\Delta t\, (\hbox{ms})}
   \;\Bigl( \dover{\emax}{\hbox{\rm 1 MeV}}\Bigr)^{\alpha-1}\,
   f(\hbox{\rm 1 MeV})\quad .
 \label{eq:Gammablob}
\end{equation}
Here \teq{f(1\,\hbox{MeV})} is just the source flux evaluated at 1 MeV,
per MeV energy interval, and therefore is proportional to \teq{{\cal
F}}.  This formula can be approximately reproduced from
Eq.~(\ref{eq:thinshell}) by setting  \teq{R_0\sim \Gamma c\Delta t}.
Alternatively the thick shell approximation in
Eq.~(\ref{eq:fullsphere}) gives more or less the same estimate if
\teq{\Delta R\sim c\Delta t/\Gamma}.  It therefore follows that the
``blob calculation'' corresponds to the boundary between thin and thick
shell cases where the line-of-sight and transverse (e.g. variability)
timescales are comparable.  Furthermore, it also coincides with the
boundary of narrow beam expansions, namely for \teq{\Thetab\sim
1/\Gamma}, as can be established by setting \teq{1-\zeta =1-\beta} in
Eq.~(\ref{eq:beam}) and choosing intermediate shell thicknesses.

We remark here that the choice of variability timescale for use in the
estimation of minimum source bulk Lorentz factors is subjective.  The
smallest \teq{\Delta t} observed in BATSE data (see Fishman et al.
1994) is of the order of milliseconds, and it is quite conceivable that
intrinsic source variability occurs on even shorter timescales.  Such
small \teq{\Delta t} were adopted in the work of Baring (1993), Baring
and Harding (1993, 1995, 1996) and Harding (1994), and lead to
light-crossing time size determinations of \teq{c\Delta t\sim 3\times
10^7}cm.  However, variability in the hard gamma-ray band, i.e. for
Comptel and EGRET data, can only be conclusively inferred from more
severely photon-limited samples on the order of 0.1 second to 1 second
timescales.  Hence a conservative approach, adopted for example by Ryan
et al. (1994, for GRB 930131) and Winkler et al. (1995, for GRB
940217), uses these longer \teq{\Delta t} values in obtaining pair
production constraints on bulk motion in bursts.  Note that the
relative timescales in the different GRB energy bands may not be
related at all to intrinsic source properties but merely reflect
current instrumental limitations in the time domain.  While
experimentalists might prefer the conservative variability values,
theorists are often motivated to attribute the shortest timescales to
regions emitting the highest energies of radiation.  This is frequently
justifiable in astrophysics, since a whole host of cosmic objects are
powered from a central region and thereby generate their most energetic
photons closer to the center; such photons would generally be expected
to couple to shorter timescales.  If a central ``powerhouse'' is indeed
responsible for gamma-ray burst emission (as suggested by Fenimore,
Madras and Nayakshin 1996), then sub-millisecond timescales might
accurately reflect the source conditions appropriate to EGRET photons.
In contrast, if bursts approximate more closely the class of fireball
models that generate the emission we see by impact on the interstellar
medium (e.g. Rees and M\'esz\'aros 1992, M\'esz\'aros and Rees 1993),
perhaps through diffusive acceleration of particles at shocks, then the
highest energy photons are produced by particles diffusing on the
largest scales of the system, and therefore might be expected to have
longer variability timescales than photons in the BATSE energy range.
To accommodate a variety of perspectives, this paper considers
\teq{\Delta t} values of \teq{1}ms and 1 second.

The solutions of Eq.~(\ref{eq:Gammablob}) for the minimum bulk Lorentz
factor \teq{\gammin} for the burst parameters of Table~\ref{Table2} are
listed in Table~\ref{Table3}, for the two different variability
timescales, and for four different GRB source distances that are
typical of galactic disk, galactic halo, and nearby and distant
cosmological populations (though the redshift \teq{z=0} is chosen for
simplicity).  The immediately obvious conclusion is that,
except for a disk origin of bursts, relativistic bulk motion is
generally inferred for the EGRET sources, because of the detection of
energetic photons.  The Lorentz factors obtained for cosmological
distance scales far exceed those inferred for extragalactic jets in
active galaxies.  Conversely, Table~\ref{Table3} indicates that
isotropic emission cannot be supported in GRBs unless they are quite
local, i.e. well within the galactic disk, a conclusion that differs
from Schmidt's (1978) early work principally because of the positive
EGRET detections in the CGRO era.  These estimates for the minimum
\teq{\Gamma} in bursts are a good first guide to constraints on bulk
motion in their emission regions; the refinements of the burst
geometries addressed in this paper only modify the estimates in
Table~\ref{Table3} by factors of at most a few, as will become evident
shortly.  We remark that previous versions of these estimates (e.g.
Baring 1993, Baring and Harding 1993, Harding 1994, Baring 1995) have
sometimes used slightly different observational parameters.  Note that
a number of entries in the 1 kpc column have \teq{\gammin =1}.  These
actually represent unphysical \teq{\Gamma <1} solutions to
Eq.~(\ref{eq:Gammablob}) that are obtained only because the assumption
\teq{\Gamma\gg 1} is used to derive Eq.~(\ref{eq:Gammablob}).  Hence
\teq{\gammin =1} entries denote regimes where this ``blob'' constraint
breaks down and refinements are needed.  It must be emphasized that
Eq.~(\ref{eq:Gammablob}) and the work of this paper implicitly assume
that the GRB spectrum extends above 511 keV in the rest frame of the
emission region, so that phase space above pair creation threshold is
non-zero.  

\placetable{Table3}

It is appropriate to remark on a caveat to these results.  Given
emission observed out to \teq{\emax}, the maximum photon energy in the
source rest frame is of the order of \teq{\emax /\Gamma}, which must
exceed unity in order to be above pair threshold.  Hence the
possibility that intrinsic cutoffs could be present in the GRB spectrum
anywhere above \teq{\emax} automatically implies that \teq{\emax}
provides a potential lower bound to \teq{\Gamma}.  In particular, when
the estimate for \teq{\gammin} obtained from Eq.~(\ref{eq:Gammablob})
exceeds \teq{\emax}, it is quite possible that pair creation never
occurs at all, since opacity for photons of energy \teq{\emax} occurs
through interactions with photons \it at even higher energies\rm , for
which there is no observational evidence.  Cases with values of
\teq{\gammin /\emax} greater than unity generally arise only for small
\teq{\Delta t} and at cosmological distances (see Table~\ref{Table3}:
GRB910601 is a perfect example).  In such instances, these cutoff
considerations become quite relevant, and it becomes necessary to take
\teq{\emax} as the estimate for the minimum Lorentz factor
\teq{\Gamma}.  Since values of \teq{\emax} between 1 and 10 MeV in the
collection of EGRET bursts are a marker of fainter or steeper spectrum
bursts, i.e. probably reflecting the observational limitations of
EGRET, it is quite realistic to develop bulk motion estimates based
Eq.~(\ref{eq:Gammablob}), in the belief that many (if not most) bursts
emit at energies much higher than 100 MeV.
 
\placefigure{fig:gammin}

\subsection{Generalized Geometries}

Generalizing from Eq.~(\ref{eq:Gammablob}) to the expansion geometries
considered in this paper, a pair production transparency condition is
obtained by setting the optical depth that is obtained from
Eqs.~(\ref{eq:tauppfin}) and~(\ref{eq:flux}) equal to unity, i.e.
effectively reading off abscissa values for chosen ordinates from
curves like those in Fig.~\ref{fig:taupp}.  Since the expressions for
the optical depth and flux from the expansion involve integrals with
integrands that depend on the bulk Lorentz factor, the solutions for
\teq{\gammin} roots to \teq{\taupp =1} must be solved iteratively: we
adopt a bisection technique.  The minimum bulk Lorentz factors
\teq{\gammin} for EGRET burst sources that result from our pair
production transparency calculations in Eqs.~(\ref{eq:tauppfin})
and~(\ref{eq:flux}) are depicted in Fig.~\ref{fig:gammin} as functions
of the fractional shell thickness \teq{\Delta R/R_0}.  This
illustration limits shell thicknesses to regimes where \teq{\Delta
R/R_0\leq 1}, since \teq{R_0} is tied to the time variability via
\teq{R_0=\Gamma c\Delta t}, and this coupling becomes inappropriate for
\teq{\Delta R\gtrsim R_0} regimes, where \teq{\Delta R\sim c\Delta t}
is a more apt choice.  Again the source parameters from
Table~\ref{Table2} are used, and results are presented for large
expansion half-angles, \teq{\Thetab=90^\circ}, variability timescales
of \teq{\Delta t=1}ms, and source distances of 100 kpc and 1 Gpc that
represent galactic halo and cosmological burst scenarios,
respectively.  Cosmological redshift modifications, which depend on the
choice of cosmology, are neglected for simplicity.  All of the twelve
curves in the four panels exhibit similar behaviour, with \teq{\gammin}
being independent of \teq{\delta =\Delta R/R_0} when
\teq{\delta\lesssim 1-\beta}, the so-called thin-shell limit, and
\teq{\gammin} declining roughly as \teq{\delta^{-1}} for \teq{\delta\gg
1-\beta} when the filled-sphere regime is realized.  These dependences
on \teq{\Delta R/R_0} appear explicitly in the asymptotic forms in
Eqs.~(\ref{eq:thinshell}) and~(\ref{eq:fullsphere}).  Domains where
\teq{\Delta R/R_0>1} (i.e. for a true filled sphere) are not depicted
since they are unlikely to be encountered in gamma-ray bursts.  The
values of \teq{\gammin} obtained when \teq{\Delta R/R_0\sim 1-\beta}
(i.e. the transition regions) are comparable to those listed in
Table~\ref{Table3} for all bursts; this reflects the broad
applicability of the constraint in Eq.~(\ref{eq:Gammablob}) that is the
hallmark of the so-called ``blob'' calculation.

The curves in Fig.~\ref{fig:gammin} generally concur with the global
trends of increasing \teq{\gammin} for higher \teq{\emax} and/or
declining spectral index \teq{\alpha}.  Yet the cross-over of two
curves in the lower left-hand panel exemplifies how the expansion
geometry can complicate trends, and generate non-monotonic behaviour in
\teq{\emax} or \teq{\alpha}.  For comparison, Fig.~\ref{fig:gammin1s}
reproduces the bottom right panel of Fig.~\ref{fig:gammin}, but for a
variability timescales of \teq{\Delta t=1}sec.  The curves resemble
those of Fig.~\ref{fig:gammin}.  However, at these cosmological
distances, \teq{\gammin} is reduced from the corresponding values in
Fig.~\ref{fig:gammin} solely because the large \teq{\Delta t} dilutes
the density of internal photons inferred for the source.  Again a
cross-over of curves arises, indicating that non-monotonicity of
\teq{\gammin} in \teq{\emax} and \teq{\alpha} does not belong
exclusively to galactic halo scenarios.  Note that the uncertainties in
the observational quantities that are listed in Table~\ref{Table2} are
as large as around 10\%.  These lead to uncertainties in the
\teq{\gammin} determinations for both these figures of the order of
20\%, which are largely masked by the ranges of \teq{\gammin} produced
by varying model parameters.  Hence further consideration of
experimental uncertainties is omitted from this paper.  Note also that
the results we have presented have assumed that the test photons in the
source start from the rear of the expansion.  This maximizes the
computed optical depths, implying that it is possible to lower our
estimates for \teq{\gammin} somewhat.  Permitting test photons to be
emitted throughout the source will perhaps lower the mean optical depth
by a factor of 2 or so, leading to a reduction of \teq{\gammin} of
around 15\%.  Hence the detailed consideration of the distribution of
\it test \rm photons is neglected in this paper as this will only have
a minor impact on the \teq{\gammin} obtained.

\placefigure{fig:gammin1s}

The results presented in Figs.~\ref{fig:gammin} and~\ref{fig:gammin1s}
clearly define the behaviour of our \teq{\gammin} solutions for large
opening angles, but obviously do not address the entire available phase
space for model parameters.  Hence in Fig.~\ref{fig:gamminthet} we
depict the variations of \teq{\gammin} with expansion opening angle
\teq{\Thetab} for fixed \teq{\Delta R/R_0}.  These curves unequivocally
demonstrate how insensitive \teq{\gammin} is to \teq{\Thetab} when
\teq{\Thetab\gtrsim 3\Delta R/R_0}, i.e. for a range from modestly
small \teq{\Thetab} up to \teq{90^\circ}.  Note that such a substantial
range is realized for the adopted values of \teq{\Delta R/R_0}, chosen
to correspond to the transition between thin-shell and thick-shell
regimes.  This insensitivity, whose important implications for
cosmological models are discussed at the end of this subsection, is a
principal conclusion of this paper.  The values of \teq{\gammin} at the
intercept with the right hand side axis of each panel in
Fig.~\ref{fig:gamminthet} are just those obtained by vertically slicing
the plots in the bottom two panels of Fig.~\ref{fig:gammin} at the
appropriate value of \teq{\Delta R/R_0}.  For \teq{\Thetab\ll\Delta
R/R_0}, power-law asymptotic behaviour of \teq{\gammin} is observed in
the Figure, with the dependence being easily deduced (for
\teq{1-\beta\ll\Delta R/R_0}) from Eq.~(\ref{eq:beam}):
\teq{\gammin\propto \Thetab^{-1/(2\alpha +3)}}.  This is a very weak
dependence (for \teq{\alpha} in the range 2--3 typical of EGRET bursts)
on \teq{\Thetab}, generated by the strong variation of the optical
depth with \teq{\Gamma}.

It is quite instructive to augment these plots by summarizing the
behaviour of the optical depth results with phase space diagrams using
well-chosen variables.  This can be achieved in an enlightening manner
for both observational and theoretical (i.e. model) phase space
parameters via contour plots, i.e.  exhibiting curves of constant
\teq{\Gamma} (\teq{=\gammin}) that satisfy the criterion \teq{\taupp
=1} for the pair production optical depth.  First of all, we focus on
the space of observational parameters given by the maximum energy
observed, \teq{\emax}, and the EGRET spectral index \teq{\alpha}, both
being listed in Table~\ref{Table2} for EGRET bursts.  Fixing the
observed flux and the source time variability at the ``canonical''
values of \teq{f(\hbox{1 MeV})=3}cm$^{-2}$ sec$^{-1}$ MeV$^{-1}$ and
\teq{\Delta t=1}ms, respectively, the resulting contour plot is shown
in Fig.~\ref{fig:gammin1s}.  These contours, which represent lower
boundaries to regions of opacity (i.e. \teq{\taupp >1}), display a
number of trends that are hallmarks of the optical depth properties of
the relativistically expanding radiation gas.

\placefigure{fig:gamminthet}

First, \teq{\emax} is an increasing function of \teq{\alpha}.  This
arises because, for the particular values of \teq{f(\hbox{1 MeV})},
\teq{\Delta t} and \teq{\Gamma} chosen, solutions with \teq{\emax
<\Gamma} are always realized.  For such solutions, the test photons at
energy \teq{\emax} interact with photons near pair production threshold
in the CM frame, i.e. with photons of energy around \teq{\emax
/\Gamma^2} (\teq{<1}MeV) in the observer's frame.  Since the optical
depth is held constant (i.e.  unity), and the flux is pinned at 1 MeV,
thereby providing a ``pivot point'' in the spectrum, increasing
\teq{\alpha} then raises the number of interacting photons (with
energies below 1 MeV) so that \teq{\emax} must correspondingly be
increased to compensate.  This trend solely finds its origin in the
realization of an \teq{\emax <\Gamma} branch of solutions.  As
\teq{\alpha\to\infty}, the energy of the interacting photons must
approach the pivot point, i.e. 1 MeV.  Hence the \teq{\emax} curves
asymptotically approach \teq{\sim\Gamma^2} as \teq{\alpha} becomes very
large, behaviour that is conspicuous in the galactic halo cases in
Fig.~\ref{fig:emaxalpha}.  In the particular examples shown,
\teq{\emax} drops off rapidly as \teq{\alpha} approaches unity,
a singularity of these curves [e.g. see Eq.~(\ref{eq:tauppfin})].  The
monotonic increase of \teq{\emax} with \teq{\alpha} can be inverted to
yield a declining tendency if an \teq{\emax >\Gamma} solution branch
can be encountered, so that interacting photons are always above the pivot
point at 1 MeV.  This occurs when the product \teq{f(\hbox{1 MeV})\, 
d^2/\Delta t} is small enough, i.e. forcing \teq{\emax} higher for given
\teq{\Gamma} and \teq{\alpha} [e.g. see Eq.~(\ref{eq:tauppfin})].
Then the contours would rise rapidly as \teq{\alpha} approached unity,
but still asymptote to roughly a \teq{\Gamma^2} dependence for
\teq{\alpha\to\infty}.  This situation is more likely to arise for
\teq{\Delta t=1}sec variability timescales.  Note that the larger
values of \teq{f(\hbox{1 MeV}) d^2/\Delta t} in the cosmological cases
in Fig.~\ref{fig:emaxalpha} yield stronger dependences of \teq{\emax}
on \teq{\alpha}, primarily because these cases generally have larger
``lever arms'' for the interacting photons (at energy
\teq{\sim\emax/\Gamma^2}) around the pivot point at 1 MeV.

\placefigure{fig:emaxalpha}

The dependence of \teq{\emax} on \teq{\Gamma} is very closely given by
the blob calculation in Eq.~(\ref{eq:Gammablob}), namely
\teq{\emax\propto \Gamma^{-(1+2\alpha )/(\alpha -1)}}.  For example,
doubling \teq{\Gamma} at \teq{\alpha =2} in either of the galactic halo
or cosmological cases yields an increase of around \teq{29.4}, which is
close to the ``blob'' estimate of \teq{32}.  Similarly, doubling
\teq{\Gamma} at \teq{\alpha =3} in either of the galactic halo or
cosmological cases yields an increase of around \teq{11.2}, which is
very close to the ``blob'' estimate of \teq{11.3}.  For fixed
\teq{\alpha}, these amplification ratios are independent of the source
distance \teq{d}, since \teq{d} just forms part of the proportionality
constant for the relationship between \teq{\emax} and \teq{\Gamma}.
Another trend that is apparent in Fig.~\ref{fig:emaxalpha} is that
\teq{\emax} declines with decreasing \teq{\delta =\Delta R/R_0}.  The
variation of \teq{\emax} with \teq{\delta} is depicted using the light
solid curves for the galactic halo scenario, with the \teq{\delta
=0.3(1-\beta )} and \teq{\delta =3(1-\beta )} cases visually defining a 
band around the \teq{\delta =1-\beta} case (the behaviour for
cosmological source distances is similar).  When \teq{\delta} is
reduced, the average density of photons within the source increases,
pushing the optical depth up.  Hence, to compensate, \teq{\emax} must
also decline with \teq{\delta}, so defining the observed trend; in
Figs.~\ref{fig:gammin} and~\ref{fig:gammin1s}, this effect forces
\teq{\gammin} to increase when \teq{\emax} is held constant.  From
those figures, it is evident that when \teq{\delta\ll 1-\beta}, the
thin-shell limit produces insensitivity of the optical depth to
\teq{\delta}, a feature that is also apparent in
Fig.~\ref{fig:emaxalpha}, for which the \teq{\delta =0.3(1-\beta )}
curves are more proximate to the \teq{\delta =1-\beta} ones than are
the \teq{\delta =3(1-\beta )} cases.  Finally, we note that the value
of \teq{f(\hbox{1 MeV})} for each of the EGRET bursts that are depicted
as points in Fig.~\ref{fig:emaxalpha} differs from the chosen canonical
value.  Hence for those sources with higher \teq{f(\hbox{1 MeV})}
(GRB 910503 and GRB 910814), the exhibited curves should be slid down
somewhat to visualize the situation (i.e. infer bulk Lorentz factors)
appropriate for these bursts.  Likewise, for the remaining EGRET
bursts, the curves should be moved upwards to deduce \teq{\Gamma}
values that are consistent with the results depicted in
Figs.~\ref{fig:gammin} and~\ref{fig:gammin1s}.

\newpage

The theoretical phase space contour plot is presented in
Fig.~\ref{fig:thetabdelta}, which exhibits curves of constant
\teq{\Gamma} that satisfy the criterion \teq{\taupp =1}, in the space
defined by the opening angle \teq{\Thetab} and the fractional shell
thickness \teq{\delta =\Delta R/R_0}.  We focus on regimes where
\teq{\Delta R/R_0\leq 1} since, as mentioned above, we have tied
\teq{R_0} to the time variability via \teq{R_0=\Gamma c\Delta t}, a
coupling that becomes inappropriate for \teq{\Delta R\gtrsim R_0}
regimes, where \teq{\Delta R\sim c\Delta t} is a more apt choice.  The
large range of \teq{\Thetab} is chosen intentionally to present the
information relating to the reduction of expansion opening angles that
is omitted from earlier considerations, such as in
Figs.~\ref{fig:gammin} and~\ref{fig:gammin1s}.  For all contours, the
maximum energy and spectral index were set at \teq{\emax =100}MeV and
\teq{\alpha=2.21}, respectively, and the variability timescale
\teq{\Delta t} and the flux \teq{f(\hbox{1 MeV)}} at 1 MeV are
representative of EGRET-like sources.  These parameters were tuned
somewhat to obtain a maximum of informational content in the figure, so
that other choices of parameters can lead to some variations in contour
shape.  Remembering the general trend of a reduction in \teq{\taupp}
with increasing \teq{\Thetab}, the contours in
Fig.~\ref{fig:thetabdelta} clearly represent upper boundaries to
regions of opacity (i.e. \teq{\taupp >1}).

A number of prominent features appear in this phase space plot.
Foremost among these are the vertical portions of the \teq{\Gamma =10}
and \teq{\Gamma =250} contours, present when the opening angles are
significant.  These define regimes where the optical depth is
independent of \teq{\Thetab}, and both thin-shell and thick-shell
regimes that are well-described by the asymptotic formulae in
Eqs.~(\ref{eq:thinshell}) and~(\ref{eq:fullsphere}) can be realized.
It is precisely this upper region of the \teq{\Thetab}-\teq{\delta}
diagram that is probed in solutions depicted in Figs.~\ref{fig:gammin}
and~\ref{fig:gammin1s}.  The virtual independence of the optical depth
to \teq{\delta} observed in those solutions for \teq{\gammin} manifests
itself in Fig.~\ref{fig:thetabdelta} as an extreme sensitivity of the
position of any vertical sections of the contours to the choice of
\teq{\Gamma} (or \teq{\emax} or \teq{\alpha}).  This sensitivity
therefore produces a low density of contours in the upper left hand
portion of Fig.~\ref{fig:thetabdelta}, so that contours possessing
vertical sections occupy a minority of cases if broad ranges of
\teq{\Thetab} are considered.  Note also that low values of
\teq{\Gamma} are obtained only in the upper right of the figure.
However, since values of \teq{\Delta R/R_0} greater than unity are
unrealistic, it becomes clear that for this choice of \teq{\emax} and
\teq{\alpha}, only values of \teq{\Gamma \gtrsim 5} are attained.  This
signifies the general property of these calculations that {\it
relativistic bulk motions are always inferred} unless \teq{\emax} is
not too much greater than 1 MeV.

\placefigure{fig:thetabdelta}

The narrow beam (i.e. \teq{\Thetab\lesssim 0.01}) portion of the
parameter space exhibits distinctive power-law dependences, with
\teq{\Thetab} rising as \teq{\sqrt{\delta}} when \teq{\delta} is very
small, and declining as \teq{1/\delta} when \teq{\delta} exceeds
\teq{1-\beta}.  This asymptotic behaviour can be deduced with the aid
of Eqs.~(\ref{eq:beam}) and~(\ref{eq:beam2}).  Table~\ref{Table1}
identifies four parameter regimes, three of which are relevant to
narrow beam considerations, the other being the domain of large
\teq{\Thetab} discussed in the previous paragraph.
Eq.~(\ref{eq:beam2}) is pertinent to the lower left hand portion of
Fig.~\ref{fig:thetabdelta}; contours of unit optical depth thereby
defining the dependence \teq{\Thetab \propto\delta^{1/2}/\Gamma^{\alpha
+1/2}} (for \teq{R_0=\Gamma R_v}).  In this limit, doubling
\teq{\Gamma} decreases \teq{\Thetab} by around a factor of \teq{6.54}
for \teq{\alpha =2.21}, regardless of the assumed distance to the
source.  The lower right of the figure is described by the thick shell
(i.e. \teq{\delta\gg 1-\beta}) limit of Eq.~(\ref{eq:beam}), which
yields contours with \teq{\Thetab\propto\delta^{-1}\Gamma^{-(2\alpha
+2)}}; doubling \teq{\Gamma} in this limit reduces \teq{\Thetab} by a
factor of \teq{343} for \teq{\alpha =2.21}, behaviour that is borne out
in Fig.~\ref{fig:thetabdelta}.  The third asymptotic domain is defined
by the thin-shell limit of Eq.~(\ref{eq:beam}), yielding contours
approaching a limit with \teq{\Thetab\propto\Gamma^{-(2\alpha +2)}},
independent of \teq{\delta}.  This domain is almost attained at the
broad peaks of the contours that remain always in the narrow beam (i.e.
lower) portion of phase space, particularly for the \teq{\Gamma =40}
and \teq{\Gamma =1000} cases in the figure.  These three limiting
forms can be written explicitly as (for \teq{\Gamma\gg 1} and
\teq{R_0=\Gamma R_v}):
\begin{equation}
  \Thetab\;\sim\; \cases{
  {\cal F}\,\dover{\sigt}{c}\,\dover{d^2}{R_v}\, {\cal H}(\alpha )\,
  \dover{3\sqrt{\pi}\,\Gamma (\alpha +1/2)}{2^{2\alpha +3}\,\Gamma (\alpha +1)}
  \;\dover{\emax^{\alpha -1}}{\Gamma^{2\alpha +4}}\;\dover{1}{\delta}\; ,
    & $\;\; \sqrt{1-\zeta}\ll 1-\beta\ll\delta\; ,\vphantom{\Biggl(}$ \cr
  {\cal F}\,\dover{\sigt}{c}\,\dover{d^2}{R_v}\, {\cal H}(\alpha )\,
  \dover{\sqrt{\pi}\,\Gamma (\alpha +1/2)}{2^{2\alpha +1}\,\Gamma (\alpha +1)}
  \;\dover{\emax^{\alpha -1}}{\Gamma^{2\alpha +2}}\; ,
    & $\;\; \sqrt{1-\zeta}\ll\delta\ll 1-\beta\; ,\vphantom{\Biggl(}$ \cr
  \biggl\{ {\cal F}\,\dover{\sigt}{c}\,\dover{d^2}{R_v}
  \,\dover{{\cal H}(\alpha )}{2^{2\alpha +1}\alpha}\biggr\}^{1/2}
  \;\dover{\emax^{(\alpha -1)/2}}{\Gamma^{\alpha +1/2}}\; \delta^{1/2}\; ,
    & $\;\; 1-\zeta\ll\delta\ll\sqrt{1-\zeta}\ll 1-\beta\; .
    \vphantom{\Biggl(}$ \cr}
 \label{eq:thetabasymp}
\end{equation}
These asymptotic formulae are depicted as thin, light, dotted lines in
Fig.~\ref{fig:thetabdelta} for the \teq{\Gamma =40} galactic halo case,
clearly indicating how the contours closely approach these in the
appropriate ranges of \teq{\delta}.  The lowest \teq{\Gamma}
examples in Fig.~\ref{fig:thetabdelta} for each of the galactic halo
and cosmological scenarios do not realize thin shell portions of phase
space without assuming significant opening angles \teq{\Thetab}.  This
feature marks the general property that low \teq{\Gamma} curves occupy
the upper right corner of the \teq{\Thetab}--\teq{\delta} diagram, a
domain where the thick shell parts of the solutions in
Figs.~\ref{fig:gammin} and~\ref{fig:gammin1s} are appropriate.  Other
trends relevant to the \teq{\Thetab}--\teq{\delta} diagram include a
general reduction of \teq{\Thetab} with increases in \teq{\alpha} or
decreases in \teq{\emax}.  

This concludes the survey of observational and theoretical parameter
space.  In view of the extensive presentation of results in this
section, it is important to highlight the implication of this work that
is most salient for gamma-ray bursts.  The principal conclusion of our
work (expounded in brief in Baring and Harding 1996, with a preliminary
version given in Harding and Baring 1994), is that \teq{\gammin} is
quite insensitive to the choice of \teq{\Thetab} when
\teq{\Thetab\gtrsim 1/\gammin}, behaviour that can be inferred from
Figs.~\ref{fig:taupp}.  This result arises because causality restricts
the available phase space for pair production interactions more
effectively than does the expansion opening angle \teq{\Thetab}, when
\teq{\Thetab\gtrsim 1/\gammin}; it has profound repercussions for
gamma-ray burst models.  For such source models, the principal
advantage (e.g.  Krolik and Pier 1991) of restricting \teq{\Thetab} to
small values like \teq{1/\Gamma} is a lower (solid angle-reduced)
luminosity at the source for a given observed flux.  However, the
number of non-repeating sources must then be a factor
\teq{\Thetab^{-2}\sim \Gamma^2} higher in order  to account for the
observed burst rate.  In the case of cosmological GRBs, this factor
could be as high as \teq{10^6} for the values of \teq{\gammin}
determined here, which is unacceptably large for many models,
particularly those that involve neutron star--neutron star or neutron
star--black hole mergers (Paczy\`nski 1986; Eichler et al. 1989;
Narayan, Piran and Shemi 1991; M\'esz\'aros \& Rees 1992), failed Type
1b supernovae (Woosley 1993) and rapid spin-down of high-field
millisecond pulsars (Usov 1992).  This defined the commonly-perceived
``number problem'' for cosmological bursts.  Clearly, in view of the
results of our analysis, this problem is a non-issue, since imposition
of small opening angles \teq{\Thetab} in order to satisfy pair
production transparency in EGRET bursts is not necessary.  Hence,
causality restrictions to the optical depth differ so little between
\teq{\Thetab=90^\circ} and \teq{\Thetab\sim 1/\Gamma} cases that burst
population statistical requirements can comfortably be satisfied
without resorting to beamed expansion geometries.  Of course, opening
up the expansion angle then amplifies the energetics requirements
accordingly, so that model development must still meet the needs of
acceptable bursting rates and energy budgets.  In such considerations,
it is evident from the work presented here that pair production
constraints will play only a secondary role in determining such model
requirements, becoming involved purely through the evaluation of
permissible Lorentz factors for bulk motion in gamma-ray bursts.

\subsection{Discussion}

The choice of coupling the scale of the expansion to the variability
timescale via transverse dimensions is subjective, though it is widely
adopted in applications of bulk relativistic motion in astrophysics.
Other choices are possible, such as using burst or subpulse durations
(e.g. Fenimore, Madras and Nayakshin 1996) and/or relating these to
longitudinal dimensions in the source.  It is fitting to outline the
reasons for adhering to our preference.  Suppose that timescales larger
than the variability time \teq{\Delta t}, for example the burst
duration \teq{T_d}, are used as the observational diagnostic of the
source size.  For uniform expansions, as we have assumed, the
geometrical appearance of the ``look-back'' volume forces the time
profile to maintain a well-defined shape (\teq{[t/T_d]^{-\alpha -2}}
for spectral index \teq{\alpha =2}) that mimics the so-named FRED (fast
rise, exponential decay) profile (Fenimore, Madras and Nayakshin
1996).  This profile necessarily has a width of the order of \teq{T_d}
under these assumptions, so that its smooth, decaying shape is
inconsistent with the vast majority of burst time histories.  Temporal
consistency can therefore be attained only if the appropriate
observational timescale is of the order of the variability time
\teq{\delta t}, so that the burst comprises a multitude of shells, or
perhaps if a single shell is ``patchy'' in the transverse dimension.
The latter possibility still produces time profiles that do not match
many burst histories, so that one is compelled to adopt the many-shell
proposition, perhaps produced by a central engine, as advocated by
Fenimore, Madras and Nayakshin (1996).  The timescale that is then
appropriate is \teq{\Delta t}, precisely our choice, though the value
of this depends on whether BATSE or EGRET variabilities are used (as
discussed in Section~3.1 above).

The issue of whether the variability should be tied to transverse or
longitudinal source dimensions remains to be addressed.  Since
\teq{\Delta t} is always close to the threshold of temporal resolution
of any of the CGRO instruments BATSE, Comptel and EGRET, it is
appropriate to assume that {\it measured} variability is actually an
upper bound to the source variability.  If we opt to relate this to the
direction transverse to the line of sight to the observer, then the
inequality \teq{R_0/\Gamma\lesssim c\Delta t} follows, and it is
customary to take the equality to specify \teq{R_0}.  By the same
token, if dimensions along the line of sight are preferred, this
inequality is replaced by \teq{R_0/\Gamma^2\lesssim c\Delta t}.  A
consistent description of the expansion can only be obtained when both
of these inequalities are satisfied, which obviously occurs when the
more-constraining \teq{R_0/\Gamma\lesssim c\Delta t} is adopted.  This
motivates our choice of coupling the variability to the transverse
dimension; tying it to the line-of-sight direction is insufficiently
restrictive.  Notwithstanding, the difference between these two choices
is merely one Lorentz factor in the optical depth [compared with around
five or six imposed by the spectrum: see Eq.~(\ref{eq:beam})], to which
the estimates of \teq{\gammin} are quite insensitive: opting for
\teq{R_0/\Gamma^2\lesssim c\Delta t} reduces \teq{\gammin} by factors
of the order of two or less.  Other possibilities for choosing the
scale of the expansion exist, such as \teq{\Delta R =c\Delta t}, and
these are discussed at length in the temporal analysis of Fenimore,
Madras and Nayakshin (1996).  Note that fixing \teq{\Delta R =c\Delta
t} with either \teq{R_0\sim\Gamma c\Delta t} or \teq{R_0\sim\Gamma^2
c\Delta t} yields \teq{\Delta R/R_0\ll 1}, comfortably in the phase
space covered by Fig.~\ref{fig:thetabdelta}.  The essential point that
should be emphasized is that these subjective alternatives probe
details of the expansion microstructure that are beyond the purpose of
this analysis, and are largely peripheral to it, primarily because of
the relative insensitivity of the \teq{\gammin} estimates to these
choices.  The principal conclusions of this paper, including the
insensitivity of the optical depth to the expansion opening angle
\teq{\Thetab}, are guaranteed regardless of such variations on our
assumptions.

One question that naturally arises when obtaining estimates for the
bulk Lorentz factors via pair production constraints is why the values
of \teq{100}--\teq{1000} obtained here for cosmological bursts are of
the same order as those obtained from fireball expansions (e.g.
Paczy\`nski 1986; Shemi \& Piran 1990; Rees \& M\'esz\'aros 1992) of
enormous initial optical depths.  This similarity is no coincidence.
The bulk motions attained by the adiabatic expansion phase of pure
electron-positron (pair) fireballs yield Lorentz factors \teq{\Gamma}
that saturate at some value corresponding more or less to the
``freeze-out'' of pair production, i.e. the epoch of free expansion is
approximately marked by the onset of pair production transparency.  For
cosmological bursts, where the luminosities can be of the order of
\teq{L\sim 10^{49}-10^{50}}erg/sec and the energy deposition can be
larger than \teq{10^{51}}ergs, the optical depth is roughly \teq{L\sigt
/(R\, m_ec^2)/\Gamma^5} (e.g., for \teq{E^{-2}} spectra) and can be
\teq{10^{10}/R_{10}} or larger for \teq{\Gamma =1}, where \teq{R_{10}}
is the size of the fireball at the end of the epoch of opacity in units
of \teq{10^{10}}cm.  This optical depth can be reduced to unity by
relativistic beaming with \teq{\Gamma} in the range of
\teq{100}--\teq{1000} when \teq{R_{10}\sim 1}.  Since the Lorentz
factor attained during the fireball ``acceleration phase'' scales
roughly as its radius (e.g. Paczy\`nski 1986; Piran, Shemi and Narayan
1993), then it follows that values of a freeze-out radius of
\teq{R_{10}\sim 1} would correspond to \teq{\Gamma\sim 100}--\teq{1000}
for fireballs initiated in regions of diameter
\teq{10^7}--\teq{10^8}cm.  These \teq{10^{10}}cm scalelengths for the
onset of expansion transparency are comparable to those used for
\teq{R_0=\Gamma c\Delta t} in this paper, thereby explaining the
similarity of our estimates for \teq{\gammin} to the Lorentz factors of
fireball-initiated relativistic expansions.

The results we have presented focus on the energy range appropriate to
EGRET detections of gamma-ray bursts.  There are now ongoing programs
for searches for bursts at TeV energies, specifically the target of
opportunity monitoring of BATSE localization error boxes by the Whipple
air \v{C}erenkov experiment, using the rapid response that is
facilitated by the BACODINE alert network (Barthelmy et al. 1995).
While these efforts have failed to provide any positive TeV detections
so far, probably due to the fact that Whipple's sensitivity threshold
still inhibits any possibility of detection for all but the very
brightest of bursts (see Connaughton et al. 1995 for a discussion of
the current Whipple sensitivity), the prospect of large field of view
monitoring of the sky by the air \v{C}erenkov water tank detector
MILAGRO (e.g. Yodh 1996) in the very near future, promotes the
extension of our bulk motion estimates  to the TeV energy range.  Such
considerations also anticipate future space missions like GLAST, which
will span the 10 MeV--200 GeV range.  Obviously, increasing \teq{\emax}
to TeV-type energies would tend to push estimates of the bulk Lorentz
factor up, in order to suppress pair creation.  To explore the
implications of TeV-emitting bursts sources for estimates of
\teq{\gammin}, we computed the infinite power-law ``blob'' calculation
solutions to Eq.~(\ref{eq:Gammablob}), and depicted them in
Fig.~\ref{fig:gamminTeV}, as a function of the spectral index
\teq{\alpha_h}.  It is sufficient to focus on this simplest of cases,
noting that the complicating effects of expansion geometry mirror those
considered at lower maximum energies.

\placefigure{fig:gamminTeV}

In the figure, the cosmological cases exhibited the expected trend of a
dramatic increase in \teq{\gammin} for flatter spectra, a consequence
of the enhanced supply of interacting photons at
\teq{\sim\Gamma^2/\emax} (generally well above 511 keV) for lower
values of \teq{\alpha_h}.  For typical EGRET source spectral indices,
in the range 2--3, \teq{\gammin} is indeed an increasing function of
\teq{\emax}.  However, for \teq{\alpha_h <1}, this behaviour is
reversed, with \teq{\gammin} declining with \teq{\emax}, because the
optical depth in Eq.~(\ref{eq:Gammablob}) is then a decreasing function
of \teq{\emax}.  Note that in the figure, the galactic halo case
(\teq{d=100}kpc) displays a comparative insensitivity of \teq{\gammin}
to \teq{\alpha_h}.  This insensitivity arises because the energies
(\teq{\sim\Gamma^2/\emax}) of the photons interacting with those at
\teq{\emax} are generally relatively near the pivotal energy of
\teq{511} KeV, where the source flux is pinned.  In
Fig.~\ref{fig:gamminTeV}, the source flux \teq{f} at 511 keV is typical
of BATSE burst detections; for this flux, the MILAGRO experiment will
be sensitive to bursts with \teq{\alpha_h\lesssim 2.6}, a dividing line
that is marked in the figure.  Hence, only bright, flat spectrum
sources like GRB 910503 provide good candidates for potential
detections at TeV energies by MILAGRO, and for that matter Whipple.
GLAST will have the capability of spanning the EGRET and sub-TeV energy
ranges.  This extension of \teq{\gammin} estimates to \teq{\emax}
values in the TeV range of course assumes that bursts intrinsically
emit at such high energies.  Note that for the cosmological case, such
intrinsic burst emission is also subject to pair creation in collisions
with photons supplied by radiation fields external to the burst, for
example the infrared background (Stecker and De Jager 1996, Mannheim,
Hartmann and Funk 1996).  This type of attenuation is strongly dependent
on the redshift of the source, so that sources at 1 Gpc would be
strongly attenuated at 1 TeV, while those at 100 Mpc would be
transparent.  Such external attenuation considerations are beyond the
scope of this paper, and are examined in detail by Mannheim, Hartmann
and Funk (1996).

The analysis of this paper has made the expedient assumption that the
burst spectra are infinite power-laws, with the spectrum matching that
observed by EGRET.  This, of course assumes that the turnovers seen at
sub-MeV energies in BATSE data are immaterial to pair production
calculations.  The relevance of spectral curvature below the EGRET
range to attenuation studies depends upon the energies of the photons
that interact with the test photons at \teq{\erg_t=\emax}, which
generally are around \teq{\Gamma^2/\emax} in the stationary observer's
reference frame for infinite power-laws, as discussed just above.
Clearly, low energy spectral paucity (relative to EGRET-range
power-laws) at around \teq{\Gamma^2/\emax} limits contributions near
threshold so that phase space near the pair production threshold is
inaccessible, and the process is pushed into the Klein-Nishina regime:
the optical depth drops accordingly.  Hence MeV and sub-MeV spectral
curvature plays a role in opacity determinations when
\teq{\Gamma^2/\emax\lesssim 1}, a situation that manifests itself at
galactic halo (or even galactic disk) distances for most of the EGRET
bursts, as is evident from an inspection of Tables~\ref{Table2}
and~\ref{Table3}.  Consequently, regimes of transparency may become
possible at high energies in galactic bursts, so that the range of
opacity becomes finite.  These issues are discussed in detail in Baring
and Harding (1997), where it is demonstrated that realistic broad-band
GRB spectra may yield broad absorption troughs (and also other spectral
forms such as shelfs) in the 1 GeV -- 1 TeV range of bursts spectra in
galactic halo sources, distinctive spectral features whose existence or
otherwise could be probed by experiments such as GLAST, Whipple and
MILAGRO.  Note however, that for cosmological sources, the inferred
\teq{\Gamma}s are so high that \teq{\Gamma^2/\emax} always exceeds
unity and the infinite power-law calculations presented here are always
quite appropriate.  As a result, the optical depth is then always
monotonically increasing in \teq{\emax}, so that only simple
(exponential or power-law) spectral turnovers are possible.  Hence,
Baring and Harding (1997) postulated that such distinguishable spectral
structure may provide a means of discriminating between a galactic or
cosmological origin for gamma-ray bursts, an enticing prospect for the
high energy astrophysics community.

\section{CONCLUSION}

In this paper, we have presented our extension of pair production
transparency calculations in relativistically expanding gamma-ray burst
sources to quite general geometries, including shells of finite
thickness and arbitrary opening angle.  This work includes an extensive
analytic reduction of the optical depth from a quintuple integral to a
single integral in the special, but quite broadly applicable, case of
observing photons only along the axis of the expansion.  Such a
reduction is extremely expedient for opacity and transparency
considerations, providing the reliable and numerically-amenable
analytic expressions in Eqs.~(\ref{eq:flux}) and~(\ref{eq:tauppfin})
that completely describe the pair production optical depth.  We
determine that the minimum bulk Lorentz factor \teq{\gammin} for the
EGRET sources to be optically thin up to the maximum energies observed,
i.e. display no spectral attenuation, is only moderately dependent on
the shell thickness and virtually independent of its opening solid
angle if $\Thetab\gtrsim 1/\gammin$.  This insensitivity to $\Thetab$,
which is a consequence of the strong impact that causality has on the
available interaction phase space, relieves the commonly-perceived
number problem for non-repeating sources at cosmological distances: it
is not necessary to invoke small $\Thetab$ to effect photon escape.
This negation of the number problem for a wide range of expansion
geometries is the principal conclusion of this paper, and is an
important result for specific cosmological burst models.  The values of
\teq{\Gamma} obtained, typically of the order of 10--30 for halo bursts
and \teq{\gtrsim 100} for sources of cosmological origin, depend only
moderately on the choice of GRB timescale used to determine the expansion
size.  Our new limits on required expansion velocity for given source
geometries will significantly aid the placing of realistic constraints
on gamma-ray burst source models.

\acknowledgements
We thank Brenda Dingus and Jennifer Catelli for many discussions about
EGRET burst data, and Ed Fenimore and Jim Ryan for numerous conversations 
concerning gamma-ray bursts.  This work was funded, in part, by the
Compton Gamma-Ray Observatory Guest Investigator Program.

\clearpage

\appendix
\section{APPENDIX}

\centerline{\eightrm PROPERTIES OF THE HYPERGEOMETRIC FUNCTION
${\cal G}_{\alpha}(z)$ \rm}\vskip 5pt

In the integrand of the expression for the pair production optical depth,
Eq.~(\ref{eq:tauppfin}), appears the hypergeometric function
\begin{equation}
   {\cal G}_{\alpha}(z)\;\equiv\;
   \int_0^1 dq\,\dover{(1-q)^{2\alpha}}{(1-z q)^{\alpha +1}}\; =\;
   \dover{1}{1+2\alpha}\, F(\alpha +1,\, 1 ;\, 2\alpha +2;\, z)\quad .
 \label{eq:Gdef1}
\end{equation}
This can be represented by the alternative hypergeometric form
\begin{equation}
   {\cal G}_{\alpha}(z)\; =\;\dover{1}{1+2\alpha}\,
   \dover{1}{1-z}\, F(\alpha +1,\, 1 ;\, 2\alpha +2;\, z/[z-1])\quad ,
 \label{eq:Gdef2}
\end{equation}
using the transformation formula 9.131.1 of Gradshteyn and Ryzhik (1980).

The argument of \teq{{\cal G}_{\alpha}(z)} is either \teq{\lambda} or
\teq{\sigma\lambda}, which can be derived from
Eq.~(\ref{eq:lambdasig}).  It is clear that for \teq{s\geq 1},
\teq{\lambda} attains values within a range \teq{-\infty <\lambda
<\lambda_{\hbox{\sevenrm max}}}, with \teq{\lambda_{\hbox{\sevenrm
max}}=\beta /(1+\beta )} (note that \teq{d\lambda /ds >0}).  Clearly
the condition \teq{\zeta <1} renders \teq{\sigma} less than unity, so
that \teq{\sigma\lambda} is also bounded by the range \teq{(-\infty ,\,
\beta /(1+\beta )]}.  It follows that there are two natural ways to
evaluate \teq{{\cal G}_{\alpha}(z)} for the purposes of this paper,
both using the series expansion (9.100 of Gradshteyn and Ryzhik,
1980):
\begin{equation}
   F(\alpha +1,\, 1 ;\, 2\alpha +2;\, z)\; =\;
   1 + \dover{\alpha + 1}{2\alpha + 2}\, z +
   \dover{(\alpha + 1)(\alpha + 2)}{(2\alpha + 2)(2\alpha + 3)}\, z^2 
   + \dots \quad .
 \label{eq:hyperseries}
\end{equation}
When \teq{\vert z\vert <1}, this series can be used directly with
convergence as rapidly as the geometric series \teq{\sum_n (z/2)^n}.
When \teq{-\infty <z <-1}, the alternative form for \teq{{\cal
G}_{\alpha}(z)} in Eq.~(\ref{eq:Gdef2}) can be used, where \teq{1/2
<z/(z-1) <1}; the series in Eq.~(\ref{eq:hyperseries}) with the
substitution \teq{z\to z/(z-1)} then also converges like the geometric
series \teq{\sum_n (z/2[z-1])^n}, i.e. with the same rapidity.  With
this scheme, computation of \teq{{\cal G}_{\alpha}(z)} to high accuracy
is quick.

Note that as \teq{s\to 1}, \teq{\lambda\to -\infty}.  Using the
representation in Eq.~(\ref{eq:Gdef2}) and also
Eq.~(\ref{eq:lambdasig}), it is clear that in this limit
\teq{(1+2\alpha ){\cal G}_{\alpha}(\lambda)} approaches \teq{ F(\alpha
+1,\, 1 ;\, 2\alpha +2;\, 1)/(1-\lambda)}, and therefore becomes
approximately proportional to \teq{s-1}.  It follows that the integrand
in Eq.~(\ref{eq:tauppfin}) is finite as \teq{s\to 1}.

Also of use in this paper, specifically in the determination of the
optical depth in the limit of filled spherical expansions, is the
integral identity
\begin{equation}
   \int^1_0 dz\, z\, {\cal G}_{\alpha}(z)\; =\; \dover{1}{\alpha} + 
   \dover{2}{\alpha -1}\,\Bigl\{\psi (2\alpha ) - \psi (\alpha )
   -1\Bigr\}\quad ,\quad \psi (x)\, =\,
   \dover{d}{dx}\Bigl\{\log_e\Gamma (x)\Bigr\}\;\; ,
 \label{eq:Gint1}
\end{equation}
for use in \teq{\beta\approx 1} situations.  This can be established
using the integral representation of \teq{{\cal G}_{\alpha}} in
Eq.~(\ref{eq:Gdef1}), then reversing the order of integration and
performing the \teq{z}-integration analytically.  An integration by
parts then enables the use of identity 3.231.5 of Gradshteyn and Ryzhik
(1980), and the result ensues.  The finite rational series for
\teq{\psi (x+n)-\psi (x)} was also used in manipulating
Eq.~(\ref{eq:Gint1}), and for integer \teq{\alpha} it can be used to
obtain rational values for these integrals.  For \teq{\beta\ll 1} cases
of initially filled spherical expansions, the integral
\begin{equation}
   \int^1_0 \dover{dz}{2-z}\, {\cal G}_{\alpha}(z)\; =\; \int_0^{\pi /2}
   d\theta\,\theta\,\cos^{2\alpha}\theta\; \approx\;\dover{1}{1+2\alpha}\,
   \Biggl\{ \, 1+(\pi -3)\biggl[\dover{3}{2(\alpha +1)}\biggr]^{9/8}\,
   \Biggr\} 
 \label{eq:Gint2}
\end{equation}
is needed.  The identity is established by using the transformation
9.134.1 in Gradshteyn and Ryzhik (1980) for the hypergeometric function
in Eq.~(\ref{eq:Gdef1}), changing to an integration variable of
\teq{[z/(2-z)]^2}, using the integral representation for general
hypergeometric functions in 9.111 of Gradshteyn and Ryzhik (1980), and
then reversing the order of integration.  The integral, when multiplied
by \teq{1+2\alpha}, is only weakly dependent on \teq{\alpha}, and the
simple approximation obtained in Eq.~(\ref{eq:Gint2}) is accurate to
better than 1\% for \teq{0<\alpha <1/2} and better than 0.1\% for
\teq{1/2<\alpha <10}.

\newpage

\begin{table}
\dummytable\label{Table1}
\end{table}

\begin{table}
\dummytable\label{Table2}
\end{table}

\begin{table}
\dummytable\label{Table3}
\end{table}

\clearpage

\centerline{}
\vskip 0.5in
   \centerline{\psfig{figure=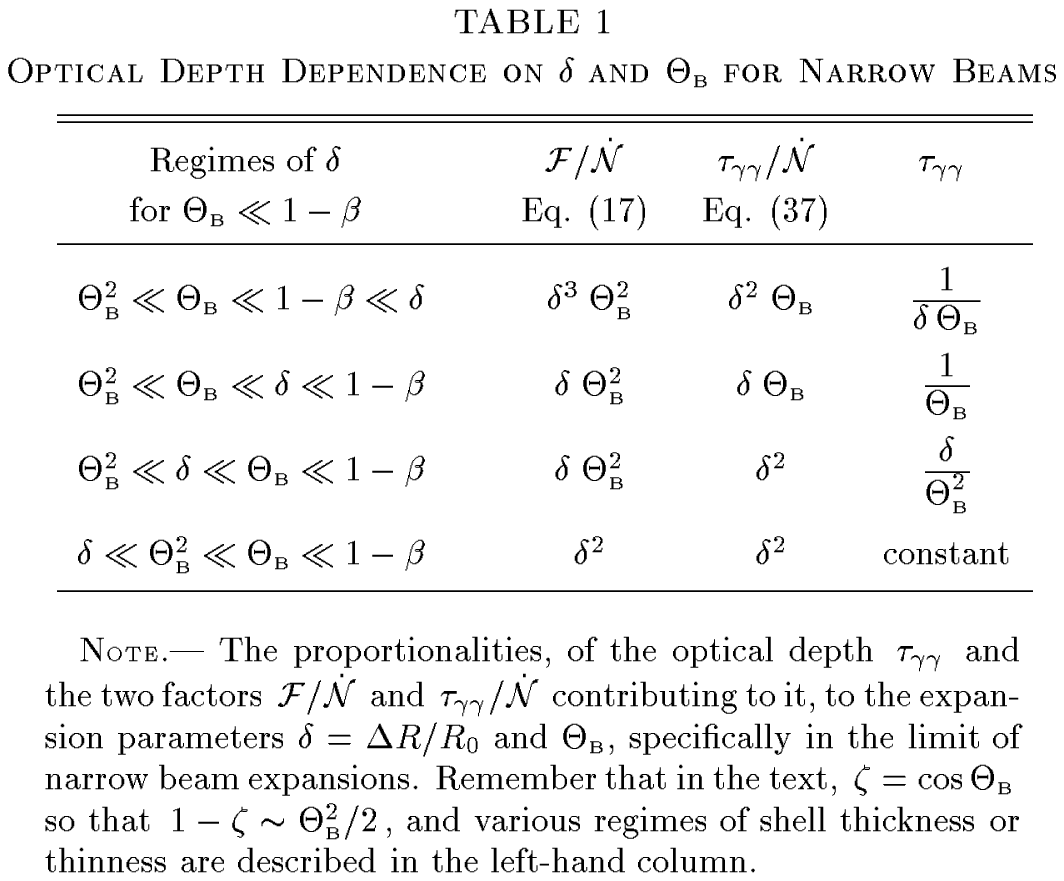,width=4.5in}}
\clearpage

\centerline{}
 \vskip 0.5in
  \centerline{\psfig{figure=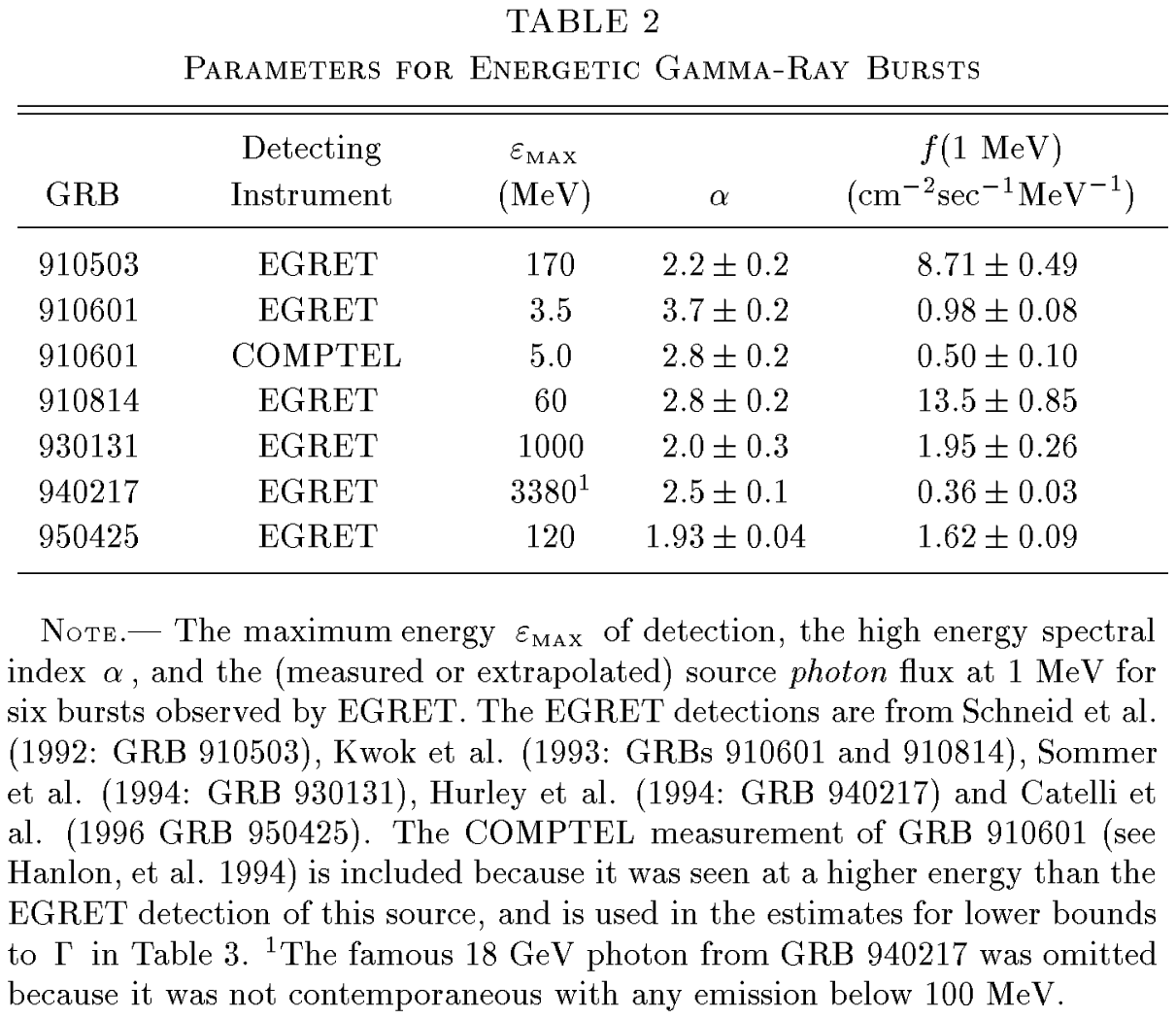,width=5.2in}}
\clearpage

\centerline{}
\vskip 0.5in
   \centerline{\psfig{figure=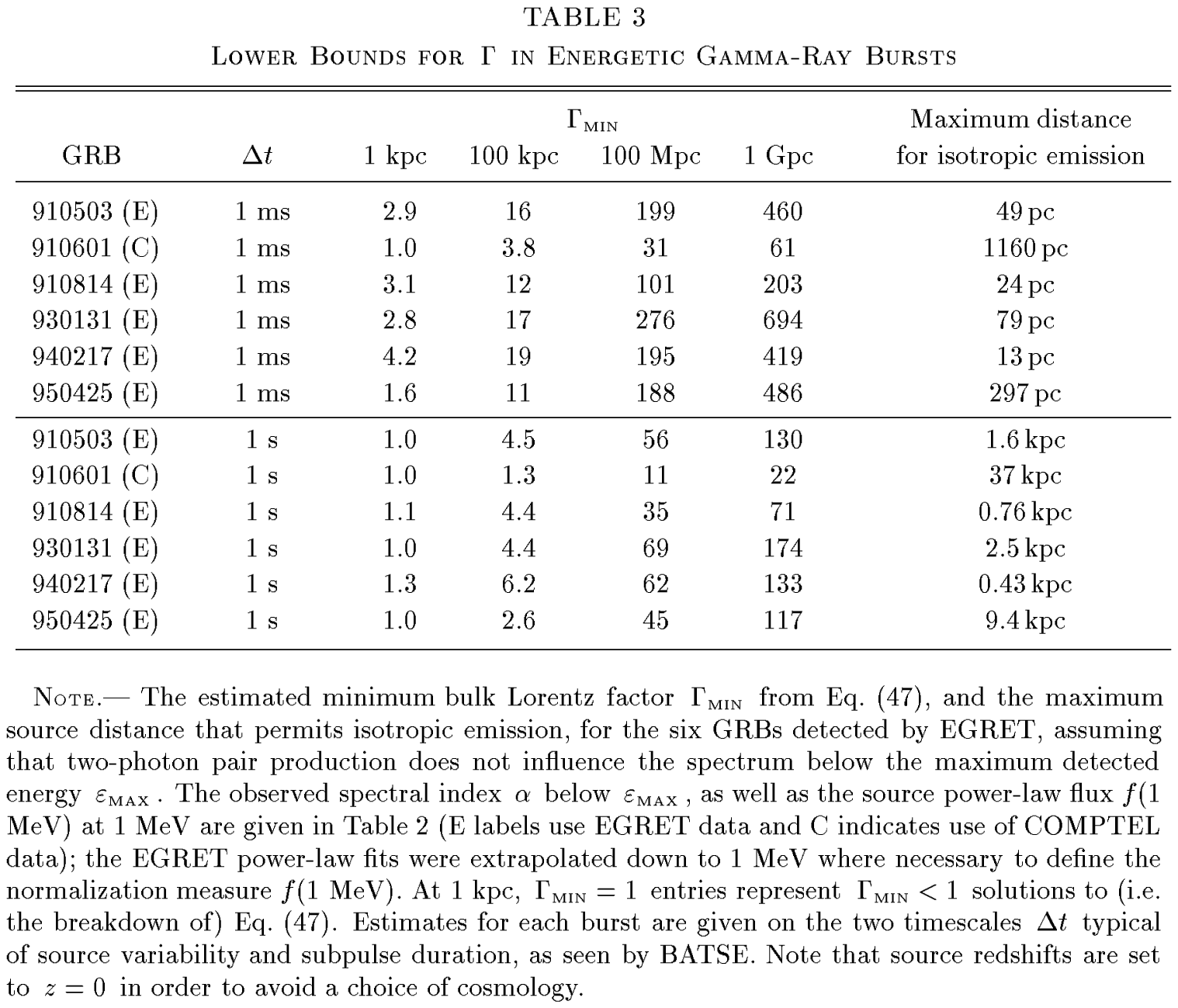,width=6.3in}}
\clearpage

%
%
%
\centerline{}
\vskip -1.0in
   \centerline{\psfig{figure=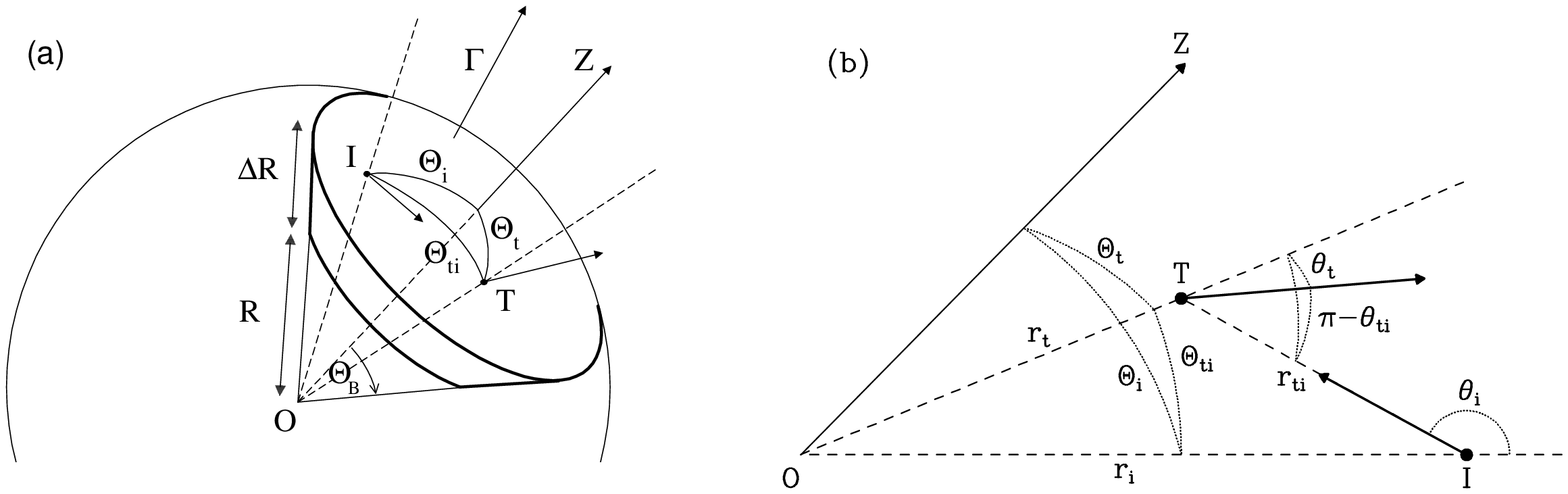,width=8.5in,angle=90}}
\vskip -2.0in
\figcaption{
   (a) A depiction of the source geometry for the relativistic expansions
of bulk Lorentz factor \teq{\Gamma} considered in this paper.  The
emission region is at all times a conical sector of a spherical shell
with half-angle \teq{\Thetab}.  At some time \teq{t} the inner radius
of the shell is given by \teq{R}, while the thickness of the shell is
always \teq{\Delta R}.  The test or observable photon is marked by T,
and it can interact via \teq{\gamma\gamma\to e^+e^-} with other
(so-called interacting) photons at typical position I.  The angles
their position vectors (from the origin O of the expansion) make with
the axis OZ of the expansion are \teq{\Theta_t} and \teq{\Theta_i},
respectively; the angle subtended by these vectors at the origin is
\teq{\Theta_{ti}}.
   (b) A depiction of the spatial and angular variables, as defined 
in the text, that are relevant to the analysis of pair production in
expanding burst sources.  OZ represents the axis of the expansion and
generally is not co-planar with the plane (OTI) formed by the positions
of the test (T) and interacting (I) photons and the origin (O) of the
expansion.  The two photon momentum vectors (heavy lines) have an angle
\teq{\theta_{ti}} between them and define a third plane that is
generally not co-planar with OTI.  The ``angles of non-radiality'' of
the test and interacting photons are \teq{\theta_t} and \teq{\theta_i},
respectively, and \teq{\Theta_{ti}} defines the angle subtended at the
origin by the positions of the two photons.  The labelled distances are
\teq{r_t=\vert\overline{OT}\vert},
\teq{r_i=\vert\overline{OI}\vert} and
\teq{r_{ti}=\vert\overline{TI}\vert}.
    \label{fig:geometry}
}
\clearpage

\figureouttwo{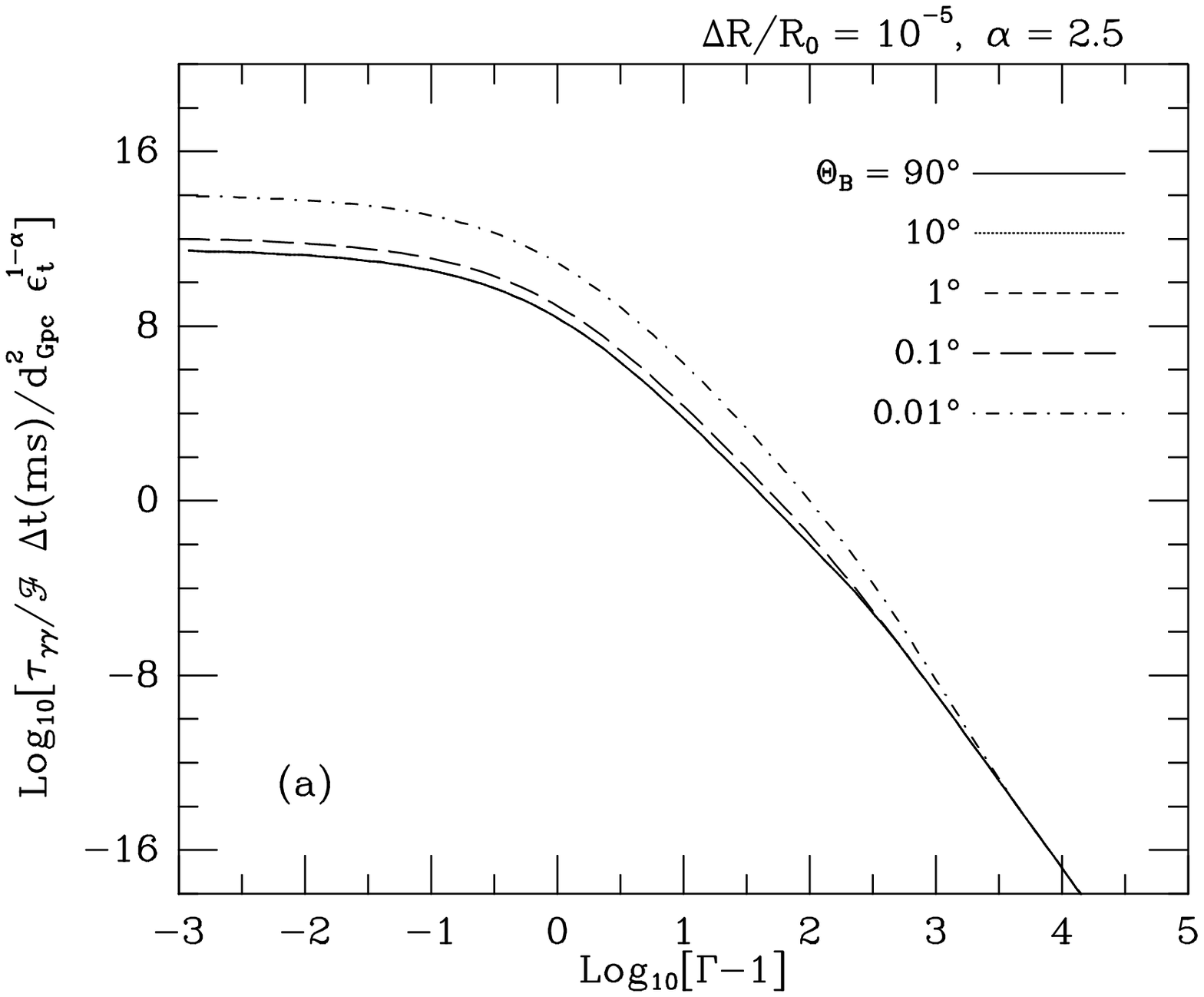}{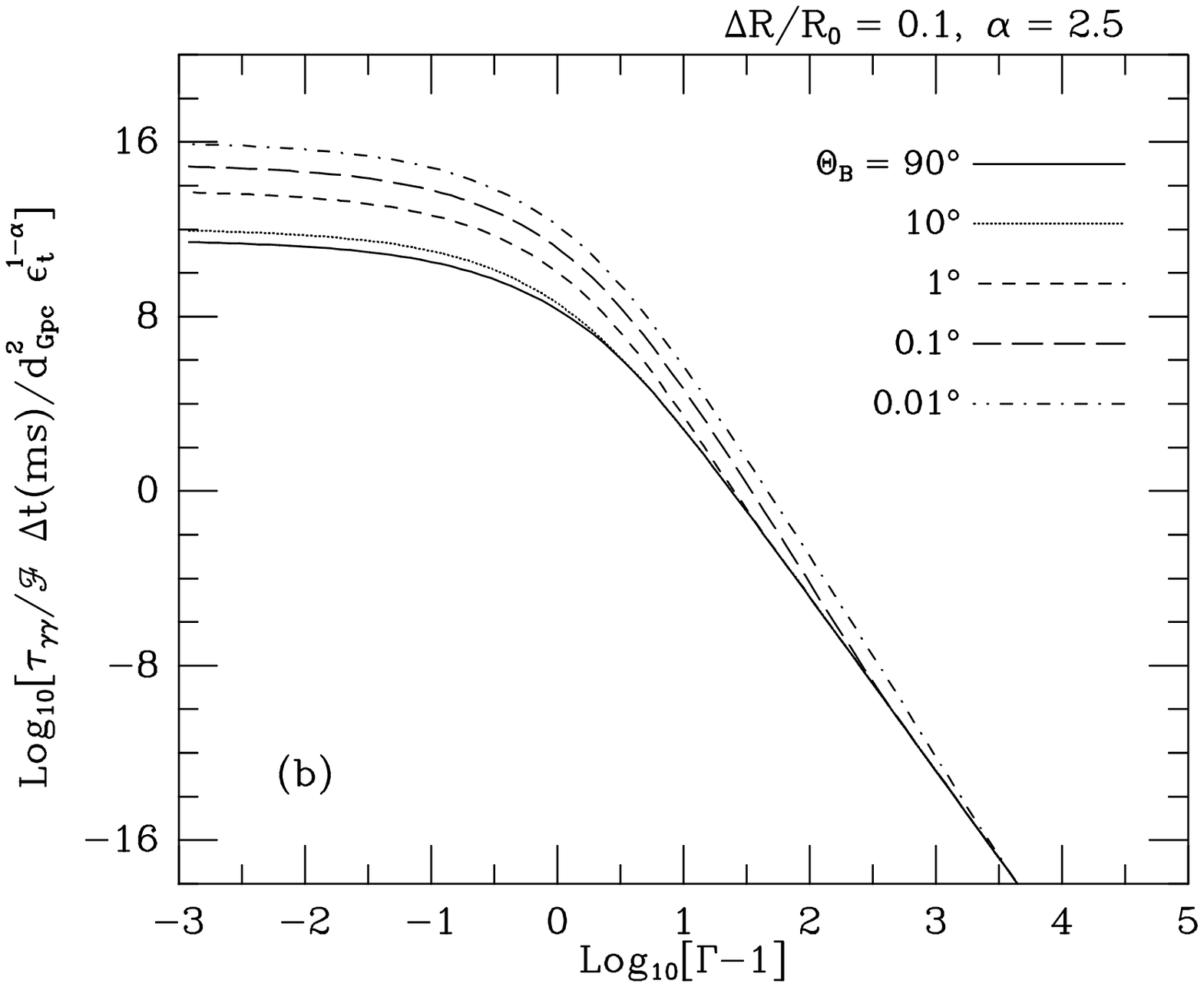}{
   The optical depth \teq{\taupp (\erg_t)} of test photons to pair
production, scaled as in Eq.~(\ref{eq:numerical}), but for redshift
\teq{z=0}, as a function of bulk Lorentz factor \teq{\Gamma} for
different opening angles \teq{\Thetab} of the expansion, as labelled,
and for shell fractional thicknesses (a) \teq{\Delta R/R_0=10^{-5}} and
(b) \teq{\Delta R/R_0=10^{-1}}.  The curves are obtained for a typical
EGRET source spectral index of \teq{\alpha =2.5} and for
\teq{R_0=\Gamma c\Delta t} (discussed in the text).  The scaling
distance is chosen to illustrate typical optical depths of cosmological
bursts, however the curves are easily translated down by around 8
orders of magnitude if a galactic halo hypothesis is preferred.
     \label{fig:taupp}
}

\figureoutsmall{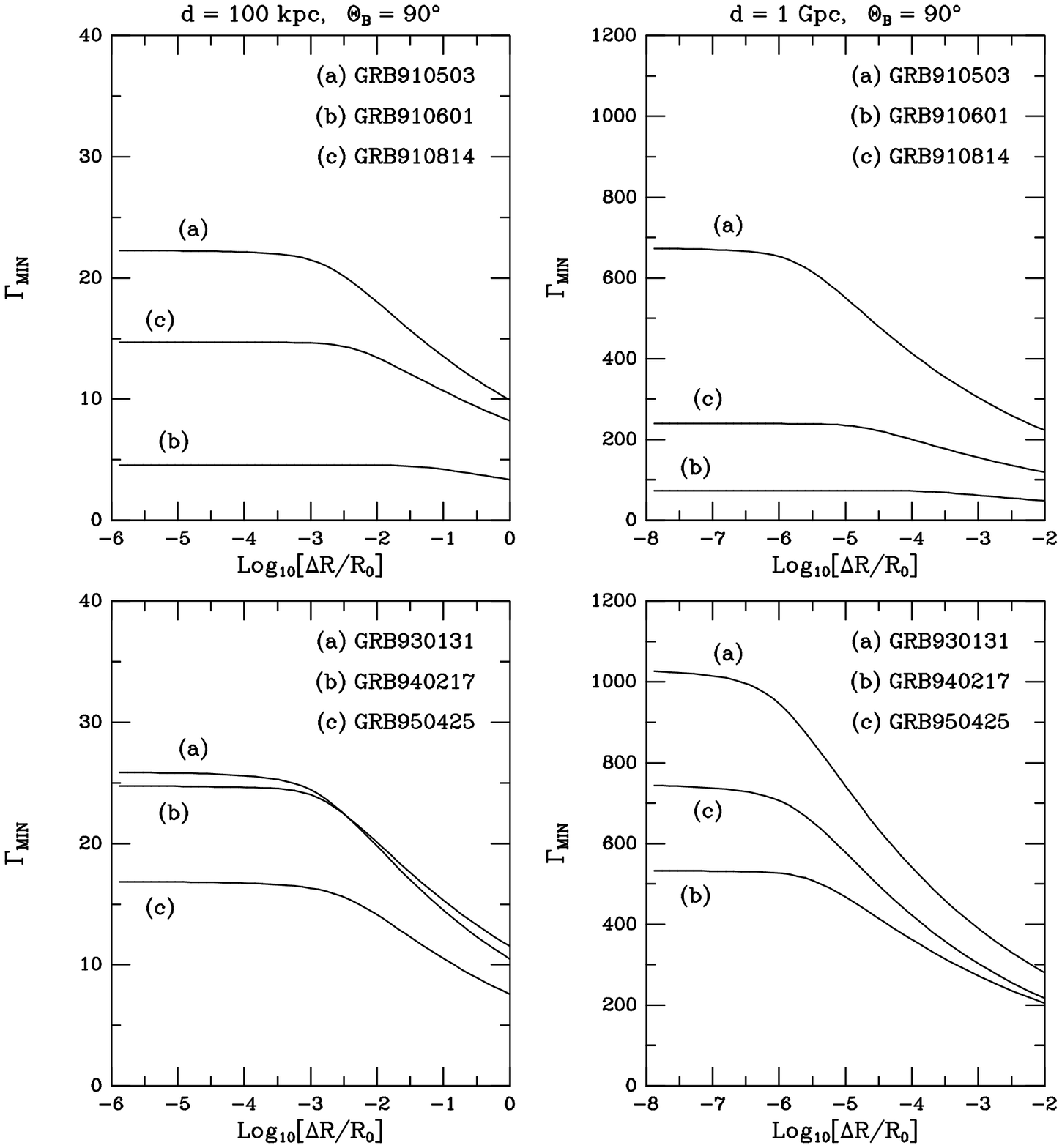}{
The minimum bulk Lorentz factor \teq{\gammin} for six EGRET GRBs, as
obtained from the pair production condition \teq{\taupp (\erg_t)=1} in
Eq.~(\ref{eq:tauppfin}).  Here \teq{\erg_t} is the maximum energy
\teq{\emax} detected by EGRET; values of \teq{\emax} and other
observational parameters are listed in Table~\ref{Table2} (Comptel data
are used for GRB 910601, as in Table~\ref{Table3}).  The top two panels
consider the first three EGRET bursts, while the bottom three are for
the most significant of more recent events.  Results are shown for two
different source distances, \teq{d=100}kpc (left-hand panels) and
\teq{d=1}Gpc (right panels), corresponding to galactic halo and
cosmological scenarios, respectively.  The expansion opening half-angle
is set to be \teq{\Thetab=90^\circ}, and the variability timescale is
\teq{\Delta t=1}ms.  Flat portions of the curves define the thin-shell
limit \teq{\Delta R/R_0\ll 1-\beta}, while the complementary sloping
portions correspond to thick-shell expansions.
     \label{fig:gammin}
}


\figureout{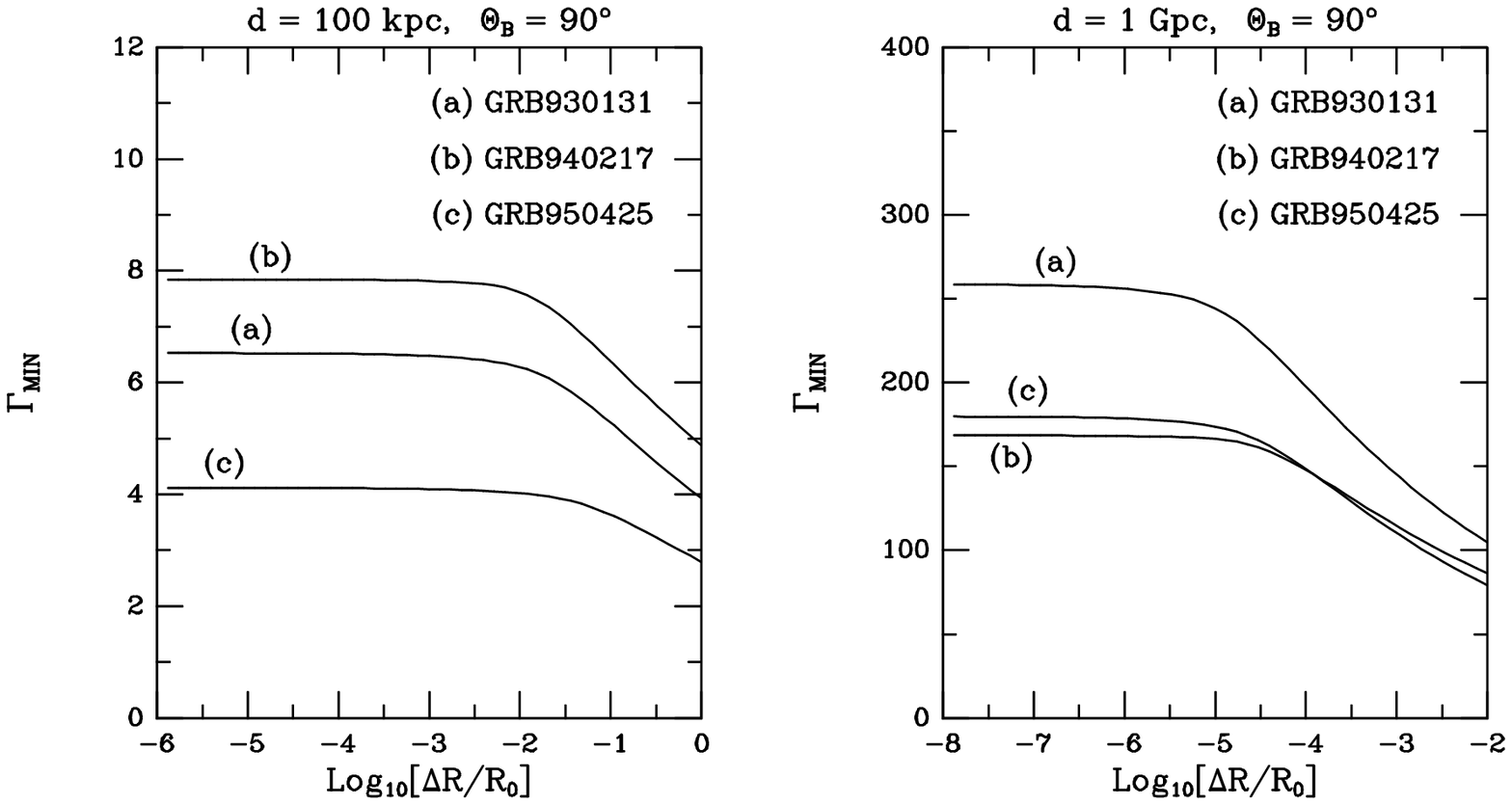}{
The minimum bulk Lorentz factor \teq{\gammin} for the most recent three
of the six EGRET bursts depicted in Fig.~\ref{fig:gammin}, but now for
\teq{\Delta t=1} second and \teq{d=100} kpc (left hand panel) and
\teq{d=1} Gpc (right hand side), to illustrate the effect of
lengthening \teq{\Delta t}.  Again, the curves represent \teq{\taupp
(\erg_t)=1} solutions to Eq.~(\ref{eq:tauppfin}), and source parameters
are taken from Table~\ref{Table2}.  Comparison with the bottom panels
of Fig.~\ref{fig:gammin} reveals that the \teq{\gammin} are reduced
significantly for these much longer variability timescales, however the
dependence on \teq{\Delta t} is nevertheless somewhat weak.
     \label{fig:gammin1s}
}

\figureout{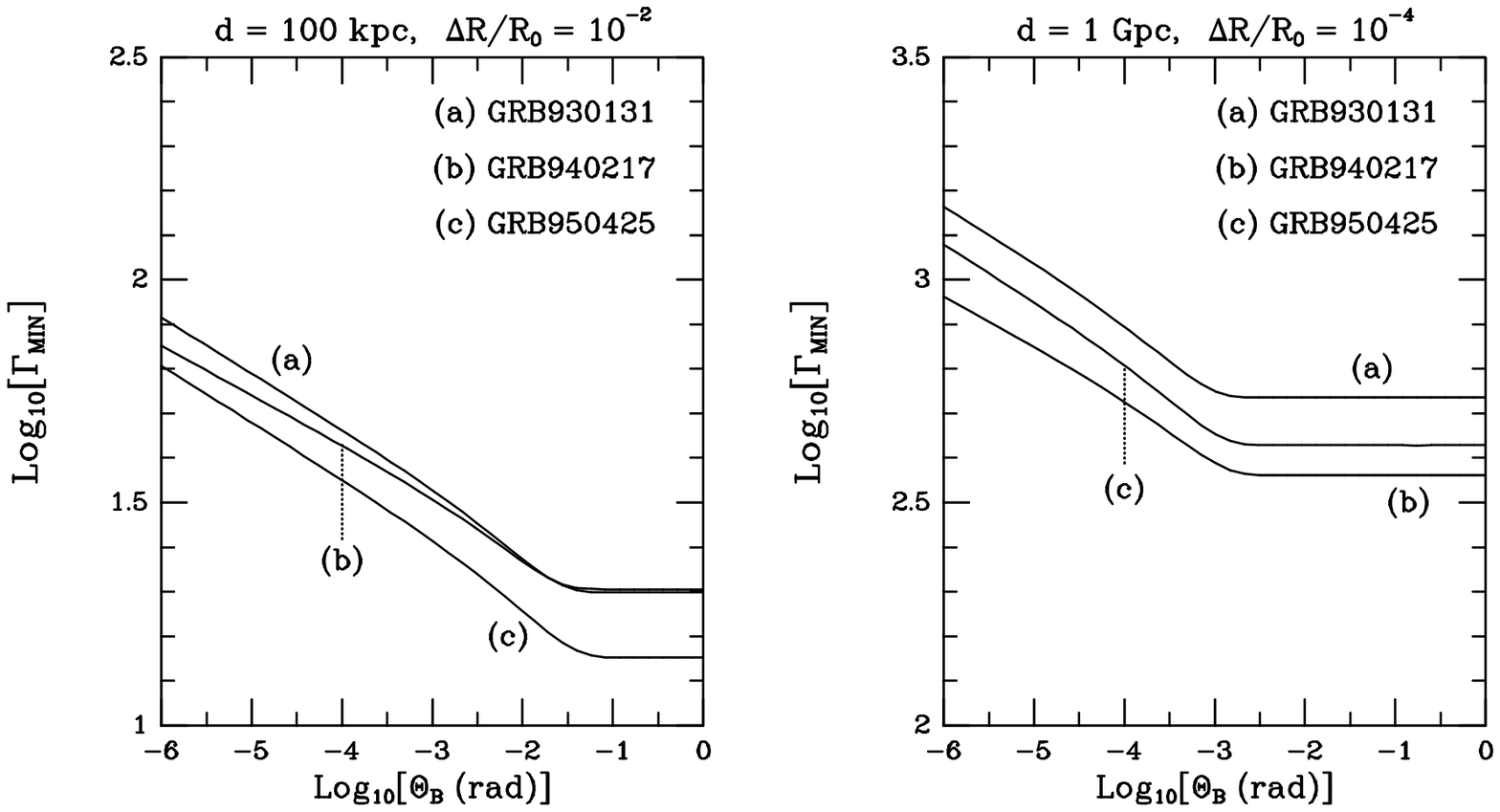}{
The minimum bulk Lorentz factor \teq{\gammin} for the most recent three
of the six EGRET bursts depicted in Fig.~\ref{fig:gammin}, for
\teq{\Delta t=1} second, but now as a function of the expansion opening
angle \teq{\Thetab}.  The left hand panel is for \teq{d=100} kpc and
\teq{\Delta R/R_0=10^{-2}} and the right hand side depicts the
cosmological case of \teq{d=1} Gpc and \teq{\Delta R/R_0=10^{-4}}; the
choices of \teq{\Delta R/R_0} correspond to transitions between the
thin and thick-shell regimes.  Again, the curves represent \teq{\taupp
(\erg_t)=1} solutions to Eq.~(\ref{eq:tauppfin}), and source parameters
are taken from Table~\ref{Table2}.  Clearly evident is the independence
of \teq{\gammin} on opening angle when \teq{\Thetab\gg\Delta R/R_0},
and a very weak power-law dependence [whose slope depends on a burst's
spectral index \teq{\alpha} via Eq.~(\ref{eq:beam})] for 
\teq{\Thetab\ll\Delta R/R_0}.
     \label{fig:gamminthet}
}

\figureout{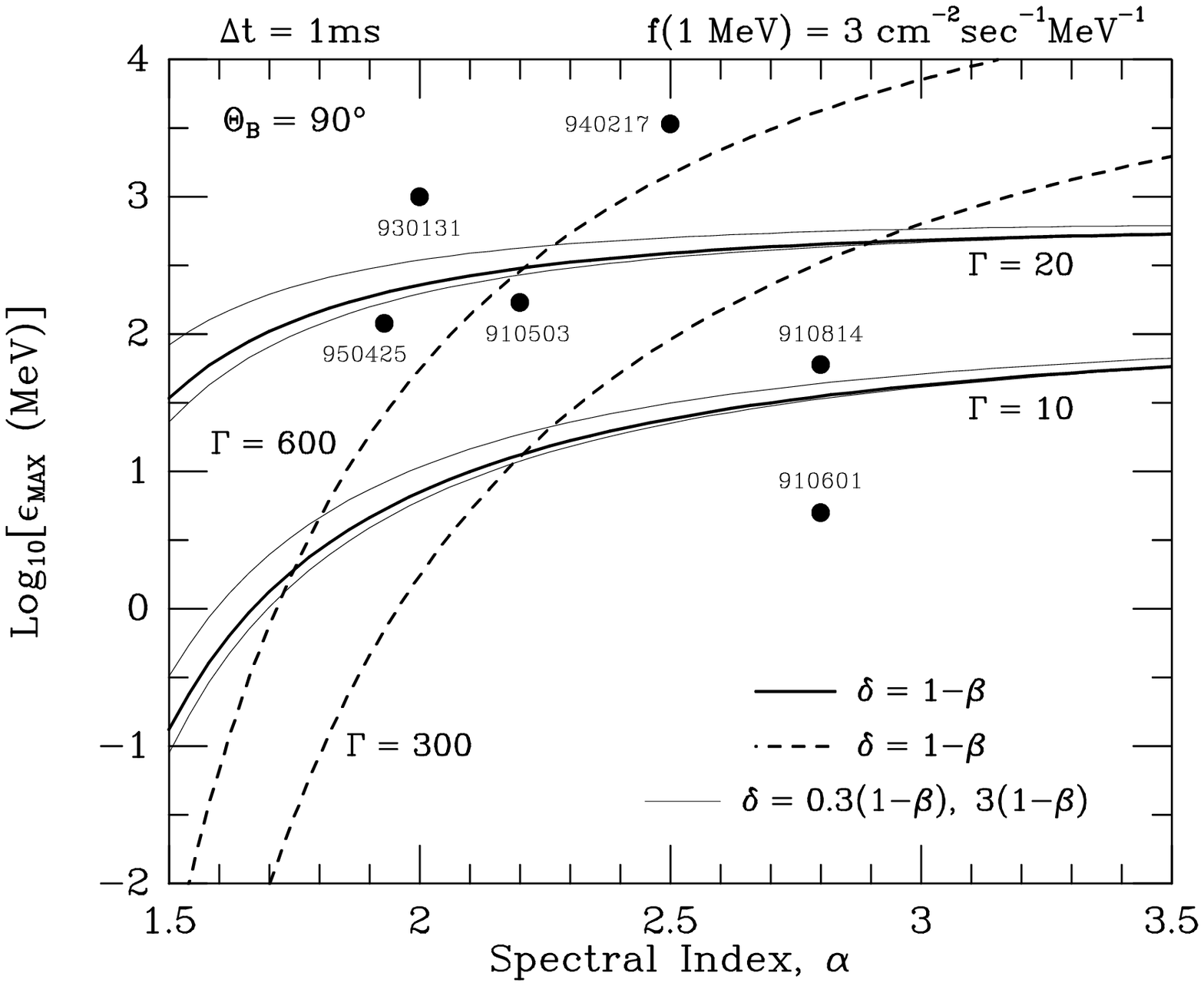}{
The phase space diagram for the observational parameters \teq{\emax}
and \teq{\alpha}, consisting of contours of constant \teq{\Gamma}, as
labelled.  The contours correspond to solutions of \teq{\taupp =1} for
specific choices of \teq{\Gamma} (\teq{=\gammin}) typical of both
galactic halo sources at \teq{d=100}kpc (three solid lines for each of
\teq{\Gamma =10, 20}) and cosmological scenarios at \teq{d=1}Gpc
(dashed lines: \teq{\Gamma =300, 600}).  The heavy contours represent
fractional thicknesses \teq{\delta\equiv \Delta R/R_0=1-\beta}, and for
the galactic halo cases of \teq{\Gamma =10, 20}, alternative
thicknesses are represented by the light contours, with the lower
curves for each \teq{\Gamma} denoting \teq{\delta =0.3(1-\beta )} and
the upper ones \teq{\delta =3(1-\beta )}.  For all contours, the
opening angle was set at \teq{\Thetab =90^\circ}, the variability
timescale set at 1 ms, and the flux \teq{f(\hbox{1 MeV)}} at 1 MeV
assumed a value typical of EGRET sources.  The observed EGRET values of
\teq{\emax} and \teq{\alpha} for six bursts (see Table~\ref{Table2})
are plotted as points; these bursts all have values of \teq{f(\hbox{1
MeV})} different from the ``canonical value'' chosen here.
     \label{fig:emaxalpha}
}

\figureout{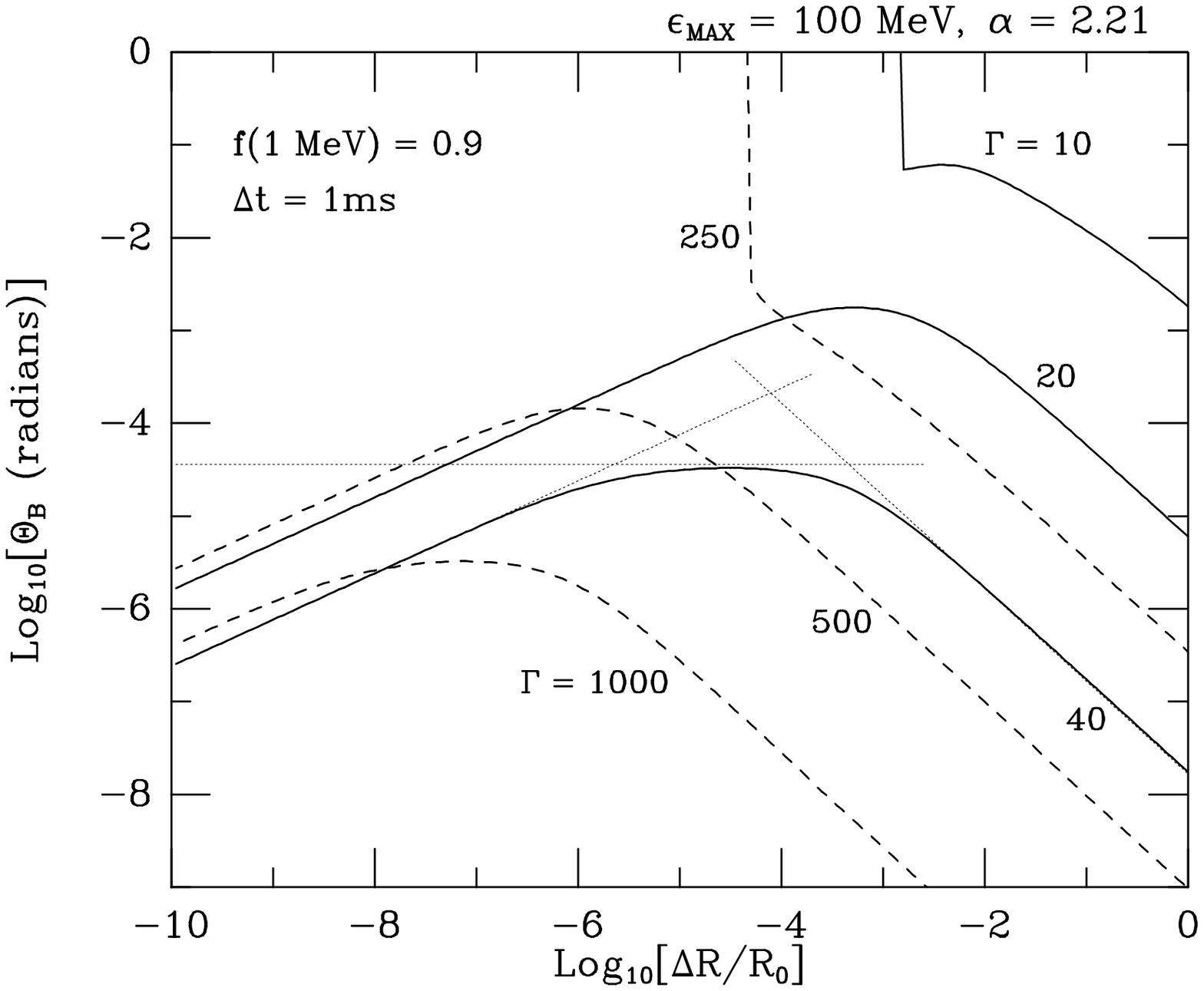}{
The phase space diagram for the theoretical parameters \teq{\Thetab}
and \teq{\delta =\Delta R/R_0}, consisting of contours of constant
\teq{\Gamma}, as labelled.  As in Fig.~\ref{fig:emaxalpha}, the
contours denote solutions of \teq{\taupp =1} for specific choices of
\teq{\Gamma} typical of both galactic halo sources at \teq{d=100}kpc
(three solid lines with \teq{\Gamma =10, 20, 40}) and cosmological
scenarios at \teq{d=1}Gpc (dashed lines: \teq{\Gamma =250, 500,
1000}).  For all contours, the maximum energy and spectral index were
set at \teq{\emax =100}MeV and \teq{\alpha=2.21}, respectively, and the
variability timescale \teq{\Delta t} and the flux \teq{f(\hbox{1 MeV)}}
at 1 MeV are representative of EGRET-like sources.  The vertical
portions of the \teq{\Gamma =10} and \teq{\Gamma =250} contours define
thin-shell/thick-shell regimes where the optical depth is independent
of \teq{\Thetab}.  The three light dotted lines are the asymptotic
limits given in Eq.~(\ref{eq:thetabasymp}).
     \label{fig:thetabdelta}
}

\figureout{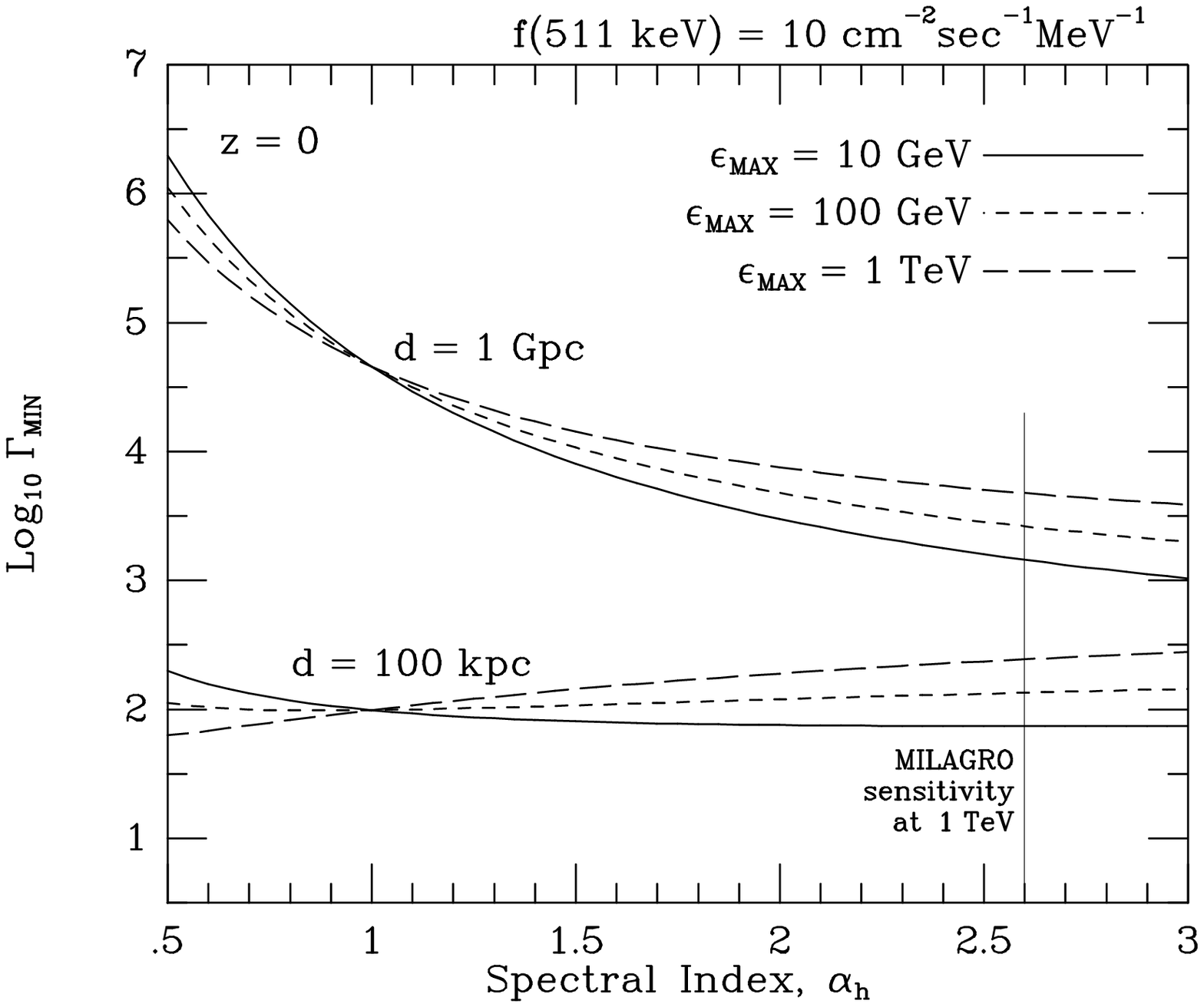}{
Solutions to Eq.~(\ref{eq:Gammablob}) for the minimum bulk Lorentz
factor \teq{\gammin} that guarantees source transparency up to energy
\teq{\emax} for two different source distances \teq{d} as labelled.
Values of \teq{\emax} are chosen to probe beyond the EGRET energy range
up to the domain of air \v{C}erenkov detection techniques.  Again infinite
power-law source spectra are assumed.  The source flux \teq{f} at 511
keV is typical of BATSE burst detections; for this flux, the MILAGRO
experiment will be sensitive to bursts with \teq{\alpha_h\lesssim 2.6}.  
The cosmological redshift was taken to be \teq{z=0} for simplicity.
     \label{fig:gamminTeV}
}


\begin{references}
%
\reference{}
Band, D., et al. 1993, \apj \vol{413}{281}
\reference{}
Baring, M.~G. 1993, \apj \vol{418}{391}
\reference{}
Baring, M.~G. 1994, \apjs \vol{90}{899}
\reference{}
Baring, M.~G. 1995, in \it Currents in High Energy Astrophysics, \rm 
   eds. Shapiro, M.~M., Silberberg, R. and Wefel, J.~P.
   (Kluwer, Dordrecht) p.~21
\reference{}
Baring, M.~G. and Harding, A.~K. 1993, in \it Proc. 23rd ICRC \rm\vol{1}{53}
\reference{}
Baring, M.~G. and Harding, A.~K. 1995, \it Adv. Space Res., \rm\vol{15(5)}{153}
\reference{}
Baring, M.~G. and Harding, A.~K. 1996, in \it Gamma-Ray Bursts, \er eds.  
  Kouveliotou, C., Briggs, M.~F., and Fishman, G.~J. (AIP Conf. Proc. 384, 
  New York) p.~724.
\reference{}
Baring, M.~G. and Harding, A.~K. 1997, \apjl\vol{481}{L85}
\reference{} 
Barthelmy, S.~D, et al. 1995, \apss \vol{231}{235}
\reference{} 
Briggs, M.~F. et al. 1996, in \it Gamma-Ray Bursts, \er eds.  
  Kouveliotou, C., Briggs, M.~F., and Fishman, G.~J. (AIP Conf. Proc. 384, 
  New York) p.~335.
\reference{} 
Catelli, J.~R. et al. 1996,in \it Gamma-Ray Bursts, \er eds.  
  Kouveliotou, C., Briggs, M.~F., and Fishman, G.~J. (AIP Conf. Proc. 384, 
  New York) p.~185.
\reference{} 
Connaughton, V. et al. 1995, in Proc. 24th ICRC (Rome), Vol. II, p.~96.
\reference{} 
Dingus, B.~L. 1995, \apss \vol{231}{187}
\reference{} 
Dingus, B.~L., et al. 1994, in \it Gamma-Ray Bursts, \er eds.  
  Fishman, G.~J., Hurley, K. and Brainerd, J.~J. (AIP Conf. Proc. 307, 
  New York) p.~22
\reference{}
Eichler, D., Livio, M., Piran, T., and Schramm, D.~N. 1989, \nat\vol{340}{126}
\reference{}  
Epstein, R.~I. 1985, \apj \vol{297}{555}
\reference{}  
Fenimore, E.~E., Epstein, R.~I. and Ho, C. 1992, in \it Gamma-Ray Bursts, \er
  eds. Paciesas, W. S. and Fishman, G. J., (AIP Conf. Proc. 265, New York) 
  p.~158
\reference{}  
Fenimore, E.~E., Epstein, R.~I. and Ho, C. 1993, \aaps \vol{97}{59}
\reference{}  
Fenimore, E.~E., Klebesadel, R.~W. \& Laros, J.~G. 1996, \apj\vol{460}{964}
\reference{}  
Fenimore, E.~E., Madras, C. \& Nayakshin, S. 1997, \apj\vol{473}{998}
\reference{}
Fishman, G.~J., et al. 1994, \apjs \vol{92}{229}
\reference{}  
Goodman, J. 1986, \apjl \vol{308}{L47}
\reference{}  
Gould, R.~J. and Schreder, G.~P. 1967, \pr \vol{155}{1404}
\reference{}
Gradshteyn, I.~S. and Ryzhik, I.~M. 1980, \it Table of Integrals, Series
  and Products\er , (Academic Press, New York)
\reference{}
Hakkila, J., et al. 1995, \apj\vol{454}{134}
\reference{}
Hanlon, L.~O., et al. 1994, \aap\vol{285}{161}
\reference{}
Harding, A.~K. 1994, in \it Proc. 2nd Compton Symp., \rm ed. Fichtel, C.,
   et al., (AIP Conf. Proc. 304, New York), p.~30
\reference{} 
Harding, A.~K. and Baring, M.~G. 1994, in \it Gamma-Ray Bursts, \er eds.  
  Fishman, G.~J., Hurley, K. and Brainerd, J.~J. (AIP Conf. Proc. 307, 
  New York) p.~520
\reference{}
Hurley, K., et al. 1994, \nat\vol{372}{652}
\reference{}
Hurley, K. 1996, in \it TeV Gamma-Ray Astrophysics, \rm eds. V\"olk, H.~J.
   and Aharonian, F.~A. (Kluwer, Dordrecht) p.~43.
\reference{}                                                  
Jauch, M.~M., and Rohrlich, F. 1980, \it The Theory of Photons and Electrons
  \rm (2nd edn. Springer, Berlin)             
\reference{}  
Krolik, J.~H. and Pier, E.~A. 1991, \apj \vol{373}{277}
\reference{}
Kwok, P.~W. et al. 1993, in \it Compton Gamma-Ray Observatory\er , 
  eds. Friedlander, M., Gehrels, N., and Macomb, D. (AIP Conf. Proc. 280,
  New York) p.~855 
\reference{}  
Mannheim, K., Hartmann, D. \& Funk, B. 1996, \apj\vol{467}{532}
\reference{}  
Meegan, C., et al. 1992, \nat \vol{355}{143}    
\reference{}  
Meegan, C., et al. 1996, \apjs\vol{106}{65}   
\reference{} 
Mitrofanov, I.~G. 1995, \apss \vol{231}{103}
\reference{}
M\'esz\'aros, P. \& Rees, M.~J. 1992, \apj\vol{397}{570}
\reference{}
M\'esz\'aros, P. \& Rees, M.~J. 1993, \apj\vol{405}{278}
\reference{}
Narayan, R., Piran, T., \& Shemi, A. 1991, \apjl\vol{379}{L17}
\reference{}
Nemiroff, R.~J., et al. 1994, \apj \vol{423}{432}
\reference{}
Nolan, P.~L. et al. 1983, in \it Positron-Electron Pairs in Astrophysics, \rm
   eds. Burns, M.~L., Harding, A.~K., and Ramaty, R., (AIP Conf. Proc. 101,
   New York) p.~59
\reference{}
Norris, J.~P., et al. 1994, \apj \vol{424}{540}
\reference{}  
Paczy\'nski, B. 1986, \apjl \vol{308}{L43}   
\reference{}
Piran, T. \& Shemi, A. 1993, \apjl\vol{403}{L67}
\reference{}
Piran, T., Shemi, A. \& Narayan, R. 1993, \mnras\vol{263}{861}
\reference{} 
Rees, M.~J. 1966, \nat \vol{211}{468}
\reference{} 
Rees, M.~J. \& M\'esz\'aros, P. 1992, \mnras \vol{258}{41P}
\reference{} 
Ryan et al. 1994, \apjl \vol{422}{L67}
\reference{} 
Rybicki, G.~B. and Lightman, A.~P. 1979, \it Radiative Processes in
  Astrophysics \rm (Wiley, New York)
\reference{}  
Schaefer, B. E., et al. 1992, \apjl \vol{393}{L51}
\reference{}  
Schaefer, B. E., et al. 1994, \apjs \vol{92}{285}
\reference{}  
Schmidt, W. K. H. 1978, \nat \vol{271}{525}
\reference{}  
Schneid, E.~J., et al. 1992, \aapl\vol{255}{L13}
\reference{}  
Schneid, E.~J., et al. 1995, \apj\vol{453}{95}
\reference{}  
Schneid, E.~J., et al. 1996, in \it Gamma-Ray Bursts, \er eds.  
  Kouveliotou, C., Briggs, M.~F., and Fishman, G.~J. (AIP Conf. Proc. 384, 
  New York) p.~253.
\reference{}
Share, G.~H., et al. 1986, \asr\vol{6(4)}{15}
\reference{}
Shemi, A. \& Piran, T. 1990, \apjl\vol{365}{L55}
\reference{}
Sommer, M., et al. 1994, \apjl \vol{422}{L63}
\reference{}  
Stecker, F.~W. \& De Jager, O.~C. 1996, \ssr\vol{75}{401}
\reference{}  
Stepney, S., and Guilbert, P.~W. 1983, \mnras\vol{204}{1269}
\reference{}  
Svensson, R. 1987, \mnras\vol{227}{403}
\reference{}
Usov, V.~V. 1992, \nat\vol{357}{472}
\reference{} 
Weaver, T.~A. 1976, \pra\vol{13}{1563}
\reference{}
Winkler, C. et al. 1993 in \it Proc. Compton Observatory
   Symposium,\rm\  eds. Friedlander, M., Gehrels, N. and Macomb, D.~J.
   (AIP Conf. Proc. 280, New York) p.~845.
\reference{}
Winkler, C. et al. 1995, \aap\vol{302}{765}
\reference{}
Woosley, S.~E. 1993, \apj\vol{405}{273}
\reference{}
Yodh, G. B. 1996, \ssr\vol{75}{199}
\reference{}  
Zdziarski, A.~A. 1984, \aap\vol{134}{301}
\end{references}
\end{document}